\documentstyle[prd,floats,aps,epsf,twocolumn]{revtex}
\input epsf

\newcommand {\abs}[1]{\mid \!\! #1 \!\! \mid}
\def \cal {\mathcal}
\def \met {{\,/\!\!\!\!E_{T}}}

\def \begineq {\begin{equation}}
\def \endeq {\end{equation}}
\def \Nmin {{N_{\rm min}}} 
\def \scriptR {\mbox{${\cal R}$}}
\def \scriptP {\mbox{${\cal P}$}}
\def \gothicP {\tilde{\scriptP}}
\def \ltapprox {\,\raisebox{-0.6ex}{$\stackrel{<}{\sim}$}\,}
\def \gtapprox {\,\raisebox{-0.6ex}{$\stackrel{>}{\sim}$}\,}
\def \Sleuth   {{Sleuth}}
\def \Sherlock {\Sleuth}
\def \hse {{\small{hse}}}
\def \hser {{\small{hser}}}
\def \ellgamma {\mbox{($\ell/\gamma$)}}
\def \jj {\,2j}
\def \jjj {\,3j}
\def \jjjj {\,4j}
\def \jjjjj {\,5j}
\def \jjjjjj {\,6j}

\newcommand{\guess}[1]
{\underline{#1}}
\def \gaugino {\tilde{\chi}}
\def \gravitino {\tilde{G}}
\def \selectron {\tilde{e}}

\def \stau {\tilde{\tau}}
\def \slepton {\tilde{\ell}}
\def \sneutrino {\tilde{\nu}}
\def \scalartop {\tilde{t}}
\def \gluino {\tilde{g}}
\def \technipi {\pi_T}
\def \technirho {\rho_T}
\def \techniomega {\omega_T}

\def \numberOfFinalStates {thirty-two}

\newcommand{\ttbar}{$t{\bar t}$}

\def \BKwidetext {\widetext}
\def \BKnarrowtext {\narrowtext}
\def \BKtwocolumn {\twocolumn}

\newcommand{\dofig}[4]
{\begin{figure}[htbp]
\begin{center}
\leavevmode
\hbox{%
\epsfxsize=#2
\epsfbox{#1}}
\caption{#3}
\label{#4}
\end{center}
\end{figure}}
\newcommand{\doxfig}[4]
{\begin{figure}[htbp]
\begin{center}
\leavevmode
\hbox{%
\epsfxsize=#2
\epsfbox{#1}}
\vspace*{.5cm}
\caption{#3}
\label{#4}
\end{center}
\end{figure}}

\def \scriptPemumet {$0.14$}
\def \scriptPemumetsigma {$+1.08$}
\def \scriptPemumetj {$0.45$}
\def \scriptPemumetjsigma {$+0.13$}
\def \scriptPemumetjj {$0.31$}
\def \scriptPemumetjjsigma {$+0.50$}
\def \scriptPemumetjjj {$0.71$}
\def \scriptPemumetjjjsigma {$-0.55$}

\def \scriptPeejj {$0.72$}
\def \scriptPeejjsigma {$-0.58$}
\def \scriptPeejjj {$0.61$}
\def \scriptPeejjjsigma {$-0.28$}
\def \scriptPeejjjj {$0.04$}
\def \scriptPeejjjjsigma {$+1.75$}

\def \scriptPeemetjj {$0.68$}
\def \scriptPeemetjjsigma {$-0.47$}
\def \scriptPeemetjjj {$0.36$}
\def \scriptPeemetjjjsigma {$+0.36$}
\def \scriptPeemetjjjj {$0.06$}
\def \scriptPeemetjjjjsigma {$+1.55$}

\def \scriptPmumujj {$0.08$}
\def \scriptPmumujjsigma {$+1.41$}
\def \scriptPmumujjj {$1.00$}

\def \scriptPZjj {$0.52$}
\def \scriptPZjjsigma {$-0.05$}
\def \scriptPZjjj {$0.71$}
\def \scriptPZjjjsigma {$-0.55$}
\def \scriptPZjjjj {$0.83$}
\def \scriptPZjjjjsigma {$-0.95$}
\def \scriptPZjjjjj {$1.00$}

\def \scriptPemetjj {$0.76$}
\def \scriptPemetjjsigma {$-0.71$}
\def \scriptPemetjjj {$0.17$}
\def \scriptPemetjjjsigma {$+0.95$}
\def \scriptPemetjjjj {$0.13$}
\def \scriptPemetjjjjsigma {$+1.13$}

\def \scriptPWjj {$0.29$}
\def \scriptPWjjsigma {$+0.55$}
\def \scriptPWjjj {$0.23$}
\def \scriptPWjjjsigma {$+0.74$}
\def \scriptPWjjjj {$0.53$}
\def \scriptPWjjjjsigma {$-0.08$}
\def \scriptPWjjjjj {$0.81$}
\def \scriptPWjjjjjsigma {$-0.88$}
\def \scriptPWjjjjjj {$0.22$}
\def \scriptPWjjjjjjsigma {$+0.77$}

\def \scriptPeeg {$0.88$}
\def \scriptPeegsigma {$-1.17$}

\def \scriptPeegmet {$0.23$}
\def \scriptPeegmetsigma {$+0.74$}
\def \scriptPZg {$0.84$}
\def \scriptPZgsigma {$-0.99$}
\def \scriptPZgj {$0.63$}
\def \scriptPZgjsigma {$-0.33$}
\def \scriptPeee {$0.89$}
\def \scriptPeeesigma {$-1.23$}

\def \scriptPegg {$0.66$}
\def \scriptPeggsigma {$-0.41$}
\def \scriptPeggj {$0.21$}
\def \scriptPeggjsigma {$+0.81$}
\def \scriptPeggjj {$0.30$}
\def \scriptPeggjjsigma {$+0.52$}
\def \scriptPWgg {$0.18$}
\def \scriptPWggsigma {$+0.92$}
\def \scriptPggg {$0.41$}
\def \scriptPgggsigma {$+0.23$}

\def \scriptPWjjjTop {$0.12$}

\def \scriptPWjjjjTop {$0.18$}

\def \scriptPWjjjjjTop {$0.37$}

\def \scriptPWjjjjjjTop {$0.09$}

\def \twiddleScriptPvalue {$0.89$}
\def \twiddleScriptPvalueSigma {$-1.23$}

\begin{document}

\onecolumn
\title{A Quasi-Model-Independent Search for New Physics at Large Transverse Momentum}

%
\author{                                                                      
B.~Abbott,$^{56}$                                                             
A.~Abdesselam,$^{11}$                                                         
M.~Abolins,$^{49}$                                                            
V.~Abramov,$^{24}$                                                            
B.S.~Acharya,$^{16}$                                                          
D.L.~Adams,$^{58}$                                                            
M.~Adams,$^{36}$                                                              
G.A.~Alves,$^{2}$                                                             
N.~Amos,$^{48}$                                                               
E.W.~Anderson,$^{41}$
% ...
\\                                                                            
\vskip 0.30cm                                                                 
\centerline{(D\O\ Collaboration)}                                             
\vskip 0.30cm                                                                 
}                                                                             
\address{                                                                     
\centerline{$^{1}$Universidad de Buenos Aires, Buenos Aires, Argentina}       
\centerline{$^{2}$LAFEX, Centro Brasileiro de Pesquisas F{\'\i}sicas,         
                  Rio de Janeiro, Brazil}                                     
\centerline{$^{3}$Universidade do Estado do Rio de Janeiro,                   
                  Rio de Janeiro, Brazil}                                     
\centerline{$^{4}$Institute of High Energy Physics, Beijing,                  
                  People's Republic of China}                                 
\centerline{$^{5}$Universidad de los Andes, Bogot\'{a}, Colombia}             
\centerline{$^{6}$Charles University, Prague, Czech Republic}                 
\centerline{$^{7}$Institute of Physics, Academy of Sciences, Prague,          
                  Czech Republic}                                             
\centerline{$^{8}$Universidad San Francisco de Quito, Quito, Ecuador}         
\centerline{$^{9}$Institut des Sciences Nucl\'eaires, IN2P3-CNRS,             
                  Universite de Grenoble 1, Grenoble, France}                 
\centerline{$^{10}$CPPM, IN2P3-CNRS, Universit\'e de la M\'editerran\'ee,     
                  Marseille, France}                                          
\centerline{$^{11}$Laboratoire de l'Acc\'el\'erateur Lin\'eaire,              
                  IN2P3-CNRS, Orsay, France}                                  
\centerline{$^{12}$LPNHE, Universit\'es Paris VI and VII, IN2P3-CNRS,         
                  Paris, France}                                              
\centerline{$^{13}$DAPNIA/Service de Physique des Particules, CEA, Saclay,    
                  France}                                                     
\centerline{$^{14}$Panjab University, Chandigarh, India}                      
\centerline{$^{15}$Delhi University, Delhi, India}                            
\centerline{$^{16}$Tata Institute of Fundamental Research, Mumbai, India}     
\centerline{$^{17}$Seoul National University, Seoul, Korea}                   
\centerline{$^{18}$CINVESTAV, Mexico City, Mexico}                            
\centerline{$^{19}$FOM-Institute NIKHEF and University of                     
                  Amsterdam/NIKHEF, Amsterdam, The Netherlands}               
\centerline{$^{20}$University of Nijmegen/NIKHEF, Nijmegen, The               
                  Netherlands}                                                
\centerline{$^{21}$Institute of Nuclear Physics, Krak\'ow, Poland}            
\centerline{$^{22}$Institute for Theoretical and Experimental Physics,        
                   Moscow, Russia}                                            
\centerline{$^{23}$Moscow State University, Moscow, Russia}                   
\centerline{$^{24}$Institute for High Energy Physics, Protvino, Russia}       
\centerline{$^{25}$Lancaster University, Lancaster, United Kingdom}           
\centerline{$^{26}$Imperial College, London, United Kingdom}                  
\centerline{$^{27}$University of Arizona, Tucson, Arizona 85721}              
\centerline{$^{28}$Lawrence Berkeley National Laboratory and University of    
                  California, Berkeley, California 94720}                     
\centerline{$^{29}$University of California, Davis, California 95616}         
\centerline{$^{30}$California State University, Fresno, California 93740}     
\centerline{$^{31}$University of California, Irvine, California 92697}        
\centerline{$^{32}$University of California, Riverside, California 92521}     
\centerline{$^{33}$Florida State University, Tallahassee, Florida 32306}      
\centerline{$^{34}$University of Hawaii, Honolulu, Hawaii 96822}              
\centerline{$^{35}$Fermi National Accelerator Laboratory, Batavia,            
                   Illinois 60510}                                            
\centerline{$^{36}$University of Illinois at Chicago, Chicago,                
                   Illinois 60607}                                            
\centerline{$^{37}$Northern Illinois University, DeKalb, Illinois 60115}      
\centerline{$^{38}$Northwestern University, Evanston, Illinois 60208}         
\centerline{$^{39}$Indiana University, Bloomington, Indiana 47405}            
\centerline{$^{40}$University of Notre Dame, Notre Dame, Indiana 46556}       
\centerline{$^{41}$Iowa State University, Ames, Iowa 50011}                   
\centerline{$^{42}$University of Kansas, Lawrence, Kansas 66045}              
\centerline{$^{43}$Kansas State University, Manhattan, Kansas 66506}          
\centerline{$^{44}$Louisiana Tech University, Ruston, Louisiana 71272}        
\centerline{$^{45}$University of Maryland, College Park, Maryland 20742}      
\centerline{$^{46}$Boston University, Boston, Massachusetts 02215}            
\centerline{$^{47}$Northeastern University, Boston, Massachusetts 02115}      
\centerline{$^{48}$University of Michigan, Ann Arbor, Michigan 48109}         
\centerline{$^{49}$Michigan State University, East Lansing, Michigan 48824}   
\centerline{$^{50}$University of Nebraska, Lincoln, Nebraska 68588}           
\centerline{$^{51}$Columbia University, New York, New York 10027}             
\centerline{$^{52}$University of Rochester, Rochester, New York 14627}        
\centerline{$^{53}$State University of New York, Stony Brook,                 
                   New York 11794}                                            
\centerline{$^{54}$Brookhaven National Laboratory, Upton, New York 11973}     
\centerline{$^{55}$Langston University, Langston, Oklahoma 73050}             
\centerline{$^{56}$University of Oklahoma, Norman, Oklahoma 73019}            
\centerline{$^{57}$Brown University, Providence, Rhode Island 02912}          
\centerline{$^{58}$University of Texas, Arlington, Texas 76019}               
\centerline{$^{59}$Texas A\&M University, College Station, Texas 77843}       
\centerline{$^{60}$Rice University, Houston, Texas 77005}                     
\centerline{$^{61}$University of Virginia, Charlottesville, Virginia 22901}   
\centerline{$^{62}$University of Washington, Seattle, Washington 98195}       
}                                                                             

\maketitle
\vskip 10pt

{\samepage
{\bf
\begin{center}
Abstract
\end{center}
}
\begin{center}
\begin{minipage}{.8\textwidth}
{\small 

We apply a quasi-model-independent strategy (``\Sleuth'') to search for new high $p_T$ physics in $\approx 100$~pb$^{-1}$ of $p\bar{p}$ collisions at $\sqrt{s} = 1.8$~TeV collected by the D\O\ experiment during 1992--1996 at the Fermilab Tevatron.  Over \numberOfFinalStates\ $e\mu X$, $W$+jets-like, $Z$+jets-like, and $\ellgamma\ellgamma\ellgamma X$ exclusive final states are systematically analyzed for hints of physics beyond the standard model.  Simultaneous sensitivity to a variety of models predicting new phenomena at the electroweak scale is demonstrated by testing the method on a particular signature in each set of final states.  No evidence of new high $p_T$ physics is observed in the course of this search, and we find that 89\% of an ensemble of hypothetical similar experimental runs would have produced a final state with a candidate signal more interesting than the most interesting observed in these data.

}
\end{minipage}
\end{center}
}

\vskip 2.0in

\BKtwocolumn

\tableofcontents

\section{Introduction}

The standard model is an impressive theory, accurately predicting, or at least accommodating, the results of nearly all particle physics experiments to date.  It is generally accepted, however, that there is good reason to believe that hints of new physics are likely to appear at or around the energy scale of 1 TeV.  

Electroweak symmetry is broken in the standard model when a scalar field (the Higgs field) acquires a vacuum expectation value.  Since the quantum corrections to the renormalized mass squared of a scalar field grow as the square of the heaviest energy scale in the theory (naively the Planck scale, of order $10^{19}$~GeV), and since the mass of the standard model Higgs boson is of the order of a few hundred GeV, a fine-tuning at the level of one part in $10^{16}$ appears to be required to keep the Higgs mass at the electroweak scale.  

Two of the most popular solutions to this hierarchy problem are supersymmetry~\cite{SusyReview} and strong dynamics~\cite{StrongDynamicsReview}.  In their most general form these classes of models are capable of ``predicting'' any of many different signatures, depending upon the values that are chosen for the model's parameters.  Previous searches for these signals have fought to strike a balance between the simultaneous desires to assume as little as possible about the signal and yet achieve ``optimal sensitivity'' to more specific signals.  These are necessarily contradictory objectives.  

Many new phenomena have been predicted in addition to those resulting from these proposed solutions to the hierarchy problem.  Among them are leptoquarks, proposed in an attempt to explain the relationship between quarks and leptons in the standard model and appearing in many grand unified theories; composite quarks and leptons, in case the ``fundamental'' particles of the standard model turn out not to be fundamental at scales $\ltapprox 10^{-18}$ meters; a fourth generation of quarks or leptons; excited quarks and leptons, in analogy to the excited states of hadrons observed at much lower energies; new heavy gauge bosons, arising from additional gauge symmetries in models extending the SU(3)$_C\times$SU(2)$_L\times$U(1)$_Y$ of the standard model; and many others.  Of course, Nature may have other ideas.  The CDF and D\O\ collaborations have performed many searches on the data collected during Run I of the Fermilab Tevatron, but have we looked in all the right places?

{\doxfig{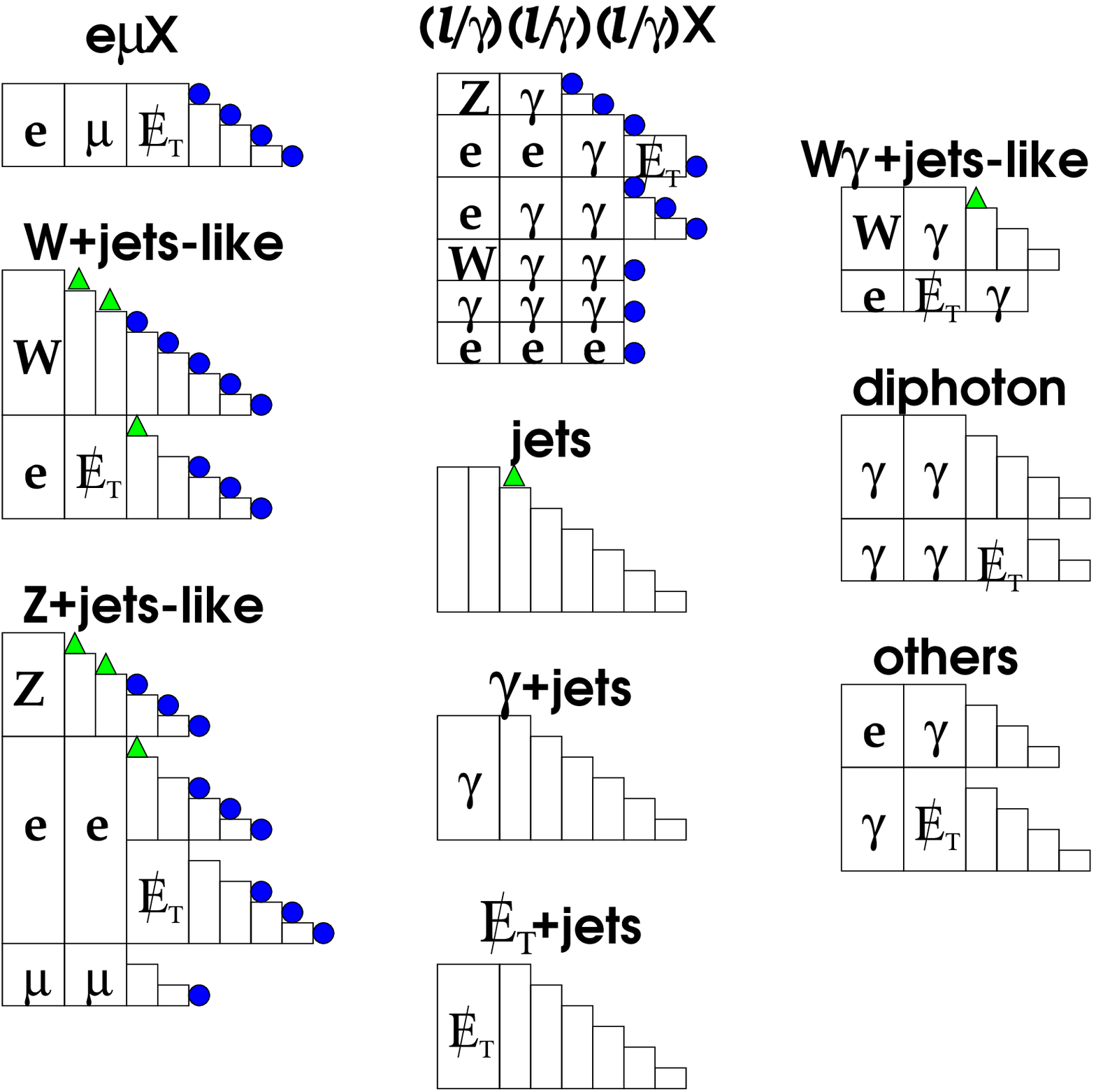}
 {3.5in} {A diagram showing the final states populated in D\O\ data in Run I.  Each row in a given column represents the final state defined by the objects in that row; to reduce clutter, jets are represented by an empty rectangle, rather than by a rectangle containing ``$j$.''  Reading down the left column are the final states $e\mu \met$, $e\mu \met j$, $e\mu \met \jj$, $e\mu \met \jjj$, $W$, $Wj$, $W\jj$, and so on.   Rows with triangles (e.g., $W$ and $Wj$) indicate final states analyzed previously by D\O\ in a manner similar to the strategy we use here, but without using \Sleuth; rows with filled circles indicate final states analyzed with \Sleuth.  The remaining rows show populated final states not discussed in this article.} {fig:SherlockFinalStatesGraph} }

Figure~\ref{fig:SherlockFinalStatesGraph}
diagrams the final states that are populated (i.e., that contain events) in the D\O\ Run I data.  In this article we undertake a systematic and quasi-model-independent analysis of many of these exclusive final states, in the hope of finding some evidence for physics beyond the standard model.  

In Refs.~\cite{SherlockPRD,KnutesonThesis} we introduced a quasi-model-independent search strategy (``\Sherlock''), designed to systematically search for new high $p_T$ physics at any collider experiment sensitive to physics at the electroweak scale, and applied it to all events in the D\O\ data containing one or more electrons and one or more muons ($e\mu X$).  Considering again Fig.~\ref{fig:SherlockFinalStatesGraph}, we see that the number of final states within $e\mu X$ is a small fraction of the total number of final states populated by the D\O\ Run I data.  If there is indeed a signal in the data, our chances of finding it grow proportionally to the number of final states considered.  

In this article we present a systematic analysis of \numberOfFinalStates\ of these final states --- those marked with a filled circle in Fig.~\ref{fig:SherlockFinalStatesGraph}.  A large number of unpopulated final states with additional objects are analyzed implicitly; e.g., $ee\mu\met$ and $e\mu\met\gamma$ are among a host of unpopulated final states analyzed within the context of $e\mu X$.

The notation we use to label final states may require explanation.  Electrons and muons are confidently identified with the D\O\ detector on an event-by-event basis, but taus are not; $\ell$ and the word ``lepton'' will therefore denote an electron ($e$) or a muon ($\mu$) in this article.  We use the composite symbol $\ellgamma$ to denote an electron, muon, or photon. $X$ will denote zero or more objects, and $(nj)$ will denote zero or more jets.  Any inclusive final state [i.e., any state whose label includes the symbol $X$ or $(nj)$] will refer to the physics objects actually reconstructed in the detector.  Thus $ee\jj (nj)$ denotes the set of all events with two electrons and two or more jets.  Any exclusive final state is defined according to the rules in Appendix~\ref{section:finalStateDefinitions}.  For example, since these rules include a prescription for identifying a $Z$ boson from two charged leptons of the same flavor, we use $ee\jj$ to denote the set of all events with two electrons and two jets having $m_{ee}$ substantially different from $M_Z$, while events with two electrons and two jets having $m_{ee}\approx M_Z$ fall within the final state $Z\jj$.

We begin in Sec.~\ref{section:Review} by providing a brief review of the \Sherlock\ search strategy and algorithm, and describing a slight change from the method advanced in Ref.~\cite{SherlockPRD}.  In Sec.~\ref{section:OtherFinalStates} we discuss eight final states already analyzed by D\O\ in a manner similar to \Sleuth, and motivate the final states to be considered in this article.  In Sec.~\ref{section:WJ} we describe the analysis of the $W$+jets-like final states --- events containing a single lepton, missing transverse energy ($\met$), and two or more jets.  In Sec.~\ref{section:ZJ} we present the analysis of the $Z$+jets-like final states --- events containing two leptons and two or more jets.  In Sec.~\ref{section:zoo} we analyze the final states containing several objects, at least three of which are either an electron, muon, or photon [$\ellgamma\ellgamma\ellgamma X$].  In Sec.~\ref{section:Summary} we present the combined results of all of these final states.  Section~\ref{section:Conclusions} contains our conclusions.

\section{\Sherlock}
\label{section:Review}

In this section we provide for completeness a brief overview of the \Sherlock\ algorithm, which is described in detail in Ref.~\cite{SherlockPRD}, and its application to the final states $e\mu X$.

\subsection{Search strategy}

We partition our data into exclusive final states, using standard identification criteria to identify electrons, muons, photons, jets, missing transverse energy, and $W$ and $Z$ bosons.  Although experimental realities will occasionally force slight modifications to these criteria, a set of standard definitions determined {\it a priori} is used wherever possible.   

The production and subsequent decay of massive, non-standard-model particles typically results in events containing objects with large transverse momentum ($p_T$).  For each exclusive final state we therefore consider the small set of variables defined by Table~\ref{tbl:VariableRules}.  In order to reduce backgrounds from QCD processes that produce extra jets from gluon radiation, or two energetic jets through a $t$-channel exchange diagram, the notation $\sum'{p_T^j}$ is shorthand for $p_T^{j_1}$ if the final state contains only one jet, $\sum_{i=2}^n{p_T^{j_i}}$ if the final state contains $n \geq 2$ jets, and $\sum_{i=3}^n{p_T^{j_i}}$ if the final state contains $n$ jets and nothing else, with $n \geq 3$.  Leptons and missing transverse energy that are reconstructed as decay products of $W$ or $Z$ bosons are not considered separately in the left-hand column.  Thus the variables corresponding to the final state $Wjj$, for example, are $p_T^W$ and $\sum'{p_T^j}$; $p_T^\ell$ and $\met$ are not used, even though the events necessarily contain a lepton and missing transverse energy, since the lepton and missing transverse energy have been combined into the $W$ boson.   Since D\O's muon momentum resolution in Run I was modest, we define $\sum{p_T^\ell} = \sum{p_T^e}$ for events with one or more electrons and one or more muons, and we determine the missing transverse energy from the transverse energy summed in the calorimeter, which includes the $p_T$ of electrons, but only a negligible fraction of the $p_T$ of muons.  When there are exactly two objects in an event (e.g., one $Z$ boson and one jet), their $p_T$ values are expected to be nearly equal, and we therefore use the average $p_T$ of the two objects.  When there is only one object in an event (e.g., a single $W$ boson), we use no variables, and simply perform a counting experiment.  We expect evidence for new physics to appear in the high tails of these distributions.

\begin{table}[htb]
\centering
\begin{tabular}{cc}
If the final state includes & then consider the variable \\ \hline
$\met$ & $\met$ \\ 
one or more charged leptons & $\sum{p_T^\ell}$ \\ 
one or more electroweak bosons & $\sum{p_T^{\gamma/W/Z}}$ \\ 
one or more jets & $\sum'{p_T^j}$ \\ 
\end{tabular}
\caption{A quasi-model-independently motivated list of interesting variables for any final state.  The set of variables to consider for any particular final state is the union of the variables in the second column for each row that pertains to that final state.}
\label{tbl:VariableRules}
\end{table}

 \def \EMMDA {39}
 \def \EMMFA {18.4$\pm$1.4}
 \def \EMMTT {0.011$\pm$0.003}
 \def \EMMDY {0.5$\pm$0.2}
 \def \EMMWW {3.9$\pm$1.0}
 \def \EMMZT {25.6$\pm$6.5}
 \def \EMMTOT {48.5$\pm$7.6}
 \def \EMMJDA {13}
 \def \EMMJFA {8.7$\pm$1.0}
 \def \EMMJTT {0.4$\pm$0.1}
 \def \EMMJDY {0.1$\pm$0.03}
 \def \EMMJWW {1.1$\pm$0.3}
 \def \EMMJZT {3$\pm$0.8}
 \def \EMMJTOT {13.2$\pm$1.5}
 \def \EMMJJDA {5}
 \def \EMMJJFA {2.7$\pm$0.6}
 \def \EMMJJTT {1.8$\pm$0.5}
 \def \EMMJJDY {0.012$\pm$0.006}
 \def \EMMJJWW {0.18$\pm$0.05}
 \def \EMMJJZT {0.5$\pm$0.2}
 \def \EMMJJTOT {5.2$\pm$0.8}
 \def \EMMJJJDA {1}
 \def \EMMJJJFA {0.4$\pm$0.2}
 \def \EMMJJJTT {0.7$\pm$0.2}
 \def \EMMJJJDY {0.005$\pm$0.004}
 \def \EMMJJJWW {0.032$\pm$0.009}
 \def \EMMJJJZT {0.07$\pm$0.05}
 \def \EMMJJJTOT {1.3$\pm$0.3}
 \def \EMMJJJJDA {0}
 \def \EMMJJJJTT {0.16$\pm$0.04}
 \def \EMMJJJJDY {0.002$\pm$0.003}
 \def \EMMJJJJWW {0.004$\pm$0.002}
 \def \EMMJJJJZT {0.02$\pm$0.03}
 \def \EMMJJJJTOT {0.2$\pm$0.2}
 \def \EMMJJJJJDA {0}
 \def \EMMJJJJJFA {0$\pm$0.2}
 \def \EMMJJJJJTT {0.025$\pm$0.007}
 \def \EMMJJJJJDY {0$\pm$0.003}
 \def \EMMJJJJJWW {0$\pm$0.0006}
 \def \EMMJJJJJZT {0$\pm$0.03}
 \def \EMMJJJJJTOT {0.025$\pm$0.2}
 \def \EMXDA {58}
 \def \EMXFA {30.2$\pm$1.8}
 \def \EMXTT {3.1$\pm$0.5}
 \def \EMXDY {0.7$\pm$0.1}
 \def \EMXWW {5.2$\pm$0.8}
 \def \EMXZT {29.2$\pm$4.5}
 \def \EMXTOT {68.3$\pm$5.7}

\def \EMMJZT {$3.0 \pm 0.8$} 
\BKwidetext
\begin{table*}[bht]
\centering
\small
\begin{tabular}{cccccccc}
Data set & Fakes & $Z\rightarrow\tau\tau$ & $\gamma^*\rightarrow\tau\tau$ & $WW$ & \ttbar & Total & Data \\ \hline
$e\mu\met$ & \EMMFA & \EMMZT & \EMMDY & \EMMWW & \EMMTT & \EMMTOT & \EMMDA \\
$e\mu\met j$ & \EMMJFA & \EMMJZT & \EMMJDY & \EMMJWW & \EMMJTT & \EMMJTOT & \EMMJDA \\
$e\mu\met \jj$ & \EMMJJFA & \EMMJJZT & \EMMJJDY & \EMMJJWW & \EMMJJTT & \EMMJJTOT & \EMMJJDA \\
$e\mu\met \jjj$ &\EMMJJJFA & \EMMJJJZT & \EMMJJJDY & \EMMJJJWW & \EMMJJJTT & \EMMJJJTOT & \EMMJJJDA \\ \hline
$e\mu X$ & \EMXFA & \EMXZT & \EMXDY & \EMXWW & \EMXTT & \EMXTOT & \EMXDA \\
\end{tabular}
\caption{The numbers of expected background events for the populated final states within $e\mu X$. The uncertainties in $e\mu X$ are smaller than in the sum of the individual background contributions obtained from Monte Carlo because of an uncertainty in the numbers of extra jets arising from initial and final state radiation in the exclusive channels.}
\label{tbl:emuXexpectedBackgrounds1}
\end{table*}
\BKnarrowtext

\subsection{Algorithm}
\label{section:SleuthAlgorithm}

Although the details of the algorithm are complicated, the concept is straightforward.  What is needed is a data sample, a set of events modeling each background process $i$, and the number of background events $\hat{b}_i \pm \delta \hat{b}_i$ from each background process expected in the data sample.  From these we determine the region of greatest excess and quantify the degree to which that excess is interesting.

The algorithm, applied to each individual final state, consists of seven steps.  
\begin{enumerate}
\item We begin by constructing a mapping from the $d$-dimensional variable space defined by Table~\ref{tbl:VariableRules} into the $d$-dimensional unit box (i.e., $[0,1]^d$) that flattens the background distribution, and we use this to map the data into the unit box.  This change of variable space greatly simplifies the subsequent analysis.  
\item Central to this algorithm is the notion of a ``region'' about a set of $1 \leq N \leq N_{\rm data}$ data points, defined as the volume within the unit box closer to one of the data points in the set than to any of the other data points in the sample.  The arrangement of data points themselves thus determines the regions.  A region containing $N$ data points is called an $N$-region.  
\item Each region $R$ contains an expected number of background events $\hat{b}_R$, equal to the volume of the region $\times$ the total number of background events expected, and an associated systematic error $\delta\hat{b}_R$, which varies within the unit box according to the systematic errors assigned to each contribution to the background estimate.  We can therefore compute the probability $p_N^R$ that the background in the region fluctuates up to or beyond the observed number of events.  This probability is our first measure of the degree of interest of a particular region.
\item The rigorous definition of regions reduces the number of candidate regions from infinity to $\approx 2^{N_{\rm data}}$.  Imposing explicit criteria on the regions that the algorithm is allowed to consider further reduces the number of candidate regions.  (See Sec.~\ref{section:regionCriteria}.)  Our assumption that new physics is most likely to appear at high $p_T$ translates to a preference for regions in a particular corner of the unit box; criteria are thus constructed to define ``reasonable'' discovery regions.  The number of remaining candidate regions is still sufficiently large that an exhaustive search is impractical, and a heuristic is employed to search for regions of excess.  In the course of this search the $N$-region $\scriptR_N$ for which $p_N^R$ is minimum is determined for each $N$, and $p_N = \min_R{(p_N^R)}$ is noted.  
\item In any reasonably-sized data set, there will always be regions in which the probability for $b_R$ to fluctuate up to or above the observed number of events is small.  The relevant issue is how often this will happen in an ensemble of {\em hypothetical similar experiments} (\hse's).  This question can be answered by performing these \hse's; i.e., generating random events drawn from the background distribution, and computing $p_N$ by following steps (1)--(4).  Generating many such \hse's, we can determine the fraction $P_N$ of \hse's in which the $p_N$ found for the \hse\ is smaller than the $p_N$ observed in the data.  
\item We define $P$ and $\Nmin$ by $P=P_\Nmin=\min_N{(P_N)}$, and identify $\scriptR = \scriptR_\Nmin$ as the most interesting region in this final state.  
\item We use a second ensemble of \hse's to determine the fraction $\scriptP$ of \hse's in which $P$ found in the \hse\ is smaller than $P$ observed in the data.  The most important output of the algorithm is this single number $\scriptP$, which may loosely be said to be the ``fraction of hypothetical similar experiments in which you would see something as interesting as what you actually saw in the data.''  $\scriptP$ takes on values between zero and one, with values close to zero indicating a possible hint of new physics.  In computing $\scriptP$ we have rigorously taken into account the many regions that have been considered within this final state.
\end{enumerate}

The smallest $\scriptP$ found in the many different final states considered ($\scriptP_{\rm min}$) determines $\gothicP$, the ``fraction of {\em hypothetical similar experimental runs} (\hser's) that would have produced an excess as interesting as actually observed in the data,'' where an \hser\ consists of one \hse\ for each final state considered.  $\gothicP$ is calculated by simulating an ensemble of hypothetical similar experimental runs, and noting the fraction of these \hser's in which the smallest $\scriptP$ found is smaller than $\scriptP_{\rm min}$.  The correspondence between $\gothicP$ and $\scriptP_{\rm min}$ is determined to zeroth order by the number of final states considered in which the expected number of background events is $\gtapprox 1$, with ``smaller'' final states contributing first order corrections.  $\gothicP$ also takes on values between zero and one, and the potential presence of new high $p_T$ physics would be indicated by finding $\gothicP$ to be small.  The difference between $\gothicP$ and $\scriptP$ is that in computing $\gothicP$ we account for the many final states that have been considered.  $\gothicP$ can be translated into units of standard deviations ($\gothicP_{[\sigma]}$) by solving the unit conversion equation
\begineq
\label{eqn:gothicPsigma}
\gothicP = \frac{1}{\sqrt{2\pi}} \int_{\gothicP_{[\sigma]}}^{\infty} {e^{-t^2/2} \, dt}
\endeq
for $\gothicP_{[\sigma]}$.  A similar equation relates $\scriptP$ and $\scriptP_{[\sigma]}$.

\subsection{${\lowercase{e}}\mu X$}
\label{section:emuX}

In Ref.~\cite{SherlockPRD} we applied \Sherlock\ to the $e\mu X$ final states, using a data set corresponding to 108$\pm$6 pb$^{-1}$ of integrated luminosity.  We summarize those results here.  Appendix~\ref{section:SignalsThatMightAppearInemuX} contains examples of the types of new physics that might be expected to appear in these final states.

Events containing one or more isolated electrons and one or more isolated muons, each with $p_T>15$~GeV, are selected.  Global cleanup cuts are applied to remove events in which there was activity in the Main Ring, the accelerator that feeds the Tevatron, reducing the total number of events by $30\%$.  The dominant standard model and instrumental backgrounds to this data set are:
\begin{itemize}
\item{top quark pair production with $t\rightarrow W b$, and with both $W$ bosons decaying leptonically, one to $e\nu$ (or to $\tau\nu\rightarrow e\nu\nu\nu$) and one to $\mu\nu$ (or to $\tau\nu\rightarrow \mu\nu\nu\nu$);}
\item{$W$ boson pair production with both $W$ bosons decaying leptonically, one to $e\nu$ (or to $\tau\nu\rightarrow e\nu\nu\nu$) and one to $\mu\nu$ (or to $\tau\nu\rightarrow \mu\nu\nu\nu$);}
\item{$Z/\gamma^*\rightarrow \tau\tau \rightarrow e\mu\nu\nu\nu\nu$; and}
\item{instrumental (``fakes''):  $W$ production with the $W$ boson decaying to $\mu\nu$ and a radiated jet or photon being mistaken for an electron, or $b\bar{b}/c\bar{c}$ production with one heavy quark producing an isolated muon and the other being mistaken for an electron~\cite{searchForTopPRD}}.
\end{itemize}
The numbers of events expected for the various samples and data sets in the populated final states within $e\mu X$ are given in Table~\ref{tbl:emuXexpectedBackgrounds1}.  

Among the systematic errors in these and other final states is an uncertainty in the modeling of additional radiated jets.  Our consideration of exclusive final states makes this error more important than if inclusive final states were considered.  An uncertainty of $\approx 20\%$ in the number of expected events, obtained by comparing the jets radiated by various Monte Carlo programs, is added in quadrature to systematic errors from other sources to obtain the total systematic error quoted in Table~\ref{tbl:emuXexpectedBackgrounds1} and elsewhere.  Because final states are analyzed independently, and because the definition of $\gothicP$ depends only on the smallest $\scriptP$ found, we can, to first order, ignore the correlations of uncertainties among different final states. 

We demonstrated \Sherlock's sensitivity to new physics by showing that the method is able to find indications of the existence of $WW$ and $t\bar{t}$ production in these final states when the backgrounds are taken to include only $Z/\gamma^*\rightarrow \tau\tau$ and fakes.  Figure~\ref{fig:hyperplanes_tt_02} shows our sensitivity to $t\bar{t}$ in an ensemble of mock data samples when the backgrounds include $WW$ in addition to $Z/\gamma^*\rightarrow \tau\tau$ and fakes.  All samples with $\gothicP_{[\sigma]}>2.0$ appear in the rightmost bin.  We see that \Sleuth, with no knowledge of the top quark's existence or characteristics, finds $\gothicP_{[\sigma]}>2.0$ in over 25\% of the mock samples.  (For mock samples containing only $Z/\gamma^*\rightarrow\tau\tau$, fakes, and $WW$, the distribution is roughly Gaussian and centered at zero with unit width.)  After performing these sensitivity checks, we added all known standard model processes to the background estimate and searched for evidence of new high $p_T$ physics.  The result of this analysis is summarized in Table~\ref{tbl:emuXresults}.  No evidence of new physics is observed.

{\dofig{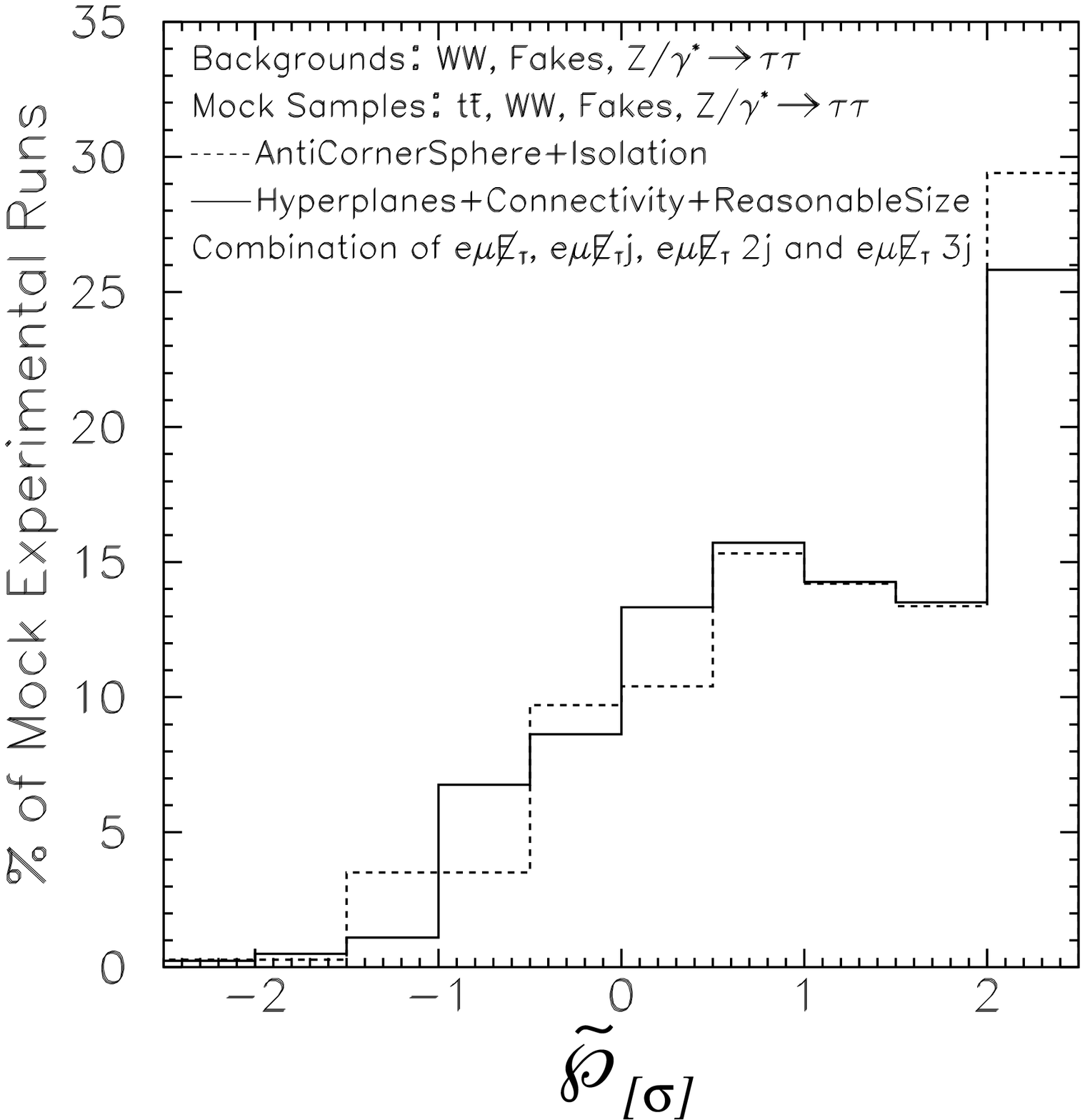} {3.5in} {Distribution of $\gothicP_{[\sigma]}$ in an ensemble of mock experimental runs on the four exclusive final states $e\mu\met$, $e\mu\met j$, $e\mu\met \jj$, and $e\mu\met \jjj$.  The background includes $Z/\gamma^*\rightarrow\tau\tau$, fakes, and $WW$.  The mock samples making up the distributions contain $t\bar{t}$ in addition to $Z/\gamma^*\rightarrow\tau\tau$, fakes, and $WW$.} {fig:hyperplanes_tt_02} }

\begin{table}[htb]
\centering
\begin{tabular}{cc}
Data set  	& $\scriptP$ 	\\ \hline
$e\mu\met$   	& 0.14	\\
$e\mu\met j$  	& 0.45	\\
$e\mu\met \jj$ 	& 0.31	\\
$e\mu\met \jjj$  & 0.71	\\ 
\end{tabular}
\caption{Summary of results on all final states within $e\mu X$ when all standard model backgrounds, including $t\bar{t}$, are included.  We note that {\em all} final states within $e\mu X$ have been analyzed, including (for example) $ee\mu\met$ and $e\mu\met\gamma$.  All final states within $e\mu X$ but not listed here are unpopulated, and have $\scriptP=1.00$.}
\label{tbl:emuXresults}
\end{table}

\subsection{Region criteria}
\label{section:regionCriteria}

Use of \Sherlock\ requires the specification of criteria that define the regions that \Sherlock\ is allowed to consider.  In the analysis of $e\mu X$ we imposed two criteria: {\em AntiCornerSphere} ($c^A$), which restricts the allowed region to be defined by those data points greater than a distance $r$ from the origin of the unit box, where $r$ is allowed to vary; and {\em Isolation} ($c^I$), which requires that there exist no data points outside the region that are closer than $\xi$ to any data point inside the region, where $\xi=1/(4N_{\rm data}^{1/d})$ is a characteristic distance between the $N_{\rm data}$ data points in the $d$-dimensional unit box.

For the analysis described in this article we use {\em Hyperplanes} ($c^H$), a criterion defined but not used in Ref.~\cite{SherlockPRD}.  Hyperplanes is less restrictive than AntiCornerSphere, in the sense that any region satisfying AntiCornerSphere will also satisfy Hyperplanes.  Hyperplanes has the advantage of allowing regions that lie in the high tails of only a subset of the variables considered.  A region $R$ in a $d$-dimensional unit box is said to satisfy Hyperplanes if, for each data point $p$ inside $R$, one can draw a $(d-1)$-dimensional hyperplane through $p$ such that all data points on the side of the hyperplane containing the point $\vec{1}$ (the ``upper right-hand corner of the unit box'') are inside $R$.  An example of a region satisfying Hyperplanes is shown in Fig.~\ref{fig:hyperplanesExample}.

{\dofig{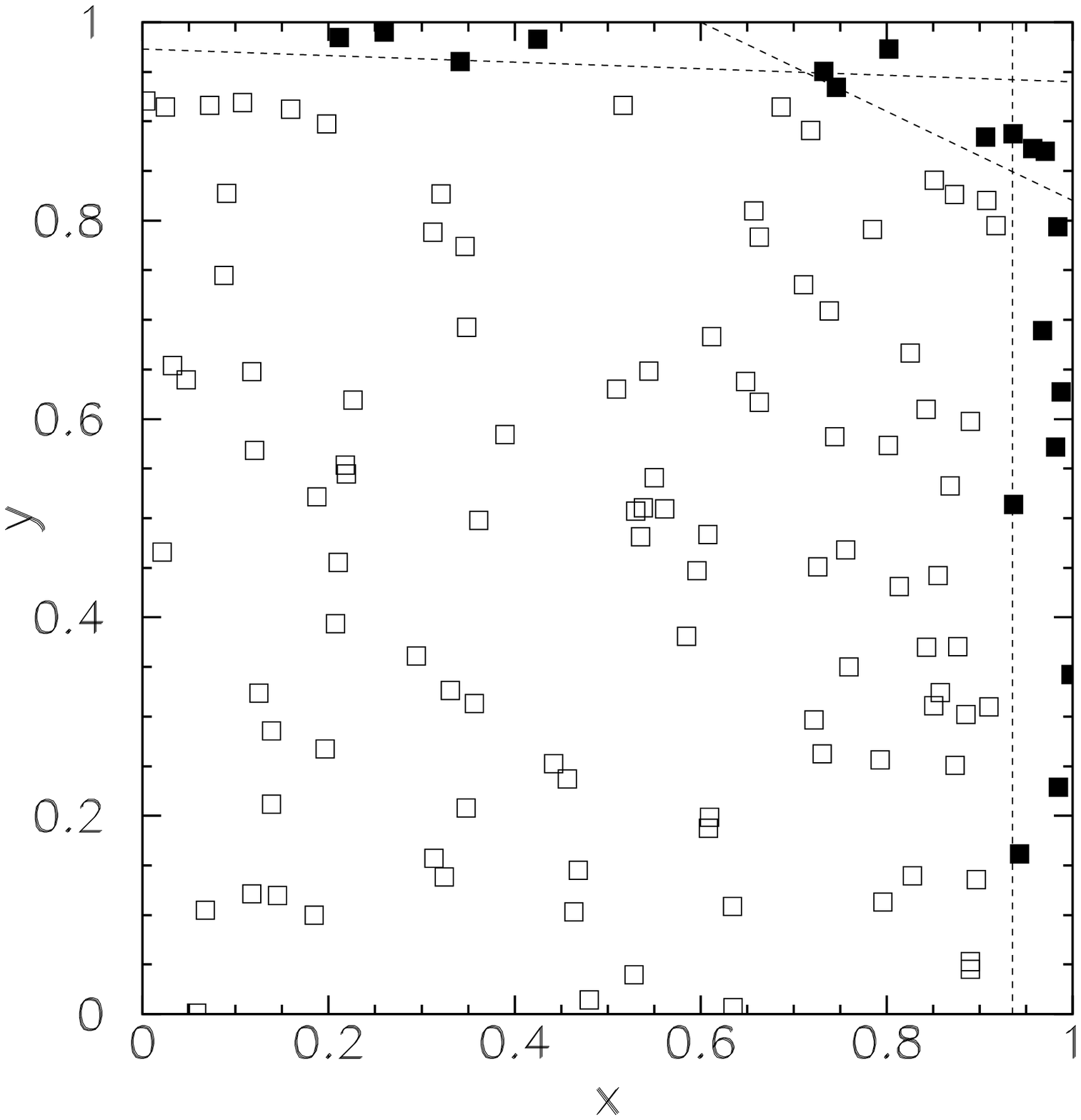} 
{3.5in} {An example of a region satisfying Hyperplanes.  The boundary of the figure is the unit box; open squares represent data points outside the region $R$; filled squares represent data points inside the region $R$.  The three dashed lines indicate hyperplanes $h_i$ (which are lines in this two-dimensional case) that can be drawn through the points at $(x,y)_i=$ $(0.34,0.96)$, $(0.74,0.95)$, and $(0.935,0.515)$ with the property that all of the data points ``up and to the right'' of $h_i$ are inside $R$.} {fig:hyperplanesExample} }

We continue this boolean criterion to the unit interval $[0,1]$ in order to ensure the continuity of the final result under small changes in the background estimate.  For each data point $i$ inside the candidate region $R$ and each hyperplane $h_i$ through $i$, we define $d_{j {h_i}}$ to be the distance between a data point $j$ lying outside $R$ and the hyperplane $h_i$.  This quantity is taken to be positive if $j$ and the point $\vec{1}$ are on the same side of $h_i$, and negative otherwise.  Letting 
\begineq
\theta(x) = \left\{ 
\begin{array}{ll}
0 & , x < 0 \nonumber \\
x & , 0 \le x \le 1 \nonumber \\
1 & , 1 < x
\end{array} \right. ,
\endeq
we define
\begineq
c_R^{H} = \prod_{i\in R}{\theta(1-\min_{h_i}{\max_{j\not\in R}{d_{j {h_i}}}}/\xi)}.
\endeq
Loosely speaking, the introduction of $c_R^H$ corresponds to widening the lines drawn in Fig.~\ref{fig:hyperplanesExample} into bands of width $\xi$, choosing $c_R^H=1$ if all data points ``up and to the right'' of these bands are inside $R$, finding $c_R^H=0$ if there is a point ``up and to the right'' that is not inside $R$, and choosing $c_R^H$ between 0 and 1 if there are one or more points not inside $R$ lying on the bands.  Note that $c_R^H$ reduces to the boolean operator of the preceding paragraph in the limit $\xi\rightarrow0$, corresponding to the squeezing of the bands back into the lines in Fig.~\ref{fig:hyperplanesExample}.

We also impose the criterion {\em Connectivity} ($c^C$) to ensure connected regions, and the criterion {\it ReasonableSize} ($c^R$) to limit the size of the regions we consider to that expected for a typical signal and to reduce the computational cost of finding the most interesting region.  A region $R$ is said to satisfy Connectivity if, given any two points $a$ and $b$ within $R$, there exists a list of points ${ p_1 = a, p_2, \ldots, p_{n-1}, p_n = b}$ such that all the $p_i$ are in $R$, and the 1-region about $p_{i+1}$ shares a border with the 1-region about $p_i$.  A region is said to satisfy ReasonableSize if it contains fewer than 50 data points.  These criteria are summarized in Table~\ref{tbl:RegionCriteria}.

\begin{table}[htb]
\centering
\begin{tabular}{ccc}
Symbol  	& Name  		& A region satisfies this criterion if \\ \hline
$c^{A}$	& AntiCornerSphere 	& {\begin{minipage}[t]{4.8cm}{One can draw a sphere centered on the origin of the unit box containing all data events outside the region and no data events inside the region.\vspace{3pt}}\end{minipage}} \\
$c^{I}$	& Isolation		& {\begin{minipage}[t]{4.8cm}{There exist no data points outside the region that are closer than $\xi$ to any data point inside the region.\vspace{3pt}}\end{minipage}} \\ \hline
$c^{H}$	& Hyperplanes		& {\begin{minipage}[t]{4.8cm}{For each data point $p$ inside $R$, one can draw a $(d-1)$-dimensional hyperplane through $p$ such that all data points on the side of the hyperplane containing the point $\vec{1}$ are inside $R$.\vspace{3pt}}\end{minipage}}  \\
$c^{C}$ 	& Connectivity	& {\begin{minipage}[t]{4.8cm}{Given any two points $a$ and $b$ within the region, there exists a list of points ${ p_1 = a, p_2, \ldots, p_{n-1}, p_n = b}$ such that all the $p_i$ are in the region and $p_{i+1}$ is a neighbor of $p_i$.\vspace{3pt}}\end{minipage}} \\
$c^{R}$	& ReasonableSize	& {\begin{minipage}[t]{4.8cm}{The region contains fewer than 50 data points.}\end{minipage}} \\
\end{tabular}
\caption{Summary of the region criteria imposed in our previous analysis of $e\mu X$ (above middle line) and those imposed in the analyses described in this article (below middle line).  $\xi=1/(4N_{\rm data}^{1/d})$ is a characteristic distance between the $N_{\rm data}$ data points in the $d$-dimensional unit box.}
\label{tbl:RegionCriteria}
\end{table}

In Ref.~\cite{SherlockPRD} we demonstrated \Sherlock's ability to find indications of $t\bar{t}$ in the $e\mu X$ final states using the criteria $c^{A}c^{I}$.  Figure~\ref{fig:hyperplanes_tt_02} shows that the combination $c^{H}c^{C}c^{R}$ (solid) performs similarly to those criteria (dashed) in this test.

\section{Charted and uncharted territory}
\label{section:OtherFinalStates}

The D\O\ experiment~\cite{D0Detector} began collecting data at $\sqrt{s}=1.8$~TeV in 1992, and completed its first series of runs in 1996.  These data have been carefully scrutinized by the D\O\ Collaboration.  Nonetheless, the incredible richness of these data, which probe fundamental physics at the highest energy scales currently achievable, allows for the possibility that something there may yet remain undiscovered.

\subsection{Final states already considered by D\O}
\label{section:finalStatesAlreadyConsidered}

Some portions of these data have been more comprehensively scrutinized than others.  In particular, there are eight final states --- those marked with triangles in Fig.~\ref{fig:SherlockFinalStatesGraph} --- that D\O\ has already analyzed in a manner similar to the \Sherlock\ prescription.  

In final states that contain only a single object (such as a $W$ or $Z$ boson), there are no non-trivial momentum variables to consider, and the \Sherlock\ search strategy reduces in this case to a counting experiment.  In final states containing exactly two objects (such as $ee$, $Zj$, or $W\gamma$), the single momentum variable available to us is the average (scalar) transverse momentum of the two objects, assuming that both are sufficiently central.  D\O\ has analyzed eight final states in these limiting cases.  These analyses do not precisely follow the \Sherlock\ prescription --- they were performed before \Sherlock\ was created --- so $\scriptP$ is not calculated for these final states.  Nonetheless, they are sufficiently close to our prescription (and therefore sufficiently quasi-model-independent) that we briefly review them here, both for completeness and in order to motivate the final states that we treat in Secs.~\ref{section:WJ}--\ref{section:zoo}.  Examples of the types of new physics that could be expected to appear in a few of these final states are provided in Appendix~\ref{section:SignalsThatMightAppearOtherFinalStates}.

\paragraph{$\jj$.}

D\O\ has performed an analysis of the dijet mass spectrum~\cite{DijetMassSpectrumAndSearchForQuarkCompositeness} and angular distribution~\cite{DijetAngularDistributionsAndSearchForQuarkCompositeness} in a search for quark compositeness.  We note that the dijet mass and the polar angle of the jet axis (in the center-of-mass frame of the system) together completely characterize these events, and that two central jets with large invariant mass also have large average $p_T$. No compelling evidence of an excess at large jet transverse momentum is seen in either case.

\paragraph{$W$.}

The \Sherlock-defined $W$ final state contains all events with either: one muon and no second charged lepton; or one electron, significant missing transverse energy, and transverse mass $30<m_T^{e\nu}<110$~GeV.  The \Sherlock\ prescription reduces to a cross section measurement in this case.  
D\O\ has measured the inclusive $W$ boson cross section~\cite{WandZCrossSection}, and finds it to be in good agreement with the standard model prediction.  

\paragraph{$e\met$.}

Events that contain one electron, no second charged lepton, substantial $\met$, and have transverse mass $m_T^{e\nu}>110$~GeV belong to the $e\met$ final state.  This final state contains two objects (the electron and the missing transverse energy), so we consider the average object $p_T$, which is approximately equal in this case to $m_T^{e\nu}/2$.  D\O\ has performed a search for right-handed $W$ bosons and heavy $W'$ bosons in 79~pb$^{-1}$ of data~\cite{RightHandedAndHeavyWs}, looking for an excess in the tail of the transverse mass distribution.  No such excess is observed.

\paragraph{$Wj$.}

In the two-object final state $Wj$, the average transverse momentum of the two objects is essentially $p_T^W$, the transverse momentum of the $W$ boson.  D\O\ has measured 
the $W$ boson $p_T$ distribution~\cite{Wpt}, and finds good agreement with the standard model.  

\paragraph{$W\gamma$.}

Similarly, the transverse momentum distribution of the photon in $W\gamma X$ events has been analyzed by D\O\ in a measurement of the $WW\gamma$ gauge boson coupling parameters~\cite{Wgamma}.  No excess at large $p_T^\gamma$ is observed.
(The Sleuth prescription for defining final states is less well satisfied in D\O's corresponding measurement of $p_T^\gamma$ in $Z\gamma X$ events~\cite{Zgamma}.)

\paragraph{$Z$.}

As in the case of the $W$ final state, our prescription reduces to a counting experiment in the $Z$ final state.  D\O\ has published a measurement of the inclusive $Z$ boson cross section~\cite{WandZCrossSection}, and finds it to be in good agreement with the standard model prediction.  

\paragraph{$ee$.}

Events containing two electrons and nothing else fall into the final state $ee$ if the invariant mass $m_{ee}$ is outside the $Z$ boson mass window of $(82,100)$~GeV.  The single variable we consider in this two-object final state is the average scalar transverse momentum of the two electrons, which is simply related to the invariant mass $m_{ee}$ for sufficiently central electrons.  D\O\ has analyzed the high mass Drell-Yan cross section in a search for indications of quark-lepton compositeness with the full data set~\cite{HighMassDrellYanCrossSectionAndLimitsOnQuarkElectronCompositeness}, and has analyzed the $ee$ invariant mass distribution in the context of a search for additional neutral gauge bosons in a subset of those data~\cite{Search For Additional Neutral Gauge Bosons}.  No discrepancy between the data and expected background is observed.

\paragraph{$Zj$.}

In the two-object final state $Zj$, the average transverse momentum of the two objects is essentially the transverse momentum of the $Z$ boson.  D\O's published measurement of the $Z$ boson $p_T$ distribution~\cite{Zpt} is in good agreement with the standard model prediction.  

\subsection{Final states considered in this article}

The decision as to which of the remaining final states should be subjected to a \Sherlock\ analysis was made on the basis of our ability to estimate the standard model and instrumental backgrounds in each final state, and the extent to which a systematic analysis for new physics is lacking in each final state.  The final states we chose to analyze arranged themselves into four ``classes'':  $e\mu X$, $W$+jets-like final states, $Z$+jets-like final states, and $\ellgamma\ellgamma\ellgamma X$.  The first of these classes has been analyzed in Ref.~\cite{SherlockPRD} and summarized in Sec.~\ref{section:emuX}.  A systematic \Sherlock\ analysis of the remaining three classes of final states is the subject of the next three sections.

\section{{\boldmath $W$}+jets-like final states}
\label{section:WJ}

In this section we analyze the $W$+jets-like final states --- events containing a single lepton, missing transverse energy, and two or more jets.   In Sec.~\ref{section:ZJdatasets} we describe the $e\met \jj(nj)$ and $\mu\met \jj(nj)$ data sets and background estimates, and in Sec.~\ref{section:WJResults} we present the results.  After this, we feign ignorance of the heaviest quark in the standard model and check the sensitivity of our method to top quark pair production in Sec.~\ref{section:topInWjjj(nj)}.  A few of the many signals that might appear in these final states are described in Appendix~\ref{section:SignalsThatMightAppearInWjj(nj)}.

\subsection{Data sets and background estimates}
\label{section:ZJdatasets}

\def \emetjjOOwjj {$6.7 \pm 1.4$}
\def \emetjjOOqcd {$3.3 \pm 0.9$}
\def \emetjjOOtop {$1.7 \pm 0.6$}
\def \emetjjOOtotal {$11.6 \pm 1.7$}
\def \emetjjOOdata {$7$}
\def \emetjjjOOwjj {$1.0 \pm 0.4$}
\def \emetjjjOOqcd {$0.48 \pm 0.22$}
\def \emetjjjOOtop {$1.0 \pm 0.4$}
\def \emetjjjOOtotal {$2.5 \pm 0.6$}
\def \emetjjjOOdata {$5$}
\def \emetjjjjOOwjj {$0.15 \pm 0.11$}
\def \emetjjjjOOqcd {$0.38 \pm 0.19$}
\def \emetjjjjOOtop {$0.26 \pm 0.09$}
\def \emetjjjjOOtotal {$0.80 \pm 0.24$}
\def \emetjjjjOOdata {$2$}
\def \emetjjjjjOOwjj {$0.030 \pm 0.020$}
\def \emetjjjjjOOqcd {$0.08 \pm 0.04$}
\def \emetjjjjjOOtop {$0.042 \pm 0.017$}
\def \emetjjjjjOOtotal {$0.15 \pm 0.05$}
\def \emetjjjjjOOdata {$0$}
\def \WjjOOwjj {$334 \pm 51$}
\def \WjjOOqcd {$12.0 \pm 2.6$}
\def \WjjOOtop {$4.0 \pm 1.4$}
\def \WjjOOtotal {$350 \pm 51$}
\def \WjjOOdata {$387$}
\def \WjjjOOwjj {$57 \pm 9$}
\def \WjjjOOqcd {$3.4 \pm 0.9$}
\def \WjjjOOtop {$6.0 \pm 2.1$}
\def \WjjjOOtotal {$66 \pm 9$}
\def \WjjjOOdata {$56$}
\def \WjjjjOOwjj {$5.9 \pm 1.3$}
\def \WjjjjOOqcd {$1.1 \pm 0.4$}
\def \WjjjjOOtop {$3.9 \pm 1.4$}
\def \WjjjjOOtotal {$10.9 \pm 1.9$}
\def \WjjjjOOdata {$11$}
\def \WjjjjjOOwjj {$0.8 \pm 0.3$}
\def \WjjjjjOOqcd {$0.19 \pm 0.12$}
\def \WjjjjjOOtop {$0.73 \pm 0.26$}
\def \WjjjjjOOtotal {$1.8 \pm 0.4$}
\def \WjjjjjOOdata {$1$}
\def \WjjjjjjOOwjj {$0.12 \pm 0.06$}
\def \WjjjjjjOOqcd {$0.030 \pm 0.015$}
\def \WjjjjjjOOtop {$0.10 \pm 0.04$}
\def \WjjjjjjOOtotal {$0.25 \pm 0.07$}
\def \WjjjjjjOOdata {$1$}
\def \WjjjjjjjOOwjj {$0.020 \pm 0.010$}
\def \WjjjjjjjOOqcd {$0.0040 \pm 0.0020$}
\def \WjjjjjjjOOtop {$0.008 \pm 0.005$}
\def \WjjjjjjjOOtotal {$0.032 \pm 0.011$}
\def \WjjjjjjjOOdata {$0$}

\BKwidetext
\begin{table*}[htb]
\centering
\begin{tabular}{lccccc}
Final State	& $W$+jets	& QCD fakes 	& $t\bar{t}$ 	& Total & Data\\ \hline
$e\met \jj$ & \emetjjOOwjj  & \emetjjOOqcd	& \emetjjOOtop & \emetjjOOtotal & \emetjjOOdata \\
$e\met \jjj$ & \emetjjjOOwjj  & \emetjjjOOqcd	& \emetjjjOOtop & \emetjjjOOtotal & \emetjjjOOdata \\
$e\met \jjjj$ & \emetjjjjOOwjj  & \emetjjjjOOqcd	& \emetjjjjOOtop & \emetjjjjOOtotal & \emetjjjjOOdata \\
$W(\rightarrow e\met) \jj$ & \WjjOOwjj & \WjjOOqcd & \WjjOOtop & \WjjOOtotal & \WjjOOdata \\
$W(\rightarrow e\met) \jjj$ & \WjjjOOwjj	& \WjjjOOqcd & \WjjjOOtop & \WjjjOOtotal & \WjjjOOdata \\
$W(\rightarrow e\met) \jjjj$ & \WjjjjOOwjj & \WjjjjOOqcd & \WjjjjOOtop & \WjjjjOOtotal & \WjjjjOOdata \\ 
$W(\rightarrow e\met) \jjjjj$ & \WjjjjjOOwjj & \WjjjjjOOqcd & \WjjjjjOOtop & \WjjjjjOOtotal & \WjjjjjOOdata \\ 
$W(\rightarrow e\met) \jjjjjj$ & \WjjjjjjOOwjj & \WjjjjjjOOqcd & \WjjjjjjOOtop & \WjjjjjjOOtotal & \WjjjjjjOOdata \\ 
\end{tabular}
\caption{Expected backgrounds to the $e\met \jj(nj)$ final states.  The final states labeled ``$W(\rightarrow e\met)$'' have $m_T^{e\nu}<110$~GeV; the final states labeled ``$e\met$'' have $m_T^{e\nu}>110$~GeV.  We have extrapolated our background estimates to final states with five or more jets.  Berends scaling and the data in this table suggest that a factor of $\approx 7$ in cross section is the price to be paid for an additional radiated jet with transverse energy above 20~GeV.}
\label{tbl:expectedBkgsEmetjjJ}
\end{table*}
\BKnarrowtext

\BKwidetext
\def \WjjOOwjjmc {$48 \pm 15$}
\def \WjjOOzjjmc {$1.6 \pm 0.4$}
\def \WjjOOwwmc {$0.5 \pm 0.3$}
\def \WjjOOtopmc {$0.42 \pm 0.14$}
\def \WjjOOtotal {$50 \pm 15$}
\def \WjjOOdata {$54$}
\def \WjjjOOwjjmc {$10 \pm 3$}
\def \WjjjOOzjjmc {$0.27 \pm 0.08$}
\def \WjjjOOwwmc {$0.41 \pm 0.26$}
\def \WjjjOOtopmc {$0.58 \pm 0.20$}
\def \WjjjOOtotal {$11 \pm 3$}
\def \WjjjOOdata {$11$}
\def \WjjjjOOwjjmc {$2.8 \pm 1.3$}
\def \WjjjjOOzjjmc {$0.022 \pm 0.011$}
\def \WjjjjOOwwmc {$-$}
\def \WjjjjOOtopmc {$0.61 \pm 0.21$}
\def \WjjjjOOtotal {$3.5 \pm 1.3$}
\def \WjjjjOOdata {$4$}

\begin{table*}[htb]
\centering
\begin{tabular}{lcccccc}
Final State	& $W$+jets	& $Z$+jets	& $WW$	& $t\bar{t}$ 	& Total & Data \\ \hline
$W(\rightarrow\mu\met) \jj$ 	& \WjjOOwjjmc & \WjjOOzjjmc & \WjjOOwwmc & \WjjOOtopmc & \WjjOOtotal & \WjjOOdata \\
$W(\rightarrow\mu\met) \jjj$	& \WjjjOOwjjmc & \WjjjOOzjjmc & \WjjjOOwwmc & \WjjjOOtopmc & \WjjjOOtotal & \WjjjOOdata \\
$W(\rightarrow\mu\met) \jjjj$	& \WjjjjOOwjjmc & \WjjjjOOzjjmc & \WjjjjOOwwmc & \WjjjjOOtopmc & \WjjjjOOtotal & \WjjjjOOdata \\ 
\end{tabular}
\caption{Expected backgrounds for the $W(\rightarrow\mu\met)\jj(nj)$ final states.}
\label{tbl:expectedBkgsMumetjjJ}
\end{table*}
\BKnarrowtext

\subsubsection{$e\met \jj(nj)$}
\label{section:emetjjJ}

The $e\met \jj(nj)$ data set~\cite{LeptoquarksToENu} comprises $115 \pm 6$~pb$^{-1}$ of collider data, collected with triggers that require the presence of an electromagnetic object, with or without jets and missing transverse energy.  Offline event selection requires: one electron with transverse energy $p_T^e > 20$~GeV and pseudorapidity $\abs{\eta_{\rm det}}<1.1$ or $1.5<\abs{\eta_{\rm det}}<2.5$~\cite{PseudoRapidity}; $\met>30$~GeV; and two or more jets with $p_T^j>20$~GeV and $\abs{\eta_{\rm det}}<2.5$.  Effects of jet energy mismeasurement are reduced by requiring the $\met$ vector to be separated from the jets by $\Delta\phi>0.25$ radians if $\met<120$~GeV.  To reduce background from a class of events in which a fake electron's energy is overestimated, leading to spurious $\met$, we reject events with $p_T^W<40$~GeV.  Events containing isolated muons appear in a sample analyzed previously with this method ($e\mu X$), and are not considered here.

The dominant standard model and instrumental backgrounds to the $e\met \jj(nj)$ final states are from:
\begin{itemize}
\item{$W$ + jets production, with $W\rightarrow e\nu$;}
\item{multijet production, with mismeasured $\met$ and one jet faking an electron; and}
\item{$t\bar{t}$ pair production, with $t\rightarrow Wb$ and with at least one $W$ boson decaying to an electron or to a tau that in turn decays to an electron.}
\end{itemize}

The $W$+jets background is simulated using {\sc vecbos}~\cite{VECBOS}, with {\sc herwig}~\cite{HERWIG} used for fragmenting the partons.  The background from multijet events containing a jet that is misidentified as an electron, and with $\met$ arising from the mismeasurement of jet energies, is modeled using multijet data.  The probability for a jet to be misidentified as an electron is estimated~\cite{LeptoquarksToEE} to be $(3.50\pm 0.35) \times 10^{-4}$.  The background from $t\bar{t}$ decays into an electron plus two or more jets is simulated using {\sc herwig} with a top quark mass of 170~GeV.  All Monte Carlo event samples are processed through the D\O\ detector simulation based on the {\sc geant}~\cite{GEANT} package.

We estimate the number of $t\bar{t}$ events in the $W$+jets-like final states to be $18\pm6$ using the measured $t\bar{t}$ production cross section of $5.5\pm 1.8$~pb~\cite{topCrossSection}.  The multijet background is estimated to be $21\pm7$ events, using a sample of events with three or more jets with $\met>30$~GeV.  This is done by multiplying the fake probability by the number of ways the events satisfy the selection criteria with one of the jets passing the electron $p_T$ and $\eta$ requirements.  After the estimated numbers of $t\bar{t}$ and multijet background events are subtracted, the number of events with transverse mass of the electron and neutrino ($m_T^{e\nu}$) below 110~GeV is used to obtain an absolute normalization for the $W$+jets background.  

Following the \Sherlock\ prescription, we combine the electron and missing transverse energy into a $W$ boson if $30<m_T^{e\nu}<110$~GeV, and reject events with $m_T^{e\nu} < 30$~GeV.  The expected numbers of background events for the exclusive final states within this $e\met \jj(nj)$ sample are provided in Table~\ref{tbl:expectedBkgsEmetjjJ}.  

\subsubsection{$\mu\met \jj (nj)$}
\label{section:mumetjjJ}

The $\mu\met \jj(nj)$ data set~\cite{LeptoquarksToMuNu} corresponds to $94 \pm 5$~pb$^{-1}$ of integrated luminosity.  The initial sample is composed of events passing any of several muon + jets triggers requiring a muon with $p_T^\mu> 5$~GeV within $\abs{\eta_{\rm det}}<1.7$ and one or more jets with $p_T^j > 8$~GeV and $\abs{\eta_{\rm det}}<2.5$.  Using standard jet and muon identification criteria, we define a final sample containing one muon with $p_T>25$~GeV and $\abs{\eta_{\rm det}}<0.95$, two or more jets with $p_T^j>15$~GeV and $\abs{\eta_{\rm det}}<2.0$ and with the most energetic jet within $\abs{\eta_{\rm det}}<1.5$, and missing transverse energy $\met>30$~GeV.  Because an energetic muon's momentum is not well measured in the detector, we are unable to separate ``$W$-like'' events from ``non-$W$-like'' events using the transverse mass, as we have done above in the electron channel.  The muon and missing transverse energy are therefore always combined into a $W$ boson.

The dominant standard model and instrumental backgrounds to these final states are from:
\begin{itemize}
\item{$W$ + jets production with $W\rightarrow\mu\nu$;}
\item{$Z$ + jets production with $Z\rightarrow\mu\mu$, where one of the muons is not detected;}
\item{$WW$ pair production with one $W$ boson decaying to a muon or to a tau that in turn decays to a muon; and}
\item{$t\bar{t}$ pair production with $t\rightarrow Wb$ and with at least one $W$ boson decaying to a muon or to a tau that in turn decays to a muon.}
\end{itemize}

Samples of $W$ + jets and $Z$ + jets events are generated using {\sc vecbos}, employing {\sc herwig} for parton fragmentation.  Background due to $WW$ pair production is simulated with {\sc pythia}~\cite{PYTHIA}.  Background from $t\bar{t}$ pair production is simulated using {\sc herwig} with a top quark mass of 170~GeV.  All Monte Carlo samples are again processed through a detector simulation program based on the {\sc geant} package.

The expected backgrounds for the exclusive final states within $\mu\met \jj (nj)$ are listed in Table~\ref{tbl:expectedBkgsMumetjjJ}.  These $W(\rightarrow\mu\met) \jj(nj)$ final states are combined with the $W(\rightarrow e\met) \jj(nj)$ final states described in Sec.~\ref{section:emetjjJ} to form the $W \jj(nj)$ final states treated in Sec.~\ref{section:WjjJ}.  For consistency in this combination, we also require $p_T^W>40$~GeV for the $W(\rightarrow \mu\nu)\jj(nj)$ final states.

\subsubsection{$W\jj (nj)$}
\label{section:WjjJ}

Combining the results in Tables~\ref{tbl:expectedBkgsEmetjjJ} and~\ref{tbl:expectedBkgsMumetjjJ} gives the expected backgrounds for the $W\jj(nj)$ final states shown in Table~\ref{tbl:expectedBkgsWjjJ}.  We note the good agreement in all final states between the total number of background events expected and the number of data events observed.  This of course is due in part to the method of normalizing the $W$+jets background.  The agreement in the final states containing additional jets is also quite good.  A more detailed comparison between data and background in the more heavily populated final states ($W\jj$, $W\jjj$, and $W\jjjj$) is provided in Appendix~\ref{section:Distributions}.

Monte Carlo programs suitable for estimating backgrounds to final states with many additional jets are not readily available.  It has been observed that the rate of a process may be related to the rate of the process with an additional radiated jet by a multiplicative factor of $1/4$--$1/7$, depending upon the $p_T$ and $\eta$ thresholds used to define a jet --- this phenomenological law is known as Berends scaling~\cite{VECBOS}.  We estimate that this factor is $\approx 1/5$ for jets with $\abs{\eta_{\rm det}}<2.5$ and $p_T>15$~GeV, and that it is $\approx 1/7$ for jets with $\abs{\eta_{\rm det}}<2.5$ and $p_T>20$~GeV.  This will be used to estimate particular background contributions to final states in which the expected background is $\ltapprox 1$.


\def \WjjOOemetwjj {$334 \pm 51$}
\def \WjjOOemetqcd {$12.0 \pm 2.6$}
\def \WjjOOemettop {$4.0 \pm 1.4$}
\def \WjjOOmumetwjjmc {$48 \pm 15$}
\def \WjjOOmumetzjjmc {$1.6 \pm 0.4$}
\def \WjjOOmumetwwmc {$0.5 \pm 0.3$}
\def \WjjOOmumettopmc {$0.42 \pm 0.14$}
\def \WjjOOtotal {$400 \pm 53$}
\def \WjjOOdata {$441$}
\def \WjjjOOemetwjj {$57 \pm 9$}
\def \WjjjOOemetqcd {$3.4 \pm 0.9$}
\def \WjjjOOemettop {$6.0 \pm 2.1$}
\def \WjjjOOmumetwjjmc {$10 \pm 3$}
\def \WjjjOOmumetzjjmc {$0.27 \pm 0.08$}
\def \WjjjOOmumetwwmc {$0.41 \pm 0.26$}
\def \WjjjOOmumettopmc {$0.58 \pm 0.20$}
\def \WjjjOOtotal {$77 \pm 10$}
\def \WjjjOOdata {$67$}
\def \WjjjjOOemetwjj {$5.9 \pm 1.3$}
\def \WjjjjOOemetqcd {$1.1 \pm 0.4$}
\def \WjjjjOOemettop {$3.9 \pm 1.4$}
\def \WjjjjOOmumetwjjmc {$2.8 \pm 1.3$}
\def \WjjjjOOmumetzjjmc {$0.022 \pm 0.011$}
\def \WjjjjOOmumetwwmc {$-$}
\def \WjjjjOOmumettopmc {$0.61 \pm 0.21$}
\def \WjjjjOOtotal {$14.3 \pm 2.3$}
\def \WjjjjOOdata {$15$}
\def \WjjjjjOOemetwjj {$0.8 \pm 0.3$}
\def \WjjjjjOOemetqcd {$0.19 \pm 0.12$}
\def \WjjjjjOOemettop {$0.73 \pm 0.26$}
\def \WjjjjjOOmumetwjjmc {$-$}
\def \WjjjjjOOmumetzjjmc {$-$}
\def \WjjjjjOOmumetwwmc {$-$}
\def \WjjjjjOOmumettopmc {$-$}
\def \WjjjjjOOtotal {$1.8 \pm 0.4$}
\def \WjjjjjOOdata {$1$}
\def \WjjjjjjOOemetwjj {$0.12 \pm 0.06$}
\def \WjjjjjjOOemetqcd {$0.030 \pm 0.015$}
\def \WjjjjjjOOemettop {$0.10 \pm 0.04$}
\def \WjjjjjjOOmumetwjjmc {$-$}
\def \WjjjjjjOOmumetzjjmc {$-$}
\def \WjjjjjjOOmumetwwmc {$-$}
\def \WjjjjjjOOmumettopmc {$-$}
\def \WjjjjjjOOtotal {$0.25 \pm 0.07$}
\def \WjjjjjjOOdata {$1$}
\def \WjjjjjjjOOemetwjj {$0.020 \pm 0.010$}
\def \WjjjjjjjOOemetqcd {$0.0040 \pm 0.0020$}
\def \WjjjjjjjOOemettop {$0.008 \pm 0.005$}
\def \WjjjjjjjOOmumetwjjmc {$-$}
\def \WjjjjjjjOOmumetzjjmc {$-$}
\def \WjjjjjjjOOmumetwwmc {$-$}
\def \WjjjjjjjOOmumettopmc {$-$}
\def \WjjjjjjjOOtotal {$0.032 \pm 0.011$}
\def \WjjjjjjjOOdata {$0$}

\begin{table}[htb]
\centering
\begin{tabular}{lcc}
Final State	& Total & Data \\ \hline
$W\jj$ & \WjjOOtotal & \WjjOOdata \\
$W\jjj$ & \WjjjOOtotal & \WjjjOOdata \\
$W\jjjj$ & \WjjjjOOtotal & \WjjjjOOdata \\ 
$W\jjjjj$ & \WjjjjjOOtotal & \WjjjjjOOdata \\ 
$W\jjjjjj$ & \WjjjjjjOOtotal & \WjjjjjjOOdata \\ 
\end{tabular}
\caption{Expected backgrounds to the $W\jj(nj)$ final states.}
\label{tbl:expectedBkgsWjjJ}
\end{table}

\subsection{Results}
\label{section:WJResults}

\begin{table}[htb]
\centering
\begin{tabular}{cc}
Data set  	& $\scriptP$ \\ \hline
$e\met \jj$ 	& \scriptPemetjj \\
$e\met \jjj$  	& \scriptPemetjjj \\ 
$e\met \jjjj$  	& \scriptPemetjjjj \\ 
$W \jj$		& \scriptPWjj	\\
$W \jjj$	& \scriptPWjjj	\\
$W \jjjj$	& \scriptPWjjjj	\\
$W \jjjjj$	& \scriptPWjjjjj \\
$W \jjjjjj$	& \scriptPWjjjjjj \\
\end{tabular}
\caption{Summary of results on $e\met \jj(nj)$ and $W\jj(nj)$.}
\label{tbl:WjjJResults}
\end{table}

The results of applying \Sherlock\ to the $e\met \jj(nj)$ and $W \jj(nj)$ data sets are summarized in Table~\ref{tbl:WjjJResults} and in Figs.~\ref{fig:data_mapping_emetjj} and~\ref{fig:data_mapping_wjj}.  Recall from Sec.~\ref{section:SleuthAlgorithm} that the positions of the data points within the unit box are determined by the background distribution, which defines the transformation from the original variable space, in addition to the location of the data points in that original space.  We observe quite good agreement with the standard model in the $W$+jets-like final states.

{\dofig{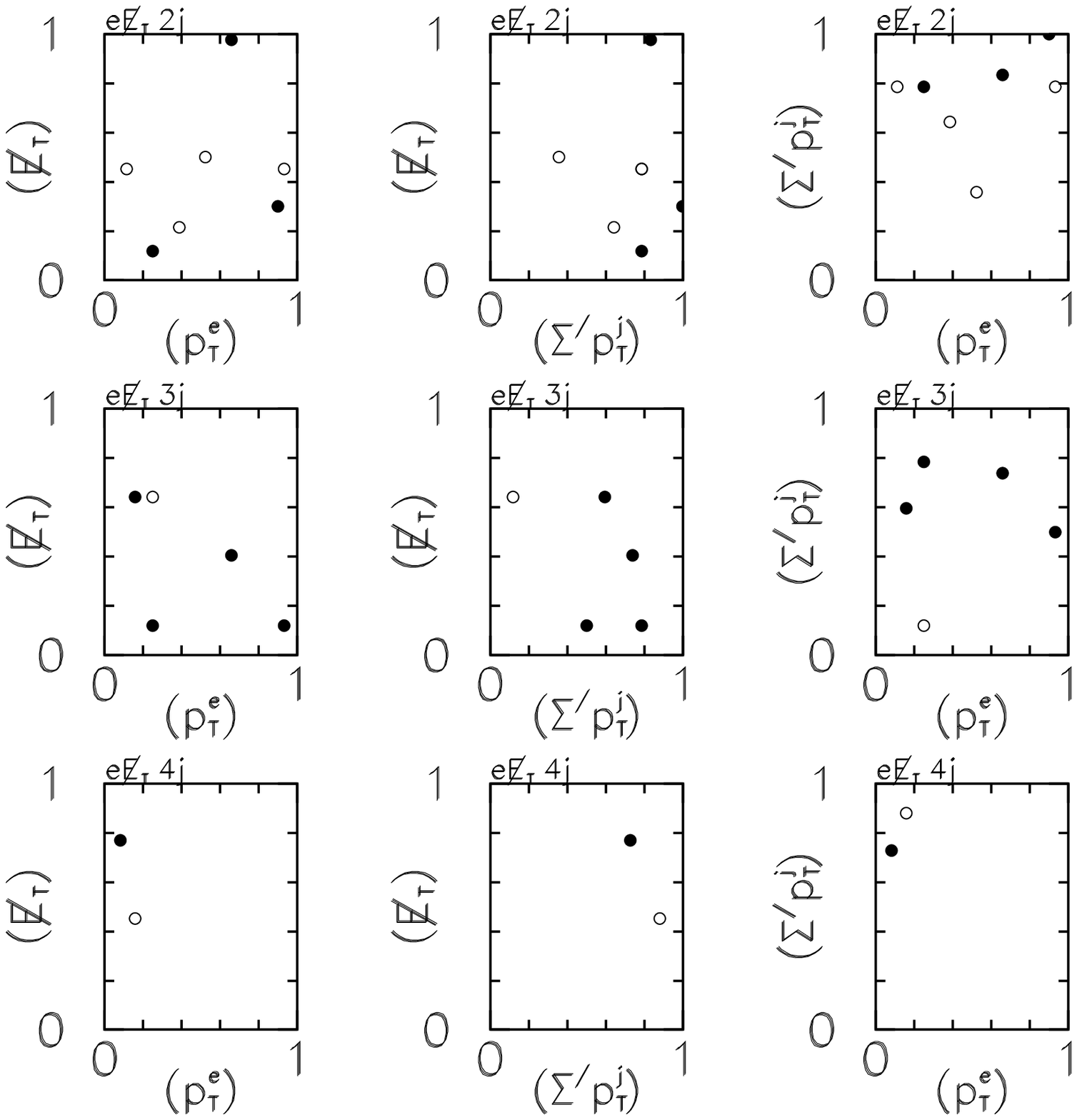} {3.5in} {The positions of the transformed data points in the final states $e\met \jj$, $e\met \jjj$, and $e\met \jjjj$.  The data points inside the region chosen by \Sherlock\ are shown as filled circles; those outside the region are shown as open circles.  For these final states the variables $p_T^e$, $\met$, and $\sum'{p_T^j}$ are considered, and the unit box is in this case a unit cube.  The two-dimensional views shown here are the projections of that cube onto three orthogonal faces.} {fig:data_mapping_emetjj}}

{\dofig{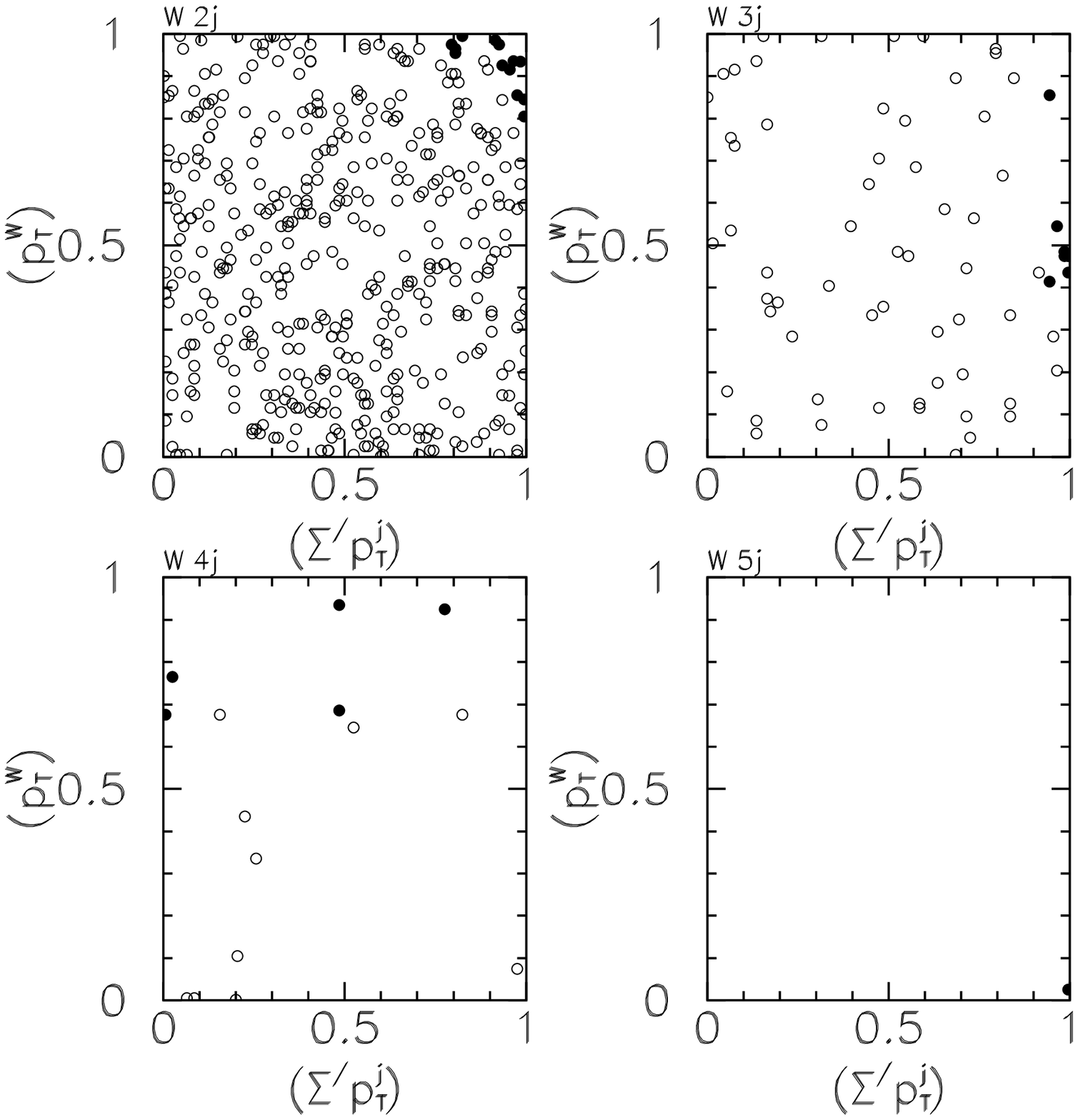} {3.5in} {The positions of the transformed data points in the final states $W\jj$, $W\jjj$, $W\jjjj$, and $W\jjjjj$.  The data points inside the region chosen by \Sherlock\ are shown as filled circles; those outside the region are shown as open circles.  The single event in the $W\jjjjj$ final state is in the lower right-hand corner of the unit square, having $\sum'{p_T^j} = 300$~GeV.} {fig:data_mapping_wjj}}

\subsection{Sensitivity check:  $t\bar{t}\rightarrow$ $\ell$+jets}
\label{section:topInWjjj(nj)}

\def \fractionOfNailedTopinWjjjXhsers {30\%}

In this section we check \Sleuth's sensitivity to $t\bar{t}$ in the final states $W\jjj$, $W\jjjj$, $W\jjjjj$, and $W\jjjjjj$.  After briefly putting this signal into context, we test \Sleuth's ability to find $t\bar{t}$ in the data, and then in an ensemble of mock experiments.

In 1997 D\O\ published a measurement of the top quark production cross section~\cite{topCrossSection} based on events in the dilepton, $\ell$+jets, $\ell$+jets($/\mu$), and ``$e\nu$'' channels, where ``$/\mu$'' indicates that one or more of the jets contains a muon, and hence is likely to be the product of a $b$ quark.  19 events with no $b$-quark tag are observed in $\ell$+jets (nine events in the electron channel, and ten events in the muon channel) with an expected background of $8.7\pm1.7$.  An additional eleven events are observed with a $b$-quark tag (five events in the electron channel, and six events in the muon channel) with an expected background of $2.5\pm0.5$ events.  Three or more jets with $p_T>15$~GeV are required in both cases.  The number of events observed in all four channels is 39 with an expected background of $13\pm2.2$ events.  The probability for $13\pm2.2$ to fluctuate up to or above 39 is $6\times10^{-7}$, or $4.8$ standard deviations.  In the $\ell$+jets channel alone, the probability that $8.7\pm1.7$ fluctuates to the 19 events observed is 0.005, corresponding to a ``significance'' of $2.6\sigma$.

{\dofig{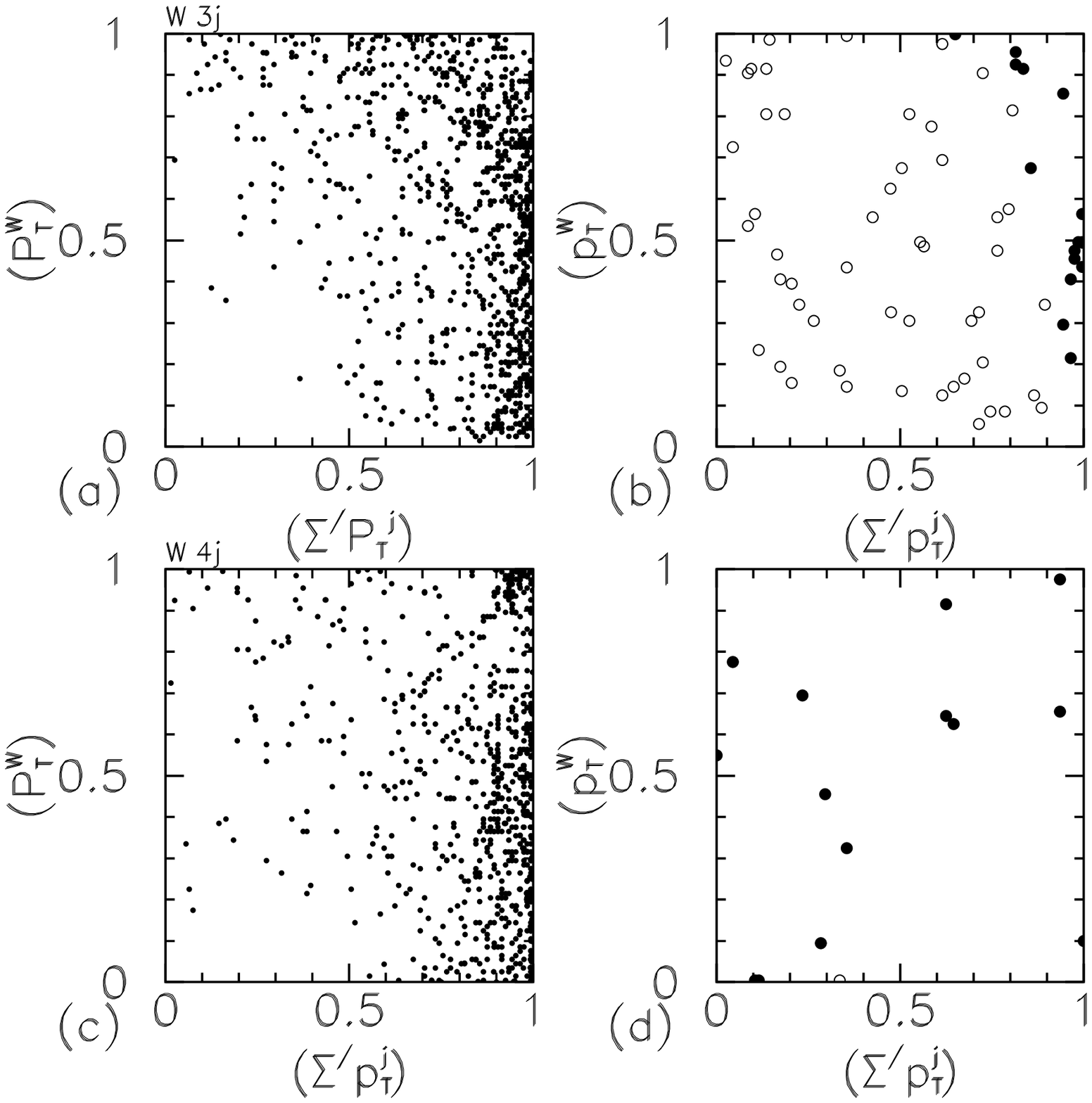} {3.5in} {Scatter plot of where $t\bar{t}$ Monte Carlo events fall in the unit box in the final states $W\jjj$ (a) and $W\jjjj$ (c).  Although top quark events appear in the high tails of $\sum'{p_T^j}$, the variable $p_T^W$ is not particularly discriminating.  The locations of the data points are shown in (b) and (d).  The backgrounds are taken to include all standard model processes except top quark pair production.} {fig:ttbar_in_wjjjx}}

Figures~\ref{fig:ttbar_in_wjjjx}(a) and~\ref{fig:ttbar_in_wjjjx}(c) show where $t\bar{t}$ Monte Carlo events fall in the unit box in the final states $W\jjj$ and $W\jjjj$.  The distribution of these events is quite diffuse in the case of $W\jjj$, since $t\bar{t}$ is similar to the background in the variables $p_T^W$ and $\sum'{p_T^j}$ in this channel.  In the $W\jjjj$ final state $t\bar{t}$ tends to populate regions of large $\sum'{p_T^j}$, but the signal is nearly indistinguishable from background in the variable $p_T^W$.  A check of \Sherlock's ability to find $t\bar{t}$ in the $W\jjj(nj)$ final states tests how well \Sherlock\ performs when the signal shows up in a subset of the variables we choose to consider.

Figures~\ref{fig:ttbar_in_wjjjx}(b) and~\ref{fig:ttbar_in_wjjjx}(d) show D\O\ data in the final states $W\jjj$ and $W\jjjj$, when $t\bar{t}$ is not included in the background estimate.  Notice that the region chosen by \Sherlock\ in the $W\jjj$ final state in Fig.~\ref{fig:ttbar_in_wjjjx}(b) is very similar to the region populated by $t\bar{t}$ in Fig.~\ref{fig:ttbar_in_wjjjx}(a).  In the $W\jjjj$ final state (d), the region chosen by \Sherlock\ is nearly the entire unit box.  Comparison with Fig.~{\protect{\ref{fig:data_mapping_wjj}}} shows how the absence of $t\bar{t}$ in the background estimate in this figure affects the transformation from the original variable space into the unit box.  Applying \Sleuth\ to these data while continuing to feign ignorance of $t\bar{t}$, we find $\scriptP_{W\jjj}=\mbox{\scriptPWjjjTop}$, $\scriptP_{W\jjjj}=\mbox{\scriptPWjjjjTop}$, $\scriptP_{W\jjjjj}=\mbox{\scriptPWjjjjjTop}$, and $\scriptP_{W\jjjjjj}=\mbox{\scriptPWjjjjjjTop}$.  Upon combining these results, we find $\scriptP_{\rm min} = \min(\scriptP_{W\jjj},\scriptP_{W\jjjj},\scriptP_{W\jjjjj},\scriptP_{W\jjjjjj})=0.09$ (1.3$\sigma$).  

Figure~\ref{fig:mock_wjjjxtop} shows a histogram of $\scriptP_{\rm min}$ for a sample of mock experimental runs in which the backgrounds include $W$+jets and QCD events, and the mock samples include $t\bar{t}$ in addition to the expected background.  The number of background and $t\bar{t}$ events in the mock samples are allowed to vary according to statistical and systematic errors.  Note that since four final states are considered, the distribution of $\scriptP_{\rm min}$ for an ensemble of experiments including background only has a median of $\approx 1\sigma$.   We see that \Sherlock\ is able to find indications of the presence of $t\bar{t}$ in these final states, returning $\scriptP_{{\rm min}\,[\sigma]}>3$ in \fractionOfNailedTopinWjjjXhsers\ of an ensemble of mock experimental runs containing $t\bar{t}$ events, compared to only $0.5\%$ of an ensemble of mock experimental runs containing background only.

{\dofig{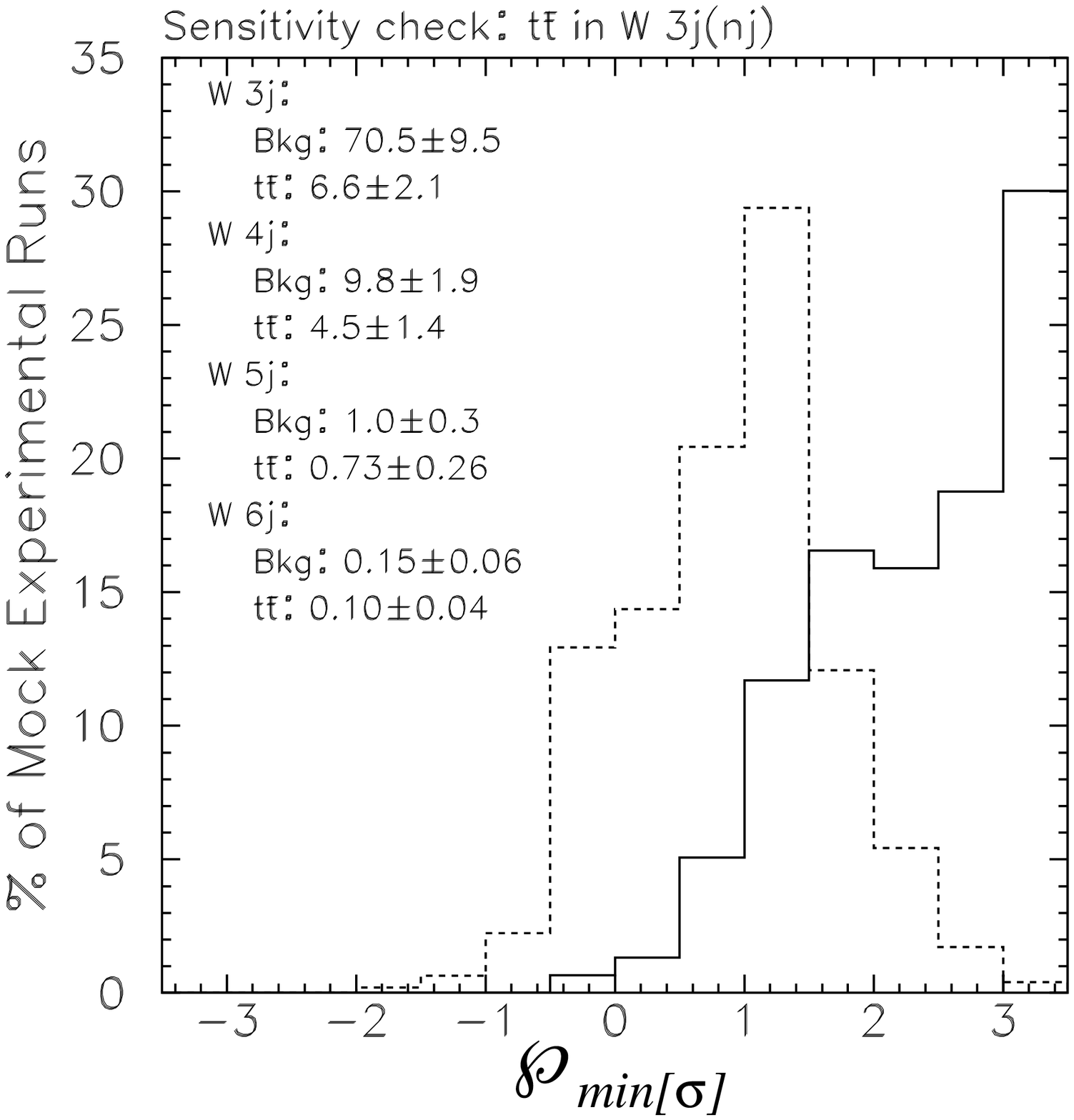} {3.5in} {Histogram of $\scriptP_{\rm min} = \min(\scriptP_{W\jjj},\scriptP_{W\jjjj},\scriptP_{W\jjjjj},$ $\scriptP_{W\jjjjjj})$ for an ensemble of mock experimental runs in which the backgrounds include $W$+jets and QCD events, and the mock samples include (solid) / do not include (dashed) $t\bar{t}$ in addition to the expected background.  All experimental runs with $\scriptP_{\rm min}>3\sigma$ are in the rightmost bin.  
} {fig:mock_wjjjxtop}}

We conclude from this sensitivity check that \Sleuth\ would not have been able to ``discover'' $t\bar{t}$ in the D\O\ $W$+jets data, but that in \fractionOfNailedTopinWjjjXhsers\ of an ensemble of mock experimental runs \Sleuth\ would have found $\scriptP_{{\rm min}\,[\sigma]}>3$.  

\section{{\boldmath $Z$}+jets-like final states}
\label{section:ZJ}

In this section we analyze the $Z$+jets-like final states.  We first describe the data sets and background estimates for the dielectron+jets channels, and we then discuss the dimuon+jets channels.  After presenting our results, we check the sensitivity of our method to the presence of first generation scalar leptoquarks.  Appendix~\ref{section:SignalsThatMightAppearInZjj(nj)} describes signals that might appear in these final states.

\subsection{Data sets and background estimates}

\subsubsection{$ee\jj(nj)$}
\label{section:eejjJ}

\def \eejjOOdy {$20 \pm 4$}
\def \eejjOOqcd {$12.2 \pm 1.8$}
\def \eejjOOtotal {$32 \pm 4$}
\def \eejjOOdata {$32$}
\def \eejjjOOdy {$2.6 \pm 0.6$}
\def \eejjjOOqcd {$1.85 \pm 0.28$}
\def \eejjjOOtotal {$4.5 \pm 0.6$}
\def \eejjjOOdata {$4$}
\def \eejjjjOOdy {$0.40 \pm 0.20$}
\def \eejjjjOOqcd {$0.24 \pm 0.04$}
\def \eejjjjOOtotal {$0.64 \pm 0.20$}
\def \eejjjjOOdata {$3$}
\def \eejjjjjOOdy {$0.030 \pm 0.015$}
\def \eejjjjjOOqcd {$0.026 \pm 0.005$}
\def \eejjjjjOOtotal {$0.056 \pm 0.016$}
\def \eejjjjjOOdata {$0$}
\def \eejjjjjjOOdy {$0.0060 \pm 0.0030$}
\def \eejjjjjjOOqcd {$0.018 \pm 0.003$}
\def \eejjjjjjOOtotal {$0.024 \pm 0.005$}
\def \eejjjjjjOOdata {$0$}
\def \eejjjjjjjOOdy {$0.0010 \pm 0.0005$}
\def \eejjjjjjjOOqcd {$0.0157 \pm 0.0031$}
\def \eejjjjjjjOOtotal {$0.017 \pm 0.003$}
\def \eejjjjjjjOOdata {$0$}
\def \eemetjjOOdy {$3.7 \pm 0.8$}
\def \eemetjjOOqcd {$-$}
\def \eemetjjOOtotal {$3.7 \pm 0.8$}
\def \eemetjjOOdata {$2$}
\def \eemetjjjOOdy {$0.45 \pm 0.13$}
\def \eemetjjjOOqcd {$-$}
\def \eemetjjjOOtotal {$0.45 \pm 0.13$}
\def \eemetjjjOOdata {$1$}
\def \eemetjjjjOOdy {$0.061 \pm 0.028$}
\def \eemetjjjjOOqcd {$-$}
\def \eemetjjjjOOtotal {$0.061 \pm 0.028$}
\def \eemetjjjjOOdata {$1$}
\def \eemetjjjjjOOdy {$0.040 \pm 0.020$}
\def \eemetjjjjjOOqcd {$-$}
\def \eemetjjjjjOOtotal {$0.040 \pm 0.020$}
\def \eemetjjjjjOOdata {$0$}
\def \eemetjjjjjjOOdy {$0.008 \pm 0.004$}
\def \eemetjjjjjjOOqcd {$-$}
\def \eemetjjjjjjOOtotal {$0.008 \pm 0.004$}
\def \eemetjjjjjjOOdata {$0$}
\def \eemetjjjjjjjOOdy {$0.0010 \pm 0.0005$}
\def \eemetjjjjjjjOOqcd {$-$}
\def \eemetjjjjjjjOOtotal {$0.0010 \pm 0.0005$}
\def \eemetjjjjjjjOOdata {$0$}
\def \ZjjOOdy {$94 \pm 19$}
\def \ZjjOOqcd {$1.88 \pm 0.28$}
\def \ZjjOOtotal {$96 \pm 19$}
\def \ZjjOOdata {$82$}
\def \ZjjjOOdy {$12.7 \pm 2.7$}
\def \ZjjjOOqcd {$0.27 \pm 0.04$}
\def \ZjjjOOtotal {$13.0 \pm 2.7$}
\def \ZjjjOOdata {$11$}
\def \ZjjjjOOdy {$1.8 \pm 0.5$}
\def \ZjjjjOOqcd {$0.034 \pm 0.006$}
\def \ZjjjjOOtotal {$1.8 \pm 0.5$}
\def \ZjjjjOOdata {$1$}
\def \ZjjjjjOOdy {$0.26 \pm 0.10$}
\def \ZjjjjjOOqcd {$0.0025 \pm 0.0009$}
\def \ZjjjjjOOtotal {$0.26 \pm 0.10$}
\def \ZjjjjjOOdata {$0$}
\def \ZjjjjjjOOdy {$0.020 \pm 0.010$}
\def \ZjjjjjjOOqcd {$0.0036 \pm 0.0011$}
\def \ZjjjjjjOOtotal {$0.024 \pm 0.010$}
\def \ZjjjjjjOOdata {$0$}
\def \ZjjjjjjjOOdy {$0.0040 \pm 0.0020$}
\def \ZjjjjjjjOOqcd {$0.0019 \pm 0.0007$}
\def \ZjjjjjjjOOtotal {$0.0059 \pm 0.0021$}
\def \ZjjjjjjjOOdata {$0$}

\BKwidetext
\begin{table*}[htb]
\centering
\begin{tabular}{lcccc}
Final State	& $Z/\gamma^*$+jets	& QCD fakes	& Total & Data \\ \hline
$ee\jj$	& \eejjOOdy & \eejjOOqcd & \eejjOOtotal & \eejjOOdata \\
$ee\jjj$ & \eejjjOOdy & \eejjjOOqcd & \eejjjOOtotal & \eejjjOOdata \\
$ee\jjjj$ & \eejjjjOOdy & \eejjjjOOqcd & \eejjjjOOtotal & \eejjjjOOdata \\
$ee\met \jj$	& \eemetjjOOdy & \eemetjjOOqcd & \eemetjjOOtotal & \eemetjjOOdata \\
$ee\met \jjj$	& \eemetjjjOOdy & \eemetjjjOOqcd & \eemetjjjOOtotal & \eemetjjjOOdata \\
$ee\met \jjjj$ & \eemetjjjjOOdy & \eemetjjjjOOqcd & \eemetjjjjOOtotal & \eemetjjjjOOdata \\
$Z(\rightarrow ee) \jj$ & \ZjjOOdy & \ZjjOOqcd & \ZjjOOtotal & \ZjjOOdata \\
$Z(\rightarrow ee) \jjj$ & \ZjjjOOdy & \ZjjjOOqcd & \ZjjjOOtotal & \ZjjjOOdata \\
$Z(\rightarrow ee) \jjjj$ & \ZjjjjOOdy & \ZjjjjOOqcd & \ZjjjjOOtotal & \ZjjjjOOdata \\
$Z(\rightarrow ee) \jjjjj$ & \ZjjjjjOOdy & \ZjjjjjOOqcd & \ZjjjjjOOtotal & \ZjjjjjOOdata \\
\end{tabular}
\caption{Expected backgrounds to the $ee\jj(nj)$, $ee\met \jj(nj)$, and $Z(\rightarrow ee)\jj(nj)$ final states.}
\label{tbl:expectedBkgsEejjJ}
\end{table*}
\BKnarrowtext

The $ee\jj(nj)$ data set~\cite{LeptoquarksToEE}, corresponding to an integrated luminosity of $123 \pm 7$~pb$^{-1}$, is collected with triggers requiring the presence of two electromagnetic objects.  Offline event selection requires two electrons passing standard identification criteria with transverse momenta $p_T^e > 20$~GeV and pseudorapidity $\abs{\eta_{\rm det}}<1.1$ or $1.5<\abs{\eta_{\rm det}}<2.5$, and two or more jets with $p_T^j>20$~GeV and $\abs{\eta_{\rm det}}<2.5$.  At least one electron is required to have a matching track in the central tracking detectors and to satisfy ionization requirements in the tracking chambers and transition radiation detector.  For these data the trigger energy threshold forces a transverse momentum cut of 20~GeV, rather than the \Sherlock-preferred requirement of 15~GeV.  We cut on a likelihood described in Appendix~\ref{section:pmet} in order to correctly identify any events with significant missing transverse energy.  Electron pairs are combined into a $Z$ boson if $82<m_{ee}<100$~GeV, unless the event contains significant $\met$ (in which case it falls within $ee\met X$, discussed in this section) or a third charged lepton (in which case it falls within $\ellgamma\ellgamma\ellgamma X$, discussed in Sec.~\ref{section:zoo}).

The dominant standard model and instrumental backgrounds to this data set are:
\begin{itemize}
\item{Drell-Yan + jets production, with $Z/\gamma^*\rightarrow ee$;}
\item{QCD multijets, with two jets faking electrons; and}
\item{$t\bar{t}$ pair production with $t\rightarrow Wb$ and with each $W$ boson decaying to an electron or to a tau that in turn decays to an electron.}
\end{itemize}

\BKwidetext
\def \ZjjOOwjjmctwo {$-$}
\def \ZjjOOzjjmctwo {$2.2 \pm 0.4$}
\def \ZjjOOwwmctwo {$-$}
\def \ZjjOOtopmctwo {$0.050 \pm 0.020$}
\def \ZjjOOtotal {$2.3 \pm 0.4$}
\def \ZjjOOdata {$3$}
\def \ZjjjOOwjjmctwo {$-$}
\def \ZjjjOOzjjmctwo {$0.24 \pm 0.05$}
\def \ZjjjOOwwmctwo {$-$}
\def \ZjjjOOtopmctwo {$0.018 \pm 0.009$}
\def \ZjjjOOtotal {$0.26 \pm 0.06$}
\def \ZjjjOOdata {$1$}
\def \ZjjjjOOwjjmctwo {$-$}
\def \ZjjjjOOzjjmctwo {$0.022 \pm 0.009$}
\def \ZjjjjOOwwmctwo {$-$}
\def \ZjjjjOOtopmctwo {$0.006 \pm 0.004$}
\def \ZjjjjOOtotal {$0.028 \pm 0.010$}
\def \ZjjjjOOdata {$0$}
\def \mumujjOOwjjmctwo {$-$}
\def \mumujjOOzjjmctwo {$0.112 \pm 0.029$}
\def \mumujjOOwwmctwo {$0.25 \pm 0.13$}
\def \mumujjOOtopmctwo {$0.14 \pm 0.05$}
\def \mumujjOOtotal {$0.50 \pm 0.15$}
\def \mumujjOOdata {$2$}
\def \mumujjjOOwjjmctwo {$-$}
\def \mumujjjOOzjjmctwo {$0.007 \pm 0.004$}
\def \mumujjjOOwwmctwo {$0.06 \pm 0.04$}
\def \mumujjjOOtopmctwo {$0.065 \pm 0.025$}
\def \mumujjjOOtotal {$0.13 \pm 0.05$}
\def \mumujjjOOdata {$0$}
\def \mumujjjjOOwjjmctwo {$-$}
\def \mumujjjjOOzjjmctwo {$0.0036 \pm 0.0027$}
\def \mumujjjjOOwwmctwo {$0.010 \pm 0.005$}
\def \mumujjjjOOtopmctwo {$0.012 \pm 0.007$}
\def \mumujjjjOOtotal {$0.025 \pm 0.009$}
\def \mumujjjjOOdata {$0$}

\begin{table*}[htb]
\centering
\begin{tabular}{lccccc}
Final State	& $Z$+jets	& $WW$	& $t\bar{t}$ 	& Total & Data \\ \hline
$\mu\mu \jj$	& \mumujjOOzjjmctwo & \mumujjOOwwmctwo & \mumujjOOtopmctwo & \mumujjOOtotal & \mumujjOOdata \\
$\mu\mu \jjj$	& \mumujjjOOzjjmctwo & \mumujjjOOwwmctwo & \mumujjjOOtopmctwo & \mumujjjOOtotal & \mumujjjOOdata \\
$Z(\rightarrow\mu\mu) \jj$ & \ZjjOOzjjmctwo & \ZjjOOwwmctwo & \ZjjOOtopmctwo & \ZjjOOtotal & \ZjjOOdata \\
$Z(\rightarrow\mu\mu) \jjj$ & \ZjjjOOzjjmctwo & \ZjjjOOwwmctwo & \ZjjjOOtopmctwo & \ZjjjOOtotal & \ZjjjOOdata \\
\end{tabular}
\caption{Expected backgrounds to the $Z(\rightarrow\mu\mu)\jj(nj)$ and $\mu\mu \jj(nj)$ final states.}
\label{tbl:expectedBkgsMumujjJtwo}
\end{table*}
\BKnarrowtext

Monte Carlo samples for the Drell-Yan events are generated using {\sc isajet}~\cite{ISAJET}.  The Drell-Yan cross section normalization is fixed by comparing the Monte Carlo events with $Z$ + $\ge2$ jets data in the $Z$ boson region.  Top quark events are generated using {\sc herwig} at a top quark mass of 170~GeV with all dilepton final states included.  The D\O\ measured $t\bar{t}$ production cross section of $5.5\pm 1.8$~pb at a top quark mass of 173.3~GeV was used~\cite{topCrossSection}.  The multijet background is estimated from a sample of events with four or more jets in which the probability for two jets or photons to be misidentified as electrons is weighted by the number of jets in the event that passed the electron $p_T$ and $\eta$ requirements.  This misidentification probability is calculated from a sample of events with three jets to be $(3.50\pm0.35)\times 10^{-4}$ for an electron with a reconstructed track and $(1.25\pm0.13)\times10^{-3}$ for an electron without a reconstructed track.  The uncertainties in these probabilities reflect a slight dependence on the jet $p_T$ and $\eta$.  The expected backgrounds for the exclusive final states within $ee \jj (nj)$ are listed in Table~\ref{tbl:expectedBkgsEejjJ}.  

\subsubsection{$\mu\mu \jj (nj)$}
\label{section:mumujjJ}

The $\mu\mu \jj(nj)$ data set~\cite{LeptoquarksToMuMu} corresponds to $94 \pm 5$~pb$^{-1}$ of integrated luminosity.  The initial sample is composed of events passing any of several muon + jets triggers requiring a muon with $p_T^\mu> 5$~GeV within $\abs{\eta_{\rm det}}<1.7$ and one or more jets with $p_T^j > 8$~GeV and $\abs{\eta_{\rm det}}<2.5$.  Using standard jet and muon identification criteria, we define a final sample containing two or more muons with $p_T>20$~GeV and $\abs{\eta_{\rm det}}<1.7$ and at least one muon in the central detector ($\abs{\eta_{\rm det}}<1.0$), and two or more jets with $p_T^j>20$~GeV and $\abs{\eta_{\rm det}}<2.5$. 

We combine a $\mu\mu$ pair into a $Z$ boson if the muon momenta can be varied within their resolutions such that $m_{\mu\mu}\approx M_Z$ and the missing transverse energy becomes negligible.  More specifically, we combine a muon pair into a $Z$ boson if
\begin{eqnarray}
\label{eqn:mumuMass}
\chi = \min_{a,b} & \left( 
\frac{1/a - 1/p^{\mu_1}}{\delta(1/p^{\mu_1})} \oplus \frac{1/b - 1/p^{\mu_2}}{\delta(1/p^{\mu_2})} \oplus \right. \nonumber \\
 &
\left. \,\,\, \frac{m_{ab}-M_Z}{\Gamma_Z} \oplus \frac{\met_{ab}}{\delta(\met)} \right) < 20,
\end{eqnarray}
where $\delta(1/p) = 0.18 (p-2)/p^2 \oplus 0.003$ is the uncertainty in the reciprocal of the muon momentum; $\delta(\met) = 0.7~{\rm GeV} \sqrt{\sum{p_T^j}/{\rm GeV}}$ is the error on the missing transverse energy measured in the calorimeter; $m_{ab}$ and $\met_{ab}$ are the muon pair invariant mass and missing transverse energy, computed taking the muons to have scalar momenta $a$ and $b$; $M_Z$ and $\Gamma_Z$ are the mass and width of the $Z$ boson; and $\oplus$ means addition in quadrature.  The cut of $\chi < 20$ is chosen so that $Z(\rightarrow \mu\mu)$ is not the dominant background to the $\mu\mu \jj(nj)$ final states.

The most significant standard model and instrumental backgrounds to this data set are
\begin{itemize}
\item{$Z$ + jets production with $Z\rightarrow\mu\mu$,}
\item{$WW$ pair production with each $W$ boson decaying to a muon or to a tau that in turn decays to a muon, and}
\item{$t\bar{t}$ pair production with $t\rightarrow Wb$ and with each $W$ boson decaying to a muon or to a tau that in turn decays to a muon.}
\end{itemize}

A sample of $Z$ + jets events was generated using {\sc vecbos}, employing {\sc herwig} for parton fragmentation.  Background due to $WW$ pair production is simulated with {\sc pythia}.  Background from $t\bar{t}$ pair production is simulated using {\sc herwig} with a top quark mass of 170~GeV.  All Monte Carlo samples are processed through a detector simulation program based on the {\sc geant} package.

The expected backgrounds for the exclusive final states within $\mu\mu \jj (nj)$ are listed in Table~\ref{tbl:expectedBkgsMumujjJtwo}.  The $Z(\rightarrow\mu\mu) \jj(nj)$ final states are combined with the $Z(\rightarrow ee) \jj(nj)$ final states described in Sec.~\ref{section:eejjJ} to form the $Z \jj(nj)$ final states treated in Sec.~\ref{section:ZjjJ}.

\subsubsection{$Z\jj (nj)$}
\label{section:ZjjJ}

Combining the results in Tables~\ref{tbl:expectedBkgsEejjJ} and~\ref{tbl:expectedBkgsMumujjJtwo} gives the expected backgrounds for the $Z\jj(nj)$ final states, shown in Table~\ref{tbl:expectedBkgsZjjJ}.  The number of dimuon events in these tables is significantly smaller than the number of dielectron events due to especially tight identification requirements on the muons.  

$Z/\gamma^*$ is the dominant background to nearly all final states discussed in this section, although other sources of background contribute significantly when the dilepton mass is outside the $Z$ boson mass window.  The agreement between the total number of events expected and the number observed in the data is quite good, even for final states with several jets.  While any analysis of $Z$+jets-like states will need to rely to some degree on an accurate $Z/\gamma^*$+jets Monte Carlo, having a reliable estimate of the jet distributions in such events is especially important when exclusive final states are considered.  We anticipate that this will become increasingly important in the next Tevatron run.  Differential agreement between data and the expected background may be seen by considering a comparison of various kinematic quantities in Appendix~\ref{section:Distributions}.

\def \ZjjOOeedy {$94 \pm 19$}
\def \ZjjOOeeqcd {$1.88 \pm 0.28$}
\def \ZjjOOmumuwjjmctwo {$-$}
\def \ZjjOOmumuzjjmctwo {$2.2 \pm 0.4$}
\def \ZjjOOmumuwwmctwo {$-$}
\def \ZjjOOmumutopmctwo {$0.050 \pm 0.020$}
\def \ZjjOOtotal {$98 \pm 19$}
\def \ZjjOOdata {$85$}
\def \ZjjjOOeedy {$12.7 \pm 2.7$}
\def \ZjjjOOeeqcd {$0.27 \pm 0.04$}
\def \ZjjjOOmumuwjjmctwo {$-$}
\def \ZjjjOOmumuzjjmctwo {$0.24 \pm 0.05$}
\def \ZjjjOOmumuwwmctwo {$-$}
\def \ZjjjOOmumutopmctwo {$0.018 \pm 0.009$}
\def \ZjjjOOtotal {$13.2 \pm 2.7$}
\def \ZjjjOOdata {$12$}
\def \ZjjjjOOeedy {$1.8 \pm 0.5$}
\def \ZjjjjOOeeqcd {$0.034 \pm 0.006$}
\def \ZjjjjOOmumuwjjmctwo {$-$}
\def \ZjjjjOOmumuzjjmctwo {$0.022 \pm 0.009$}
\def \ZjjjjOOmumuwwmctwo {$-$}
\def \ZjjjjOOmumutopmctwo {$0.006 \pm 0.004$}
\def \ZjjjjOOtotal {$1.9 \pm 0.5$}
\def \ZjjjjOOdata {$1$}
\def \ZjjjjjOOeedy {$0.26 \pm 0.10$}
\def \ZjjjjjOOeeqcd {$0.0025 \pm 0.0009$}
\def \ZjjjjjOOmumuwjjmctwo {$-$}
\def \ZjjjjjOOmumuzjjmctwo {$-$}
\def \ZjjjjjOOmumuwwmctwo {$-$}
\def \ZjjjjjOOmumutopmctwo {$-$}
\def \ZjjjjjOOtotal {$0.26 \pm 0.10$}
\def \ZjjjjjOOdata {$0$}
\def \ZjjjjjjOOeedy {$0.020 \pm 0.010$}
\def \ZjjjjjjOOeeqcd {$0.0036 \pm 0.0011$}
\def \ZjjjjjjOOmumuwjjmctwo {$-$}
\def \ZjjjjjjOOmumuzjjmctwo {$-$}
\def \ZjjjjjjOOmumuwwmctwo {$-$}
\def \ZjjjjjjOOmumutopmctwo {$-$}
\def \ZjjjjjjOOtotal {$0.024 \pm 0.010$}
\def \ZjjjjjjOOdata {$0$}
\def \ZjjjjjjjOOeedy {$0.0040 \pm 0.0020$}
\def \ZjjjjjjjOOeeqcd {$0.0019 \pm 0.0007$}
\def \ZjjjjjjjOOmumuwjjmctwo {$-$}
\def \ZjjjjjjjOOmumuzjjmctwo {$-$}
\def \ZjjjjjjjOOmumuwwmctwo {$-$}
\def \ZjjjjjjjOOmumutopmctwo {$-$}
\def \ZjjjjjjjOOtotal {$0.0059 \pm 0.0021$}
\def \ZjjjjjjjOOdata {$0$}

\begin{table}[htb]
\centering
\begin{tabular}{lcc}
Final State	& Total & Data \\ \hline
$Z\jj$  & \ZjjOOtotal & \ZjjOOdata \\
$Z\jjj$ & \ZjjjOOtotal & \ZjjjOOdata \\
$Z\jjjj$	& \ZjjjjOOtotal & \ZjjjjOOdata \\ 
$Z\jjjjj$ & \ZjjjjjOOtotal & \ZjjjjjOOdata \\ 
\end{tabular}
\caption{Expected backgrounds to the $Z\jj(nj)$ final states.}
\label{tbl:expectedBkgsZjjJ}
\end{table}

\subsection{Results}

The results of applying \Sherlock\ to the $Z \jj(nj)$ and $\ell\ell \jj(nj)$ data sets are summarized in Table~\ref{tbl:ZjjJResults} and Figs.~\ref{fig:data_mapping_eejj} and~\ref{fig:data_mapping_zjj}.  Figure~\ref{fig:data_mapping_eejj} shows the location of the data within the unit box for those final states in which the two leptons are not combined into a $Z$ boson, while Fig.~\ref{fig:data_mapping_zjj} displays the data for those final states in which a $Z$ boson has been identified.  Large $\scriptP$'s are found for most final states, as expected.  The smallest $\scriptP$'s in this class of final states are observed in the $ee\jjjj$ and $ee\met \jjjj$ final states.  Although the number of events is small, it is interesting to compare the number of events observed in the $Z$ + 2, 3, and 4 jet final states (showing good agreement with expected backgrounds) with the number of events observed in the $ee$ + 2, 3, and 4 jet and $ee\met$ + 2, 3, and 4 jet final states.  There is a small but statistically insignificant excess in final states with four jets --- we find in Sec.~\ref{section:Summary} that we expect to find at least one $\scriptP\ltapprox$\scriptPeejjjj\ in the analysis of so many final states.  Additionally, one of the three $ee\jjjj$ events has an $ee$ invariant mass barely outside the $Z$ boson mass window.  The kinematics of the events in the $ee\jjjj$ and $ee\met \jjjj$ final states are provided in Appendix~\ref{section:InterestingEvents}.

\begin{table}[htbp]
\centering
\begin{tabular}{cc}
Data set  	& $\scriptP$ \\ \hline
$ee \jj$ 	& \scriptPeejj	\\
$ee \jjj$ 	& \scriptPeejjj	\\
$ee \jjjj$ 	& \scriptPeejjjj \\
$ee\met \jj$ 	& \scriptPeemetjj \\
$ee\met \jjj$ 	& \scriptPeemetjjj \\
$ee\met \jjjj$ 	& \scriptPeemetjjjj \\
$\mu\mu \jj$ 	& \scriptPmumujj \\
$\mu\mu \jjj$ 	& \scriptPmumujjj \\
$Z \jj$		& \scriptPZjj \\
$Z \jjj$	& \scriptPZjjj \\
$Z \jjjj$	& \scriptPZjjjj \\
$Z \jjjjj$	& \scriptPZjjjjj \\
\end{tabular}
\caption{Summary of results on the $Z$+jets-like final states.}
\label{tbl:ZjjJResults}
\end{table}

{\dofig{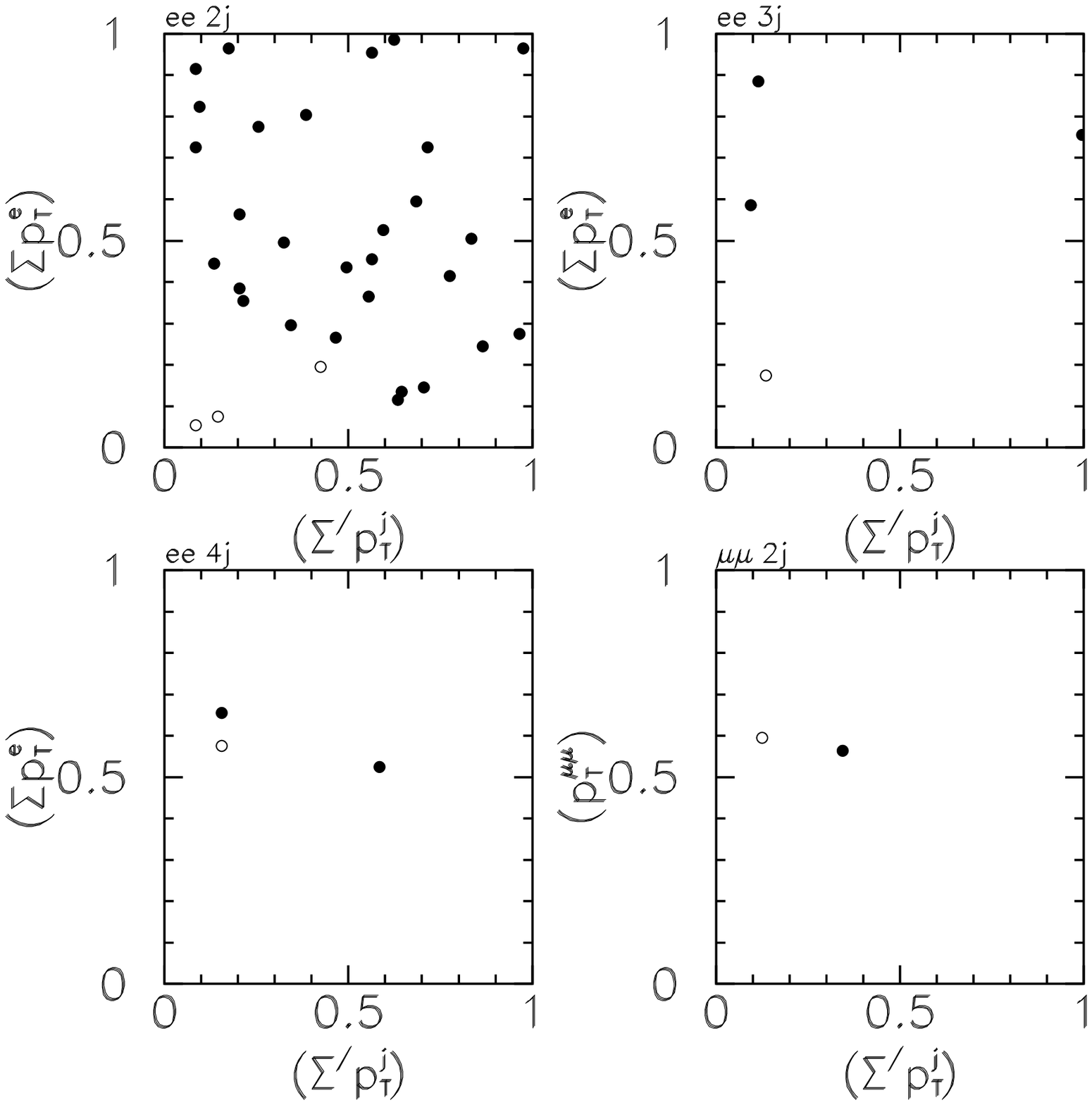} {3.5in} {The positions of the transformed data points in the final states $ee\jj$, $ee\jjj$, $ee\jjjj$, and $\mu\mu \jj$.  The data points inside the region chosen by \Sherlock\ are shown as filled circles; those outside the region are shown as open circles.} {fig:data_mapping_eejj}}

{\dofig{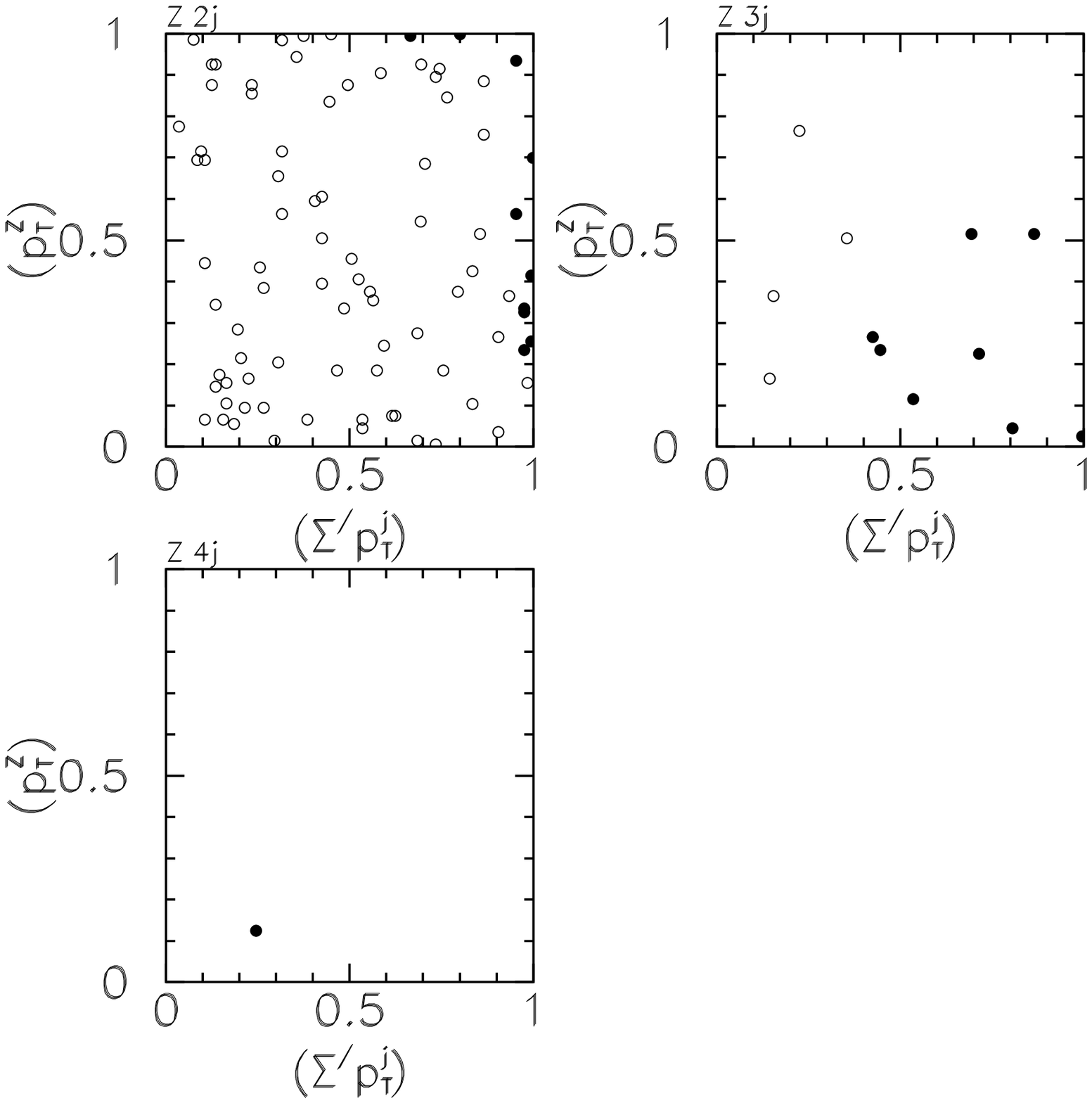} {3.5in} {The positions of the transformed data points in the final states $Z\jj$, $Z\jjj$, and $Z\jjjj$.  The data points inside the region chosen by \Sherlock\ are shown as filled circles; those outside the region are shown as open circles.} {fig:data_mapping_zjj}}


\subsection{Sensitivity check:  leptoquarks}

{\dofig{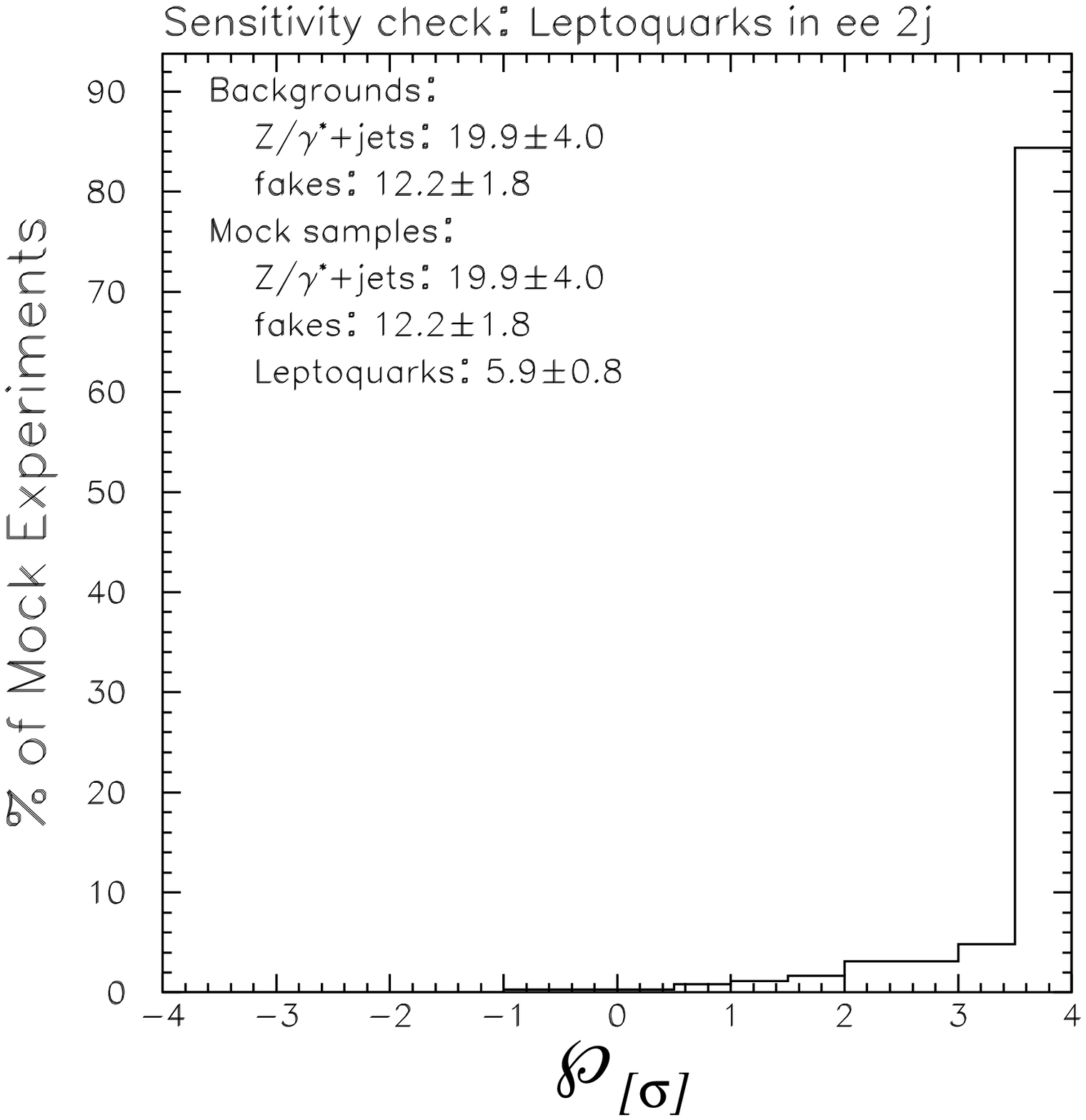} {3.5in} {Histogram of $\scriptP$ for an ensemble of mock experiments in which the backgrounds include $Z/\gamma^*$+jets and QCD fakes, and the mock samples include leptoquark pair production (with an assumed leptoquark mass of 170~GeV and $\beta=1$) in addition to the expected background.  All samples with $\scriptP>3.5\sigma$ are in the rightmost bin.  \Sherlock\ finds $\scriptP$ larger than 3.5 standard deviations in over 80\% of these mock samples.} {fig:mock_eejjlq}}

As a sensitivity check in the $Z$+jets-like final states we consider a scalar, first generation leptoquark~\cite{LQref1} of mass $m_{LQ}=170$~GeV, and assume a branching fraction to charged leptons of $\beta=1.0$.  The cross section for the process $q\bar{q}\rightarrow LQ\overline{LQ}$ with these parameters is $0.54$~pb.  The overall efficiency for this type of event is $24\pm4\%$~\cite{LeptoquarksToEE}, including trigger and object requirement efficiencies and geometric and kinematic acceptances.  If such a leptoquark were to exist, we would expect $11.2\pm1.5$ events of signal in the inclusive sample $ee\jj X$, of which $5.9\pm0.8$ events would fall in the exclusive final state $ee\jj$, on a background of \eejjOOtotal\ events.  Figure~\ref{fig:mock_eejjlq} shows the result of \Sherlock\ applied to an ensemble of mock experiments in this final state.  We see that \Sherlock\ finds $\scriptP$ larger than 3.5 standard deviations in over 80\% of these mock samples.

\section{{\boldmath $\ellgamma\ellgamma\ellgamma X$}}
\label{section:zoo}

In this section we analyze the $\ellgamma\ellgamma\ellgamma X$ final states.  After describing the data sets and background estimates, we provide the results obtained by applying \Sleuth\ to these channels.  We conclude the section with a sensitivity check [$X'\rightarrow\ellgamma\ellgamma\ellgamma X$] that is more general in nature than those provided for the $e\mu X$, $W$+jets-like, and $Z$+jets-like final states above.  Examples of a few of the many signals that might appear in these final states are provided in Appendix~\ref{section:SignalsThatMightAppearInlllX}.


\subsection{Data sets and background estimates}
\label{section:lllDataAndBkg}
\def \eeeOOzg {$2.6 \pm 1.0$}
\def \eeeOOegg {$-$}
\def \eeeOOggg {$-$}
\def \eeeOOzj {$-$}
\def \eeeOOwgg {$-$}
\def \eeeOOwz {$-$}
\def \eeeOOtotal {$2.6 \pm 1.0$}
\def \eeeOOdata {$1$}
\def \eemumetOOzg {$-$}
\def \eemumetOOegg {$-$}
\def \eemumetOOggg {$-$}
\def \eemumetOOzj {$-$}
\def \eemumetOOwgg {$-$}
\def \eemumetOOwz {$0.10 \pm 0.05$}
\def \eemumetOOtotal {$0.10 \pm 0.05$}
\def \eemumetOOdata {$0$}
\def \emumuOOzg {$-$}
\def \emumuOOegg {$-$}
\def \emumuOOggg {$-$}
\def \emumuOOzj {$-$}
\def \emumuOOwgg {$-$}
\def \emumuOOwz {$0.040 \pm 0.020$}
\def \emumuOOtotal {$0.040 \pm 0.020$}
\def \emumuOOdata {$0$}
\def \mumumuOOzg {$-$}
\def \mumumuOOegg {$-$}
\def \mumumuOOggg {$-$}
\def \mumumuOOzj {$-$}
\def \mumumuOOwgg {$-$}
\def \mumumuOOwz {$0.020 \pm 0.010$}
\def \mumumuOOtotal {$0.020 \pm 0.010$}
\def \mumumuOOdata {$0$}
\def \eegOOzg {$-$}
\def \eegOOegg {$-$}
\def \eegOOggg {$-$}
\def \eegOOzj {$-$}
\def \eegOOwgg {$-$}
\def \eegOOwz {$-$}
\def \eegOOtotal {$-$}
\def \eegOOdata {$1$}
\def \eegmetOOzg {$-$}
\def \eegmetOOegg {$-$}
\def \eegmetOOggg {$-$}
\def \eegmetOOzj {$-$}
\def \eegmetOOwgg {$-$}
\def \eegmetOOwz {$-$}
\def \eegmetOOtotal {$-$}
\def \eegmetOOdata {$1$}
\def \ZgOOzg {$-$}
\def \ZgOOegg {$-$}
\def \ZgOOggg {$-$}
\def \ZgOOzj {$-$}
\def \ZgOOwgg {$-$}
\def \ZgOOwz {$-$}
\def \ZgOOtotal {$-$}
\def \ZgOOdata {$3$}
\def \ZgjOOzg {$-$}
\def \ZgjOOegg {$-$}
\def \ZgjOOggg {$-$}
\def \ZgjOOzj {$-$}
\def \ZgjOOwgg {$-$}
\def \ZgjOOwz {$-$}
\def \ZgjOOtotal {$-$}
\def \ZgjOOdata {$1$}
\def \eggOOzg {$-$}
\def \eggOOegg {$10.7 \pm 2.1$}
\def \eggOOggg {$-$}
\def \eggOOzj {$-$}
\def \eggOOwgg {$-$}
\def \eggOOwz {$-$}
\def \eggOOtotal {$10.7 \pm 2.1$}
\def \eggOOdata {$6$}
\def \eggjOOzg {$2.3 \pm 0.7$}
\def \eggjOOegg {$-$}
\def \eggjOOggg {$-$}
\def \eggjOOzj {$-$}
\def \eggjOOwgg {$-$}
\def \eggjOOwz {$-$}
\def \eggjOOtotal {$2.3 \pm 0.7$}
\def \eggjOOdata {$4$}
\def \eggjjOOzg {$0.37 \pm 0.15$}
\def \eggjjOOegg {$-$}
\def \eggjjOOggg {$-$}
\def \eggjjOOzj {$-$}
\def \eggjjOOwgg {$-$}
\def \eggjjOOwz {$-$}
\def \eggjjOOtotal {$0.37 \pm 0.15$}
\def \eggjjOOdata {$1$}
\def \eggmetOOzg {$-$}
\def \eggmetOOegg {$-$}
\def \eggmetOOggg {$-$}
\def \eggmetOOzj {$-$}
\def \eggmetOOwgg {$0.026 \pm 0.010$}
\def \eggmetOOwz {$0.11 \pm 0.05$}
\def \eggmetOOtotal {$0.14 \pm 0.05$}
\def \eggmetOOdata {$1$}
\def \WggOOzg {$-$}
\def \WggOOegg {$-$}
\def \WggOOggg {$-$}
\def \WggOOzj {$-$}
\def \WggOOwgg {$0.045 \pm 0.015$}
\def \WggOOwz {$0.16 \pm 0.08$}
\def \WggOOtotal {$0.21 \pm 0.08$}
\def \WggOOdata {$1$}
\def \mumugOOzg {$3.9 \pm 0.9$}
\def \mumugOOegg {$-$}
\def \mumugOOggg {$-$}
\def \mumugOOzj {$-$}
\def \mumugOOwgg {$-$}
\def \mumugOOwz {$-$}
\def \mumugOOtotal {$3.9 \pm 0.9$}
\def \mumugOOdata {$0$}
\def \mumugjOOzg {$0.64 \pm 0.20$}
\def \mumugjOOegg {$-$}
\def \mumugjOOggg {$-$}
\def \mumugjOOzj {$-$}
\def \mumugjOOwgg {$-$}
\def \mumugjOOwz {$-$}
\def \mumugjOOtotal {$0.64 \pm 0.20$}
\def \mumugjOOdata {$0$}
\def \mumugjjOOzg {$0.13 \pm 0.04$}
\def \mumugjjOOegg {$-$}
\def \mumugjjOOggg {$-$}
\def \mumugjjOOzj {$-$}
\def \mumugjjOOwgg {$-$}
\def \mumugjjOOwz {$-$}
\def \mumugjjOOtotal {$0.13 \pm 0.04$}
\def \mumugjjOOdata {$0$}
\def \mumugjjjOOzg {$0.025 \pm 0.010$}
\def \mumugjjjOOegg {$-$}
\def \mumugjjjOOggg {$-$}
\def \mumugjjjOOzj {$-$}
\def \mumugjjjOOwgg {$-$}
\def \mumugjjjOOwz {$-$}
\def \mumugjjjOOtotal {$0.025 \pm 0.010$}
\def \mumugjjjOOdata {$0$}
\def \gggOOzg {$-$}
\def \gggOOegg {$-$}
\def \gggOOggg {$2.5 \pm 0.5$}
\def \gggOOzj {$-$}
\def \gggOOwgg {$-$}
\def \gggOOwz {$-$}
\def \gggOOtotal {$2.5 \pm 0.5$}
\def \gggOOdata {$2$}

\def \ZOOzg {$8.9 \pm 1.8$}
\def \ZOOzj {$0.021 \pm 0.006$}
\def \ZOOwz {$-$}
\def \ZOOtotal {$8.9 \pm 1.8$}
\def \ZOOdata {$5$}
\def \ZjOOzg {$2.1 \pm 0.6$}
\def \ZjOOzj {$0.0047 \pm 0.0014$}
\def \ZjOOwz {$-$}
\def \ZjOOtotal {$2.1 \pm 0.6$}
\def \ZjOOdata {$0$}
\def \ZjjOOzg {$-$}
\def \ZjjOOzj {$0.00054 \pm 0.00025$}
\def \ZjjOOwz {$-$}
\def \ZjjOOtotal {$0.00054 \pm 0.00025$}
\def \ZjjOOdata {$0$}
\def \ZgOOzg {$3.3 \pm 0.7$}
\def \ZgOOzj {$0.99 \pm 0.27$}
\def \ZgOOwz {$-$}
\def \ZgOOtotal {$4.3 \pm 0.7$}
\def \ZgOOdata {$3$}
\def \ZgjOOzg {$0.80 \pm 0.30$}
\def \ZgjOOzj {$0.23 \pm 0.06$}
\def \ZgjOOwz {$-$}
\def \ZgjOOtotal {$1.03 \pm 0.31$}
\def \ZgjOOdata {$1$}
\def \ZgjjOOzg {$0.10 \pm 0.05$}
\def \ZgjjOOzj {$0.033 \pm 0.009$}
\def \ZgjjOOwz {$-$}
\def \ZgjjOOtotal {$0.13 \pm 0.05$}
\def \ZgjjOOdata {$0$}
\def \ZgjjjOOzg {$0.020 \pm 0.010$}
\def \ZgjjjOOzj {$0.0048 \pm 0.0014$}
\def \ZgjjjOOwz {$-$}
\def \ZgjjjOOtotal {$0.025 \pm 0.010$}
\def \ZgjjjOOdata {$0$}
\def \ZgjjjjOOzg {$0.0040 \pm 0.0020$}
\def \ZgjjjjOOzj {$0.0009 \pm 0.0004$}
\def \ZgjjjjOOwz {$-$}
\def \ZgjjjjOOtotal {$0.0049 \pm 0.0020$}
\def \ZgjjjjOOdata {$0$}
\def \eegOOzg {$2.1 \pm 0.4$}
\def \eegOOzj {$0.13 \pm 0.04$}
\def \eegOOwz {$-$}
\def \eegOOtotal {$2.2 \pm 0.4$}
\def \eegOOdata {$1$}
\def \eegjOOzg {$0.50 \pm 0.25$}
\def \eegjOOzj {$0.033 \pm 0.009$}
\def \eegjOOwz {$-$}
\def \eegjOOtotal {$0.53 \pm 0.25$}
\def \eegjOOdata {$0$}
\def \eegjjOOzg {$0.10 \pm 0.05$}
\def \eegjjOOzj {$0.0046 \pm 0.0014$}
\def \eegjjOOwz {$-$}
\def \eegjjOOtotal {$0.10 \pm 0.05$}
\def \eegjjOOdata {$0$}
\def \eegjjjOOzg {$-$}
\def \eegjjjOOzj {$-$}
\def \eegjjjOOwz {$-$}
\def \eegjjjOOtotal {$-$}
\def \eegjjjOOdata {$0$}
\def \eegmetOOzg {$0.010 \pm 0.005$}
\def \eegmetOOzj {$0.024 \pm 0.007$}
\def \eegmetOOwz {$0.23 \pm 0.10$}
\def \eegmetOOtotal {$0.26 \pm 0.10$}
\def \eegmetOOdata {$1$}
\def \eegjmetOOzg {$0.0020 \pm 0.0010$}
\def \eegjmetOOzj {$0.012 \pm 0.003$}
\def \eegjmetOOwz {$0.045 \pm 0.020$}
\def \eegjmetOOtotal {$0.059 \pm 0.020$}
\def \eegjmetOOdata {$0$}
\def \eegjjmetOOzg {$-$}
\def \eegjjmetOOzj {$0.0035 \pm 0.0011$}
\def \eegjjmetOOwz {$-$}
\def \eegjjmetOOtotal {$0.0035 \pm 0.0011$}
\def \eegjjmetOOdata {$0$}
\def \eegjjjmetOOzg {$-$}
\def \eegjjjmetOOzj {$0.0012 \pm 0.0005$}
\def \eegjjjmetOOwz {$-$}
\def \eegjjjmetOOtotal {$0.0012 \pm 0.0005$}
\def \eegjjjmetOOdata {$0$}

The $\ellgamma\ellgamma\ellgamma X$ data set corresponds to an integrated luminosity of $123 \pm 7$~pb$^{-1}$.  Global cleanup cuts are imposed as above.  In this section we strictly adhere to standard particle identification criteria.  All objects (electrons, photons, muons, and jets) are required to have transverse momentum $\geq 15$~GeV, to be isolated, to be within the fiducial volume of the detector, and to be central.  For electrons and photons the fiducial requirement is $\abs{\eta_{\rm det}}<1.1$ or $1.5<\abs{\eta_{\rm det}}<2.5$; for muons it is $\abs{\eta_{\rm det}}<1.7$.  For the case of hadronic jets our centrality requirement of $\abs{\eta}<2.5$ is more stringent than the fiducial requirement of $\abs{\eta_{\rm det}} \ltapprox 4$.  We require electrons, photons, and muons to be separated by at least $0.4$ in $\Delta R = \sqrt{(\Delta\eta)^2 + (\Delta\phi)^2}$.  $\met$ is identified as an object if its magnitude is larger than 15~GeV.  The selection of events is facilitated by use of the database described in Ref.~\cite{BrownDatabase}.  

\def \fakeProbOOeTOe {$0.61 \pm 0.04$}
\def \fakeProbOOeTOg {$0.28 \pm 0.03$}
\def \fakeProbOOgTOe {$0.16 \pm 0.016$}
\def \fakeProbOOgTOg {$0.73 \pm 0.012$}
\def \fakeProbOOjTOe {$0.00035 \pm 0.000035$}
\def \fakeProbOOjTOg {$0.00125 \pm 0.00013$}

\begin{table}[htb]
\centering
\begin{tabular}{ccc}
 	& $e$ 	& $\gamma$ \\ \hline
$e$	& \fakeProbOOeTOe~\cite{LeptoquarksToENu} & \fakeProbOOeTOg~\cite{Monopole} \\
$\gamma$& \fakeProbOOgTOe~\cite{Monopole} & \fakeProbOOgTOg~\cite{Monopole} \\
$j$	& \fakeProbOOjTOe~\cite{LeptoquarksToENu} & \fakeProbOOjTOg~\cite{LeptoquarksToENu} \\
\end{tabular}
\caption{(Mis)identification probabilities.  The number at (row $i$, column $j$) is the probability that the object labeling row $i$ will be reconstructed as the object labeling column $j$.}
\label{tbl:misidentificationProbabilities}
\end{table}

We make frequent use of the (mis)identification probabilities determined for these identification criteria, which are summarized in Table~\ref{tbl:misidentificationProbabilities}.  

\subsubsection{$ee\gamma X$}
\label{section:eeg}

\BKwidetext
\def \ZOOzg {$8.9 \pm 1.8$}
\def \ZOOzj {$0.021 \pm 0.006$}
\def \ZOOwz {$-$}
\def \ZOOtotal {$8.9 \pm 1.8$}
\def \ZOOdata {$5$}
\def \ZjOOzg {$2.1 \pm 0.6$}
\def \ZjOOzj {$0.0047 \pm 0.0014$}
\def \ZjOOwz {$-$}
\def \ZjOOtotal {$2.1 \pm 0.6$}
\def \ZjOOdata {$0$}
\def \ZjjOOzg {$-$}
\def \ZjjOOzj {$0.00054 \pm 0.00025$}
\def \ZjjOOwz {$-$}
\def \ZjjOOtotal {$0.00054 \pm 0.00025$}
\def \ZjjOOdata {$0$}
\def \ZgOOzg {$3.3 \pm 0.7$}
\def \ZgOOzj {$0.99 \pm 0.27$}
\def \ZgOOwz {$-$}
\def \ZgOOtotal {$4.3 \pm 0.7$}
\def \ZgOOdata {$3$}
\def \ZgjOOzg {$0.80 \pm 0.30$}
\def \ZgjOOzj {$0.23 \pm 0.06$}
\def \ZgjOOwz {$-$}
\def \ZgjOOtotal {$1.03 \pm 0.31$}
\def \ZgjOOdata {$1$}
\def \ZgjjOOzg {$0.10 \pm 0.05$}
\def \ZgjjOOzj {$0.033 \pm 0.009$}
\def \ZgjjOOwz {$-$}
\def \ZgjjOOtotal {$0.13 \pm 0.05$}
\def \ZgjjOOdata {$0$}
\def \ZgjjjOOzg {$0.020 \pm 0.010$}
\def \ZgjjjOOzj {$0.0048 \pm 0.0014$}
\def \ZgjjjOOwz {$-$}
\def \ZgjjjOOtotal {$0.025 \pm 0.010$}
\def \ZgjjjOOdata {$0$}
\def \ZgjjjjOOzg {$0.0040 \pm 0.0020$}
\def \ZgjjjjOOzj {$0.0009 \pm 0.0004$}
\def \ZgjjjjOOwz {$-$}
\def \ZgjjjjOOtotal {$0.0049 \pm 0.0020$}
\def \ZgjjjjOOdata {$0$}
\def \eegOOzg {$2.1 \pm 0.4$}
\def \eegOOzj {$0.13 \pm 0.04$}
\def \eegOOwz {$-$}
\def \eegOOtotal {$2.2 \pm 0.4$}
\def \eegOOdata {$1$}
\def \eegjOOzg {$0.50 \pm 0.25$}
\def \eegjOOzj {$0.033 \pm 0.009$}
\def \eegjOOwz {$-$}
\def \eegjOOtotal {$0.53 \pm 0.25$}
\def \eegjOOdata {$0$}
\def \eegjjOOzg {$0.10 \pm 0.05$}
\def \eegjjOOzj {$0.0046 \pm 0.0014$}
\def \eegjjOOwz {$-$}
\def \eegjjOOtotal {$0.10 \pm 0.05$}
\def \eegjjOOdata {$0$}
\def \eegjjjOOzg {$-$}
\def \eegjjjOOzj {$-$}
\def \eegjjjOOwz {$-$}
\def \eegjjjOOtotal {$-$}
\def \eegjjjOOdata {$0$}
\def \eegmetOOzg {$0.010 \pm 0.005$}
\def \eegmetOOzj {$0.024 \pm 0.007$}
\def \eegmetOOwz {$0.23 \pm 0.10$}
\def \eegmetOOtotal {$0.26 \pm 0.10$}
\def \eegmetOOdata {$1$}
\def \eegjmetOOzg {$0.0020 \pm 0.0010$}
\def \eegjmetOOzj {$0.012 \pm 0.003$}
\def \eegjmetOOwz {$0.045 \pm 0.020$}
\def \eegjmetOOtotal {$0.059 \pm 0.020$}
\def \eegjmetOOdata {$0$}
\def \eegjjmetOOzg {$-$}
\def \eegjjmetOOzj {$0.0035 \pm 0.0011$}
\def \eegjjmetOOwz {$-$}
\def \eegjjmetOOtotal {$0.0035 \pm 0.0011$}
\def \eegjjmetOOdata {$0$}
\def \eegjjjmetOOzg {$-$}
\def \eegjjjmetOOzj {$0.0012 \pm 0.0005$}
\def \eegjjjmetOOwz {$-$}
\def \eegjjjmetOOtotal {$0.0012 \pm 0.0005$}
\def \eegjjjmetOOdata {$0$}

\begin{table*}[htb]
\centering
\begin{tabular}{cccccc}
Final State	& $Z\gamma$ 	& $Zj$		& $WZ$		& Total 	& Data \\ \hline
$Z\gamma$	& \ZgOOzg	& \ZgOOzj	& \ZgOOwz 	& \ZgOOtotal 	& \ZgOOdata	\\
$ee\gamma$	& \eegOOzg	& \eegOOzj	& \eegOOwz 	& \eegOOtotal 	& \eegOOdata \\
$Z\gamma j$	& \ZgjOOzg	& \ZgjOOzj	& \ZgjOOwz 	& \ZgjOOtotal 	& \ZgjOOdata	\\
$ee\gamma j$	& \eegjOOzg	& \eegjOOzj	& \eegjOOwz 	& \eegjOOtotal 	& \eegjOOdata \\
$ee\gamma\met$	& \eegmetOOzg	& \eegmetOOzj	& \eegmetOOwz 	& \eegmetOOtotal 	& \eegmetOOdata \\
\end{tabular}
\caption{Expected backgrounds for the $ee\gamma X$ final states.}
\label{tbl:expectedBkgseegX}
\end{table*}
\BKnarrowtext

The dominant background to $ee\gamma X$ is the standard model process $Z/\gamma^*(\rightarrow ee)\gamma$.  We use a matrix element Monte Carlo~\cite{BaurZgammaMC} to estimate this background.  The $p\bar{p}\rightarrow Z/\gamma^*(\rightarrow ee)\gamma$ cross section, multiplied by our kinematic and geometric acceptance, is $0.50\pm0.05$~pb.  From Table~\ref{tbl:misidentificationProbabilities}, the probability for two true electrons and one true photon to be reconstructed as two electrons and one photon is $0.33$.  From these numbers we estimate the expected background from this process into the $ee\gamma X$ final states to be $14.3\pm2.9$ events.  Of these, $7.6\pm1.5$ events have $m_{ee}<82$~GeV or $m_{ee}>100$~GeV, and $82 < m_{ee\gamma} < 100$~GeV.  Following the prescription in Appendix~\ref{section:finalStateDefinitions}, such events are placed in the $Z$ final state, and are not considered in this section.

A smaller background in these final states is $Z$+jets production, with the jet faking a photon.  From Ref.~\cite{Zpt}, we expect $1100\pm200$ $Z(\rightarrow ee)$+1 jet events in our data; the probability that this jet will fake a photon is given in Table~\ref{tbl:misidentificationProbabilities}.  We therefore expect \ZgOOzj\ events of background in $Z\gamma$ from this source, and \eegOOzj\ events (roughly $10\%$ of the number expected in $Z\gamma$, determined by {\sc pythia}) in $ee\gamma$.  

The dominant background to the $ee\gamma\met$ final state comes from $W(\rightarrow e\nu)Z(\rightarrow ee)$, in which one of the three electrons is reconstructed as a photon.  The $WZ$ production cross section in the standard model is calculated to be 2.5~pb~\cite{AnomalousWWandWZ}; D\O's geometric acceptance for these events is determined using {\sc pythia}.  Using the (mis)identification probabilities in Table~\ref{tbl:misidentificationProbabilities}, we estimate the contribution from standard model $WZ$ production to this final state to be \eegmetOOwz\ events.

The numbers of expected background events in final states with additional jets are obtained by multiplying by a factor of $1/5$ for each additional jet.  The number of events expected in each final state, together with the number of events observed in the data, is given in Table~\ref{tbl:expectedBkgseegX}.  We find good agreement between the expected background and the numbers of events observed in the data.

\subsubsection{$\mu\mu\gamma X$}

The dominant background to the $\mu\mu\gamma X$ final states is standard model $Z/\gamma^*(\rightarrow \mu\mu)\gamma$.  The matrix element Monte Carlo used to estimate the backgrounds to $ee\gamma X$ is also used for this final state.  The normalization is determined by multiplying the number of expected $Z/\gamma^*(\rightarrow ee)\gamma$ events by the square of the ratio of efficiency $\times$ acceptance for muons and electrons.  For muons, the efficiency $\times$ acceptance is roughly $0.5\times0.5$; for electrons, the number is approximately $0.6\times0.8$.  The number of expected events in $\mu\mu\gamma$ is thus \mumugOOtotal.  No events are seen in this final state.  The probability of seeing zero events when \mumugOOtotal\ are expected is 2.8\%.

\subsubsection{$e\gamma\gamma X$}

The dominant background to $e\gamma\gamma X$ is the standard model process $Z/\gamma^*(\rightarrow ee)\gamma$, where one of the electrons is reconstructed as a photon.  From Table~\ref{tbl:misidentificationProbabilities} and the $Z(\rightarrow ee)\gamma$ estimate in Sec.~\ref{section:eeg}, we determine the number of expected events in the $e\gamma\gamma$ final state to be
$10.7\pm2.1$ events.  Twelve $e\gamma\gamma X$ events are seen in the data, appearing in the final states shown in Table~\ref{tbl:egammagammaX_population}.  We model the $e\gamma\gamma$ backgrounds with the Monte Carlo used for the $ee\gamma X$ final states above.  

Three of the events in the $e\gamma\gamma j$ final state have $m_{e\gamma_1\gamma_2}=95.8$~GeV, $m_{e\gamma_1\gamma_2}=85.9$~GeV, and $m_{e\gamma_1}=97.9$~GeV, respectively, and are consistent with $Z\gamma$ production with a radiated jet.  The invariant masses of the objects in the fourth event all lie substantially outside the $Z$ boson mass window.  Lacking an adequate $Z(\rightarrow ee)\gamma j$ Monte Carlo, we simply calculate the probability that the expected background fluctuates up to or above the observed number of events in this final state.  The single event in the $e\gamma\gamma \jj$ final state has $m_{e\gamma_1\gamma_2}=92.4$~GeV; this appears to be a $Z$ boson produced in association with two jets.  

One event in this sample contains significant $\met$ in addition to one electron and two photons.  In this event $m_{e\gamma_1}=95.9$~GeV, but the missing transverse energy in the event is large, and directly opposite the electron in $\phi$.  The transverse mass $m_T^{e\nu}=71.9$~GeV, so this event falls in the $W\gamma\gamma$ final state.  The dominant background to this final state is $W(\rightarrow e\nu)Z(\rightarrow ee)$, in which two electrons are reconstructed as photons; the number of such events expected in this final state is determined to be \eggmetOOwz.  $W(\rightarrow e\nu)\gamma\gamma$ is a slightly smaller but comparable background to this final state, which we estimate using a matrix element Monte Carlo~\cite{BaurWggMC}.  The total cross section for $W(\rightarrow e\nu)\gamma\gamma$ with all three detected objects in the fiducial region of the detector and $\met>15$~GeV is determined to be $0.77\pm0.08$~fb.  The number of $W(\rightarrow e\nu)\gamma\gamma$ events in our data is therefore expected to be \eggmetOOwgg.  Backgrounds from $W\gamma j$ and $W\jj$, where the jets fake photons, are comparable but smaller.  This event will be combined in the next section with any events containing one muon and two photons to form the $W\gamma\gamma$ final state.

\begin{table}[htb]
\centering
\begin{tabular}{ccc}
Final state & Bkg & Data \\ \hline
$e\gamma\gamma$ & \eggOOtotal  & \eggOOdata \\
$W(\rightarrow e\nu)\gamma\gamma$ & \eggmetOOtotal & \WggOOdata \\
$e\gamma\gamma j$ & \eggjOOtotal & \eggjOOdata \\
$e\gamma\gamma \jj$ & \eggjjOOtotal & \eggjjOOdata \\
\end{tabular}
\caption{Population of final states within $e\gamma\gamma X$.}
\label{tbl:egammagammaX_population}
\end{table}

\subsubsection{$\mu\met\gamma\gamma X$}

The dominant backgrounds to the $\mu\met\gamma\gamma X$ final states, like those from the $e\met\gamma\gamma X$ final states, come from $WZ$ and from a $W$ boson produced in association with two photons.  The number of expected events from $WZ$ is determined as above to be $0.05\pm0.02$.  The background from standard model $W\gamma\gamma$ is estimated by multiplying the number of expected $W(\rightarrow e\nu)\gamma\gamma$ events above by the ratio of efficiency $\times$ acceptance for electrons and muons.  

Adding the number of events expected from $W(\rightarrow e\nu)\gamma\gamma$ to the number of events expected from $W(\rightarrow \mu\nu)\gamma\gamma$, we find the total number of expected background events in the $W\gamma\gamma$ final state to be \WggOOtotal.  No events are seen in the muon channel, so the only event in this final state is the event in the electron channel described above.  

\subsubsection{$\gamma\gamma\gamma X$}

The dominant background to $\gamma\gamma\gamma$ is the standard model process $Z/\gamma^*(\rightarrow ee)\gamma$, where both of the electrons are reconstructed as photons.  Taking the probability of an electron faking a photon from Table~\ref{tbl:misidentificationProbabilities} and using the number of $Z/\gamma^*(\rightarrow ee)\gamma$ events determined above, we find the number of expected events in this final state from this process to be \gggOOtotal\ events.  The contributions from $\jjj$, $\gamma \jj$, and $\gamma\gamma j$ are smaller by an order of magnitude.  

Two events are seen in the data, both in the final state $\gamma\gamma\gamma$.  One of these events has a three-body invariant mass $m_{\gamma\gamma\gamma}=100.4$~GeV, consistent with the expectation that it is truly a $Z\gamma$ event.  The other has a three-body invariant mass $m_{\gamma\gamma\gamma}=153$~GeV, but two photons may be chosen whose two-body invariant mass is $m_{\gamma\gamma}=90.3$~GeV.  This event also appears to fit the $Z\gamma$ hypothesis.  

\subsubsection{$eeeX$}

The dominant background to the final state $eee$ is again $Z/\gamma^*(\rightarrow ee)\gamma$, where this time the photon is reconstructed as an electron.  The cross section quoted above for $Z/\gamma^*(\rightarrow ee)\gamma$, folded with the (mis)identification probabilities from Table~\ref{tbl:misidentificationProbabilities}, predicts \eeeOOtotal\ events expected in the final state $eee$.  One event is seen in the data.  The $eee$ invariant mass in this event is 87.6~GeV, consistent with the standard model process $Z/\gamma^*(\rightarrow ee)\gamma$, where the photon is reconstructed as an electron.

\subsubsection{$\mu\mu\mu X$}

The dominant background to $\mu\mu\mu$ is standard model $WZ$ production.  We use the $WZ$ production cross section above and take our efficiency $\times$ acceptance for picking up all three muons in the event to be roughly $(0.5\times0.5)^3=0.02$.  The total number of expected background events in $\mu\mu\mu$ from $WZ$ production is thus \mumumuOOwz\ events.  Zero events are seen in the data.

\vskip 0.5cm

The only populated final states within $\gamma\gamma\gamma X$, $eeeX$, and $\mu\mu\mu X$ are $\gamma\gamma\gamma$ and $eee$; these are summarized in Table~\ref{tbl:threeLikeObjectFinalStates_population}.

\begin{table}[htb]
\centering
\begin{tabular}{ccc}
Final state & Bkg & Data \\ \hline
$\gamma\gamma\gamma$ & \gggOOtotal  & \gggOOdata \\
$eee$ & \eeeOOtotal  & \eeeOOdata \\
\end{tabular}
\caption{Population of final states with three like objects.}
\label{tbl:threeLikeObjectFinalStates_population}
\end{table}

\subsection{Results}

Having estimated the backgrounds to each of these final states, we proceed to apply \Sleuth\ to the data.  Large $\scriptP$'s are determined for all final states, indicating no hints of new physics within $\ellgamma\ellgamma\ellgamma X$.  Table~\ref{tbl:zooResults} summarizes the results.  We note that {\em all} final states within $\ellgamma\ellgamma\ellgamma X$ have been analyzed, including (for example) $ee\gamma\gamma\met$ and $\mu\mu\gamma\gamma \jj$.  All final states within $\ellgamma\ellgamma\ellgamma X$ not listed in Table~\ref{tbl:zooResults} are unpopulated, and have $\scriptP=1.00$.

\begin{table}[htb]
\centering
\begin{tabular}{cc}
Data set  & $\scriptP$ \\ \hline
$\gamma\gamma\gamma$ & \scriptPggg \\
$eee$ & \scriptPeee \\
$Z\gamma$ & \scriptPZg \\
$Z\gamma j$ & \scriptPZgj \\
$ee\gamma$ & \scriptPeeg \\
$ee\gamma\met$	& \scriptPeegmet \\
$e\gamma\gamma$ & \scriptPegg \\
$e\gamma\gamma j$ & \scriptPeggj \\
$e\gamma\gamma \jj$ & \scriptPeggjj \\
$W\gamma\gamma$ & \scriptPWgg \\
\end{tabular}
\caption{Summary of results on the $\ellgamma\ellgamma\ellgamma X$ final states.}
\label{tbl:zooResults}
\end{table}

\subsection{\mbox{Sensitivity check:  $X'\rightarrow\ellgamma\ellgamma\ellgamma X$}}
\label{section:lllSensitivityCheck}

The backgrounds to the $\ellgamma\ellgamma\ellgamma X$ final states are sufficiently small that a signal present even at the level of one or two events can be significant.  Due to the variety of final states treated in this section and the many processes that could produce signals in one or more of these final states, our sensitivity check for this section is the general process $X'\rightarrow\ellgamma\ellgamma\ellgamma X$, rather than a specific process such as $p\bar{p}\rightarrow \gaugino^0_2 \gaugino^{\pm}_1 \rightarrow \ell\ell\ell'\met$.  We (pessimistically) take the kinematics of the final state particles to be identical to the kinematics of the standard model background.  In reality the final state objects in the signal are expected to have significantly larger momenta than those in the backgrounds, and the calculated $\scriptP$ will be correspondingly smaller.  With this minimal assumption about the kinematics of the signal, the details of the \Sherlock\ algorithm are irrelevant, and $\scriptP$ is given on average by the probability that the background fluctuates up to or above the number of expected background events plus the number of expected signal events.  

The quantity $\gothicP$ obtained by combining the $\scriptP$'s calculated in all final states is a very different measure of ``significance'' than the measure familiar to most high energy physicists.  The fact that a ``significance'' of five standard deviations is unofficially but generally accepted as the threshold for a discovery results from a rough collective accounting of the number of different places such an effect could appear.  We can better understand this accounting by first noting that five standard deviations corresponds to a (one-sided) probability of $3\times10^{-7}$.  We then estimate that there are {\em at least} $5\times10^3$ distinct regions in the many variable spaces that are considered in a multipurpose experiment such as D\O\ in which one could realistically claim to see a signal.  A probability of $1.5\times10^{-3}$, in turn, corresponds to three standard deviations.  We can therefore understand the desire for a ``5$\sigma$ effect'' in our field to really be a desire for a ``3$\sigma$ effect'' (one time in one thousand), after a rigorous accounting for the number of places that such an effect might appear.  

One of the advantages of \Sherlock\ is that this rigorous accounting is explicitly performed.  The final output of \Sherlock\ takes the form of single number, $\gothicP$, which is ``the fraction of hypothetical similar experimental runs in which you would see something as interesting as what you actually saw in the data.''  The discussion in the preceding paragraph suggests that finding $\gothicP\geq3\sigma$ is as improbable (if not more so) as finding a ``5$\sigma$ effect.'' 

The number of final states that we consider, together with the number of background events expected in each, defines the mapping between $\scriptP_{\rm min}$ (the smallest $\scriptP$ found in any final state) and $\gothicP$.  For the final states that we have considered in this article, this mapping is shown in Fig.~\ref{fig:addtwiddle}.  We see that finding $\gothicP\geq3\sigma$ requires finding $\scriptP\geq4.2\sigma$ in some final state.

{\dofig{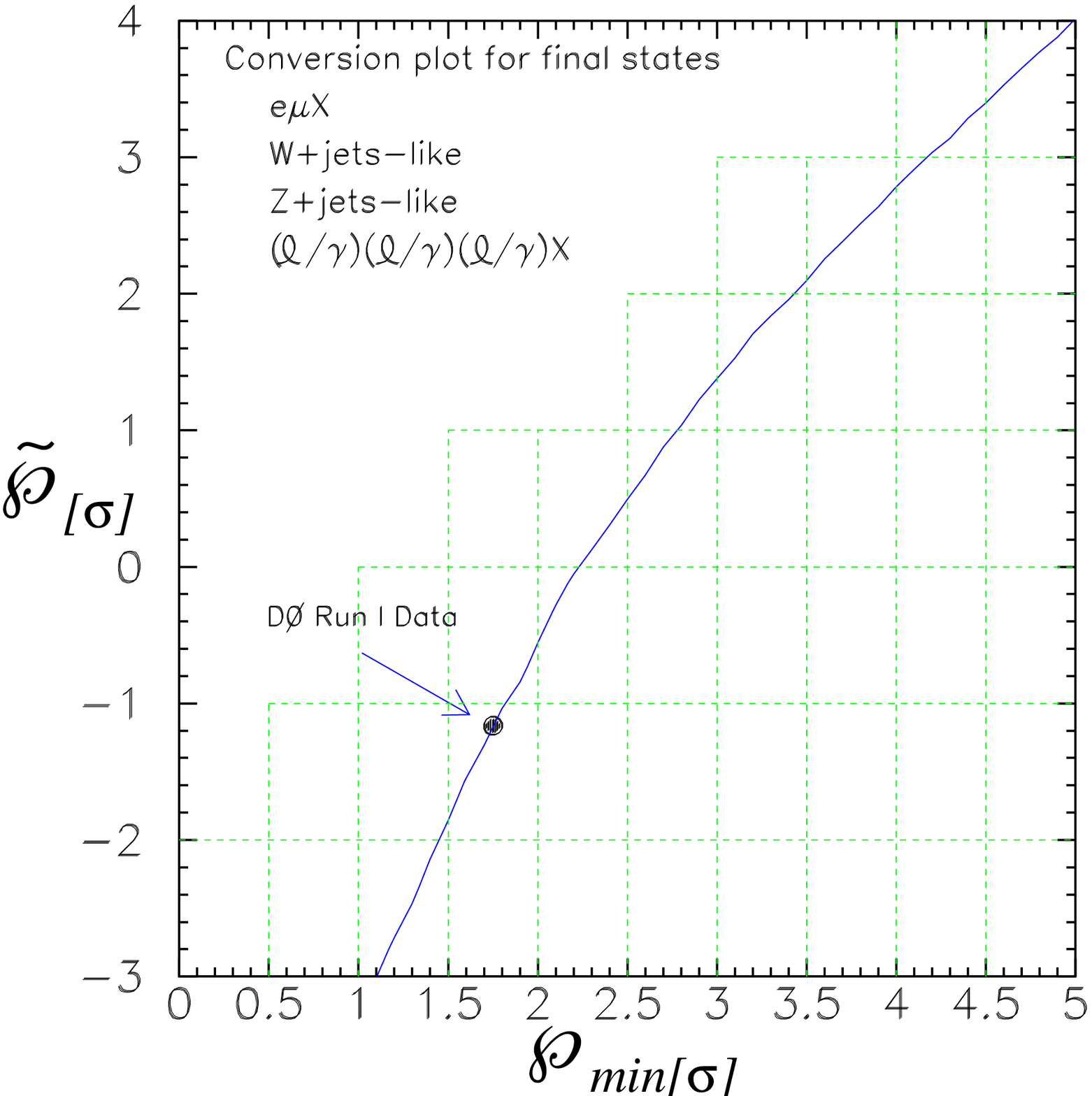} {3.5in} {Correspondence between $\scriptP_{\rm min}$ and $\gothicP$ for the final states we have considered.} {fig:addtwiddle} }

Let $N_Y$ be the smallest integer for which the probability that the background in the final state $Y$ fluctuates up to or above the expected background $\hat{b}$ plus $N_Y$ is $\leq 1.5\times10^{-5}$ ($4.2\sigma$).  This is the number of events which, if observed in $Y$, would correspond to a discovery.  This number can be related to the most probable cross section $\sigma_\vartheta$ of the new process $\vartheta$ into the final state $Y$ through 
\begineq
\sigma_\vartheta = \frac{N_Y}{a_\vartheta \epsilon_Y {\cal L}},
\endeq
where $a_\vartheta$ are the appropriate kinematic and geometric acceptance factors for the process $\vartheta$ and the D\O\ detector, $\epsilon_Y$ is the probability that the objects in the true final state $Y$ will be correctly reconstructed (which can be determined using Table~\ref{tbl:misidentificationProbabilities}), and ${\cal L}\approx85$~pb$^{-1}$ is the effective luminosity of the D\O\ data after application of global cleanup cuts.  The numbers $N_Y$ for some of the final states within $\ellgamma\ellgamma\ellgamma X$ are given in Table~\ref{tbl:NumberOfEventsRequiredToMakeADiscovery}.  (These final states are all unpopulated in the D\O\ data.)  Even with our pessimistic assumptions, using the \Sherlock\ strategy but setting aside the sophisticated \Sherlock\ algorithm, we see that a discovery could have been made had even a few signal events populated one of these channels.

\def \ZOOzg {$8.9 \pm 1.8$}
\def \ZOOzj {$0.021 \pm 0.006$}
\def \ZOOwz {$-$}
\def \ZOOtotal {$8.9 \pm 1.8$}
\def \ZOOdata {$5$}
\def \ZjOOzg {$2.1 \pm 0.6$}
\def \ZjOOzj {$0.0047 \pm 0.0014$}
\def \ZjOOwz {$-$}
\def \ZjOOtotal {$2.1 \pm 0.6$}
\def \ZjOOdata {$0$}
\def \ZjjOOzg {$-$}
\def \ZjjOOzj {$0.00054 \pm 0.00025$}
\def \ZjjOOwz {$-$}
\def \ZjjOOtotal {$0.00054 \pm 0.00025$}
\def \ZjjOOdata {$0$}
\def \ZgOOzg {$3.3 \pm 0.7$}
\def \ZgOOzj {$0.99 \pm 0.27$}
\def \ZgOOwz {$-$}
\def \ZgOOtotal {$4.3 \pm 0.7$}
\def \ZgOOdata {$3$}
\def \ZgjOOzg {$0.80 \pm 0.30$}
\def \ZgjOOzj {$0.23 \pm 0.06$}
\def \ZgjOOwz {$-$}
\def \ZgjOOtotal {$1.03 \pm 0.31$}
\def \ZgjOOdata {$1$}
\def \ZgjjOOzg {$0.10 \pm 0.05$}
\def \ZgjjOOzj {$0.033 \pm 0.009$}
\def \ZgjjOOwz {$-$}
\def \ZgjjOOtotal {$0.13 \pm 0.05$}
\def \ZgjjOOdata {$0$}
\def \ZgjjjOOzg {$0.020 \pm 0.010$}
\def \ZgjjjOOzj {$0.0048 \pm 0.0014$}
\def \ZgjjjOOwz {$-$}
\def \ZgjjjOOtotal {$0.025 \pm 0.010$}
\def \ZgjjjOOdata {$0$}
\def \ZgjjjjOOzg {$0.0040 \pm 0.0020$}
\def \ZgjjjjOOzj {$0.0009 \pm 0.0004$}
\def \ZgjjjjOOwz {$-$}
\def \ZgjjjjOOtotal {$0.0049 \pm 0.0020$}
\def \ZgjjjjOOdata {$0$}
\def \eegOOzg {$2.1 \pm 0.4$}
\def \eegOOzj {$0.13 \pm 0.04$}
\def \eegOOwz {$-$}
\def \eegOOtotal {$2.2 \pm 0.4$}
\def \eegOOdata {$1$}
\def \eegjOOzg {$0.50 \pm 0.25$}
\def \eegjOOzj {$0.033 \pm 0.009$}
\def \eegjOOwz {$-$}
\def \eegjOOtotal {$0.53 \pm 0.25$}
\def \eegjOOdata {$0$}
\def \eegjjOOzg {$0.10 \pm 0.05$}
\def \eegjjOOzj {$0.0046 \pm 0.0014$}
\def \eegjjOOwz {$-$}
\def \eegjjOOtotal {$0.10 \pm 0.05$}
\def \eegjjOOdata {$0$}
\def \eegjjjOOzg {$-$}
\def \eegjjjOOzj {$-$}
\def \eegjjjOOwz {$-$}
\def \eegjjjOOtotal {$-$}
\def \eegjjjOOdata {$0$}
\def \eegmetOOzg {$0.010 \pm 0.005$}
\def \eegmetOOzj {$0.024 \pm 0.007$}
\def \eegmetOOwz {$0.23 \pm 0.10$}
\def \eegmetOOtotal {$0.26 \pm 0.10$}
\def \eegmetOOdata {$1$}
\def \eegjmetOOzg {$0.0020 \pm 0.0010$}
\def \eegjmetOOzj {$0.012 \pm 0.003$}
\def \eegjmetOOwz {$0.045 \pm 0.020$}
\def \eegjmetOOtotal {$0.059 \pm 0.020$}
\def \eegjmetOOdata {$0$}
\def \eegjjmetOOzg {$-$}
\def \eegjjmetOOzj {$0.0035 \pm 0.0011$}
\def \eegjjmetOOwz {$-$}
\def \eegjjmetOOtotal {$0.0035 \pm 0.0011$}
\def \eegjjmetOOdata {$0$}
\def \eegjjjmetOOzg {$-$}
\def \eegjjjmetOOzj {$0.0012 \pm 0.0005$}
\def \eegjjjmetOOwz {$-$}
\def \eegjjjmetOOtotal {$0.0012 \pm 0.0005$}
\def \eegjjjmetOOdata {$0$}

\begin{table}[htb]
\centering
\begin{tabular}{ccc}
Final State	& $\hat{b}$ 	& $N$	\\ \hline
$ee\gamma j\met$& \eegjmetOOtotal & 4 	\\
$ee\gamma \jj$	& \eegjjOOtotal & 4 	\\
$Z\gamma \jj$	& \ZgjjOOtotal  & 5 	\\
$Z\gamma \jjj$	& \ZgjjjOOtotal & 3	\\
$Z\gamma \jjjj$	& \ZgjjjjOOtotal & 3 	\\
$ee\mu\met$	& \eemumetOOtotal & 4 	\\
$e\mu\mu$	& \emumuOOtotal & 4	\\
$\mu\mu\mu$	& \mumumuOOtotal & 3	\\
$W\gamma\gamma$ & \WggOOtotal 	& 5	\\
\end{tabular}
\caption{The number of signal events $N$ required in some of the final states within $\ellgamma\ellgamma\ellgamma X$ in order to find $\gothicP\geq 3\sigma$ (see the discussion in the text).  This number is pessimistic, as it assumes that the signal is distributed identically to the backgrounds in the variables of interest.  Most tenable models predict events containing final state objects that are significantly more energetic than the backgrounds, and in this case $N$ decreases accordingly.}
\label{tbl:NumberOfEventsRequiredToMakeADiscovery}
\end{table}

\section{Summary}
\label{section:Summary}

Table~\ref{tbl:FinalResults} summarizes the values of $\scriptP$ obtained for all populated final states analyzed in this article.  Taking into account the many final states (both populated and unpopulated) that have been considered in this analysis, we find $\gothicP=\mbox{\twiddleScriptPvalue}$ (\twiddleScriptPvalueSigma $\sigma$).  Figure~\ref{fig:histogram_of_scriptps} shows a histogram of the $\scriptP$'s computed for the populated final states analyzed in this article, together with the distribution expected from a simulation of many mock experimental runs.  Good agreement is observed.  

\begin{table}[htb]
\centering
\begin{tabular}{ll} 
Data set  	& \multicolumn{1}{c}{$\scriptP$} \\ \hline
\multicolumn{2}{c}{$e\mu X$} \\
$e\mu\met$   	& \scriptPemumet\ 	(\scriptPemumetsigma$\sigma$) \\
$e\mu\met j$  	& \scriptPemumetj\	(\scriptPemumetjsigma$\sigma$)\\
$e\mu\met \jj$ 	& \scriptPemumetjj\ 	(\scriptPemumetjjsigma$\sigma$)\\
$e\mu\met \jjj$  & \scriptPemumetjjj\ 	(\scriptPemumetjjjsigma$\sigma$)\\
\multicolumn{2}{c}{$W$+jets-like} \\
$W \jj$		& \scriptPWjj\		(\scriptPWjjsigma$\sigma$)\\
$W \jjj$	& \scriptPWjjj\		(\scriptPWjjjsigma$\sigma$)\\
$W \jjjj$	& \scriptPWjjjj\		(\scriptPWjjjjsigma$\sigma$)\\
$W \jjjjj$	& \scriptPWjjjjj\ 	(\scriptPWjjjjjsigma$\sigma$)\\
$W \jjjjjj$	& \scriptPWjjjjjj\ 	(\scriptPWjjjjjjsigma$\sigma$)\\
$e\met \jj$ 	& \scriptPemetjj\ 	(\scriptPemetjjsigma$\sigma$)\\
$e\met \jjj$  	& \scriptPemetjjj\ 	(\scriptPemetjjjsigma$\sigma$)\\ 
$e\met \jjjj$  	& \scriptPemetjjjj\ 	(\scriptPemetjjjjsigma$\sigma$)\\ 
\multicolumn{2}{c}{$Z$+jets-like} 	\\
$Z \jj$		& \scriptPZjj\ 	(\scriptPZjjsigma$\sigma$)\\
$Z \jjj$	& \scriptPZjjj\ 	(\scriptPZjjjsigma$\sigma$)\\
$Z \jjjj$	& \scriptPZjjjj\ 	(\scriptPZjjjjsigma$\sigma$)\\
$ee \jj$ 	& \scriptPeejj\		(\scriptPeejjsigma$\sigma$)\\
$ee \jjj$ 	& \scriptPeejjj\		(\scriptPeejjjsigma$\sigma$)\\
$ee \jjjj$ 	& \scriptPeejjjj\ 	(\scriptPeejjjjsigma$\sigma$)\\
$ee\met \jj$ 	& \scriptPeemetjj\ 	(\scriptPeemetjjsigma$\sigma$)\\
$ee\met \jjj$ 	& \scriptPeemetjjj\ 	(\scriptPeemetjjjsigma$\sigma$)\\
$ee\met \jjjj$ 	& \scriptPeemetjjjj\ 	(\scriptPeemetjjjjsigma$\sigma$)\\
$\mu\mu \jj$ 	& \scriptPmumujj\ 	(\scriptPmumujjsigma$\sigma$)\\
\multicolumn{2}{c}{$\ellgamma\ellgamma\ellgamma X$} \\
$eee$ & \scriptPeee\ 	(\scriptPeeesigma$\sigma$)\\
$Z\gamma$ & \scriptPZg\ 	(\scriptPZgsigma$\sigma$)\\
$Z\gamma j$ & \scriptPZgj\ 	(\scriptPZgjsigma$\sigma$)\\
$ee\gamma$ & \scriptPeeg\ 	(\scriptPeegsigma$\sigma$)\\
$ee\gamma\met$	& \scriptPeegmet\ 	(\scriptPeegmetsigma$\sigma$)\\
$e\gamma\gamma$ & \scriptPegg\ 	(\scriptPeggsigma$\sigma$)\\
$e\gamma\gamma j$ & \scriptPeggj\ 	(\scriptPeggjsigma$\sigma$)\\
$e\gamma\gamma \jj$ & \scriptPeggjj\ 	(\scriptPeggjjsigma$\sigma$)\\
$W\gamma\gamma$ & \scriptPWgg\ 	(\scriptPWggsigma$\sigma$)\\
$\gamma\gamma\gamma$ & \scriptPggg\ 	(\scriptPgggsigma$\sigma$)\\
\hline
\raisebox{-.6ex}{$\gothicP$} & \twiddleScriptPvalue\ (\twiddleScriptPvalueSigma$\sigma$)\\ 
\end{tabular}
\caption{Summary of results for populated final states.  The most interesting final state is found to be $ee\jjjj$, with $\scriptP=$ \scriptPeejjjj.  Upon taking into account the many final states we have considered using the curve in Fig.~\ref{fig:addtwiddle}, we find $\gothicP=$ \twiddleScriptPvalue.  The values of $\scriptP$ obtained in these final states are histogrammed in Fig.~\ref{fig:histogram_of_scriptps}, and compared to the distribution we expect from an ensemble of mock experimental runs.  No evidence for new high $p_T$ physics is observed in these data.}
\label{tbl:FinalResults}
\end{table}

{\dofig{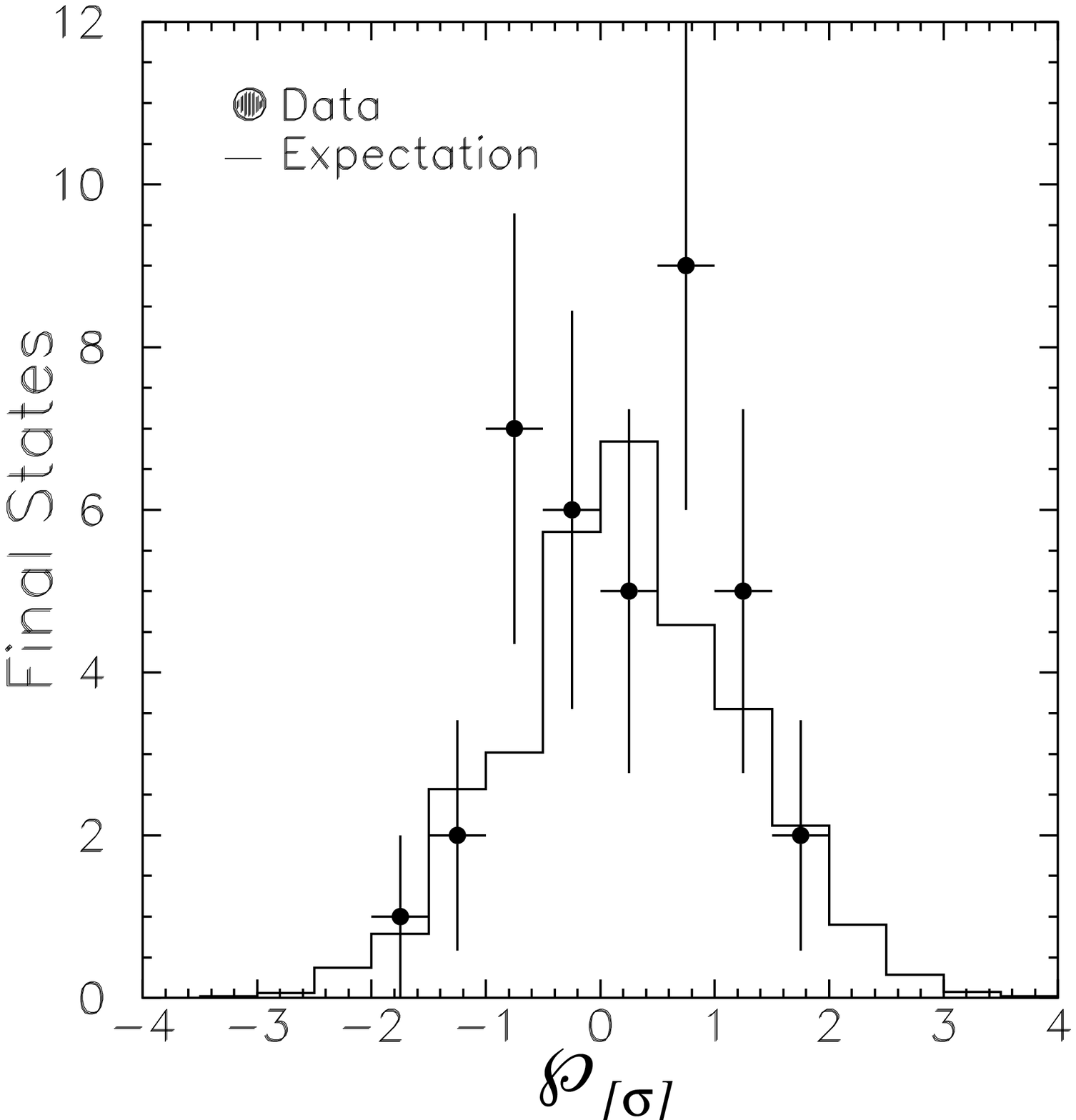} {3.5in} {Histogram of the $\scriptP$'s computed for the populated final states considered in this article.  The distribution agrees well with the expectation.} {fig:histogram_of_scriptps} }

Although no statistically significant indications of new physics are observed in this analysis, some final states appear to hold greater promise than others.  The smallest $\scriptP$'s (\scriptPeejjjj\ and \scriptPeemetjjjj) are found in the final states $ee\jjjj$ and $ee\met \jjjj$.  The kinematics of the events in these final states are provided in Appendix~\ref{section:InterestingEvents}.

It is very difficult to quantify the sensitivity of \Sherlock\ to arbitrary new physics, since the sensitivity necessarily depends on the characteristics of that new physics.  We have provided examples of \Sherlock's performance on ``typical,'' particular signatures.  This function is served by the sensitivity checks provided at the end of each of Secs.~\ref{section:WJ}--\ref{section:zoo}.  In the analysis of the $e\mu X$ data in Ref.~\cite{SherlockPRD}, our signal was first $WW$ and $t\bar{t}$ together, and then only $t\bar{t}$.  This was a difficult signal to find, for although both $WW$ and $t\bar{t}$ cluster in the upper right-hand corner of the unit box, as desired, we expect only 3.9 $WW$ events in $e\mu\met$ (with a background of 45.6 events), and 1.8 $t\bar{t}$ events in $e\mu\met \jj$ (with a background of 3.4 events).  We were able to consistently find indications of the presence of $WW$ and $t\bar{t}$ in an ensemble of mock experiments, but we would not have been sufficiently sensitive to claim a discovery.

In the $W$+jets-like final states we again chose $t\bar{t}$ for our sensitivity check.  This was both a natural sequel to the sensitivity check in $e\mu X$ and a test of \Sherlock's performance when the signal populates the high tails of only a subset of the variables considered.  We find $\scriptP_{\rm min}>3\sigma$ in \fractionOfNailedTopinWjjjXhsers\ of an ensemble of mock experimental runs containing $t\bar{t}$ events on the final states $W\jjj$, $W\jjjj$, $W\jjjjj$, and $W\jjjjjj$, compared with only $0.5\%$ of an ensemble of mock experimental runs containing background only.

In the $Z$+jets-like final states we considered a leptoquark signal.  This is in many ways an ideal signature ---  a relatively large number of events (about six) are predicted, and the signal appears in the high tails of both variables under consideration.  \Sherlock\ finds $\scriptP>3.5\sigma$ in over 80\% of the mock experiments performed.

Finally, in the final states $\ellgamma\ellgamma\ellgamma X$ we introduced the mapping between $\scriptP_{\rm min}$ and $\gothicP$ and briefly discussed its interpretation.  The generic sensitivity check we considered [$X'\rightarrow \ellgamma\ellgamma\ellgamma X$] demonstrates the advantages of considering exclusive final states.  While the other sensitivity checks rely heavily upon the \Sherlock\ algorithm, this check shows that a careful and systematic definition of final states by itself can lead to a discovery with only a few events.
\section{Conclusions}
\label{section:Conclusions}

We have applied the \Sherlock\ algorithm to search for new high $p_T$ physics  in data spanning over \numberOfFinalStates\ exclusive final states collected by the D\O\ experiment during Run I of the Fermilab Tevatron.  A quasi-model-independent, systematic search of these data has produced no evidence of physics beyond the standard model.

\section*{Acknowledgments}
%
We thank the staffs at Fermilab and collaborating institutions, 
and acknowledge support from the 
Department of Energy and National Science Foundation (USA),  
Commissariat  \` a L'Energie Atomique and 
CNRS/Institut National de Physique Nucl\'eaire et 
de Physique des Particules (France), 
Ministry for Science and Technology and Ministry for Atomic 
   Energy (Russia),
CAPES and CNPq (Brazil),
Departments of Atomic Energy and Science and Education (India),
Colciencias (Colombia),
CONACyT (Mexico),
Ministry of Education and KOSEF (Korea),
CONICET and UBACyT (Argentina),
The Foundation for Fundamental Research on Matter (The Netherlands),
PPARC (United Kingdom),
A.P. Sloan Foundation,
and the A. von Humboldt Foundation.

\appendix
\section{Definitions of final states}
\label{section:finalStateDefinitions}

This appendix reviews the definitions of final states provided in Ref.~\cite{SherlockPRD}.  The specification of the final states is based on the notions of exclusive channels and standard particle identification.  We partition the data into exclusive final states because the presence of an extra object (electron, photon, muon, \ldots) in an event often qualitatively changes the probable interpretation of the event and the variables that naturally characterize the final state, and because using inclusive final states can lead to ambiguities when different channels are combined.

We attempt to label these exclusive final states as completely as possible while maintaining a high degree of confidence in the label.  We consider a final state to be described by the number of isolated electrons, muons, photons, and jets observed in the event, and whether there is a significant imbalance in transverse momentum.  We treat $\met$ as an object in its own right, which must pass certain quality criteria.  In Run I D\O\ was unable to efficiently differentiate among jets arising from $b$ quarks, $c$ quarks, light quarks, and hadronic tau decays.  We consider final states that are related through global charge conjugation to be equivalent in $p\bar{p}$ or $e^+e^-$ (but not $pp$) collisions.  Thus in principle $e^+e^-\gamma$ is a different final state than $e^+e^+\gamma$, but $e^+e^+\gamma$ and $e^-e^-\gamma$ together make up a single final state.  D\O\ lacked a central magnetic field in Run I, so we choose not to distinguish between $e^+/e^-$ or $\mu^+/\mu^-$.  In events containing two same-flavor leptons, we assume that they are of opposite charge.

We combine an $e^+e^-$ pair into a $Z$ boson if their invariant mass $m_{e^+e^-}$ falls within a $Z$ boson mass window ($82 \leq m_{e^+e^-} \leq 100$~GeV) and the event contains neither significant $\met$ nor a third charged lepton.  A $\mu^+\mu^-$ pair is combined into a $Z$ boson if the event can be fit to the hypothesis that the two muons are decay products of a $Z$ boson and that the $\met$ in the event is negligible, and if the event contains no additional charged lepton.  If the event contains exactly one photon in addition to a $\ell^+\ell^-$ pair, and contains neither significant $\met$ nor a third charged lepton, and if $m_{\ell^+\ell^-}$ does not fall within the $Z$ boson mass window, but $m_{\ell^+\ell^- \gamma}$ does, then the $\ell^+\ell^-\gamma$ triplet becomes a $Z$ boson.  An electron and $\met$ become a $W$ boson if the transverse mass $m^T_{e\met}$ is within a $W$ boson mass window ($30 \leq m^T_{e\met} \leq 110$~GeV) and the event contains no second charged lepton.  A muon and $\met$ in an event with no second charged lepton are always combined into a $W$ boson; due to our more modest muon momentum resolution, no mass window is imposed.  Because the $W$ boson mass window is so much wider than the $Z$ boson mass window, no attempt is made to identify radiative $W$ boson decays.

\section{Examples of signals that might appear}
\label{section:SignalsThatMightAppear}

In this section we provide a few examples of signals that might have been discovered in the course of this analysis.  This discussion is provided to give the reader a taste of the many processes that might appear in the final states we have analyzed, and is by no means intended to be complete.  The possibility that the correct answer is ``none of the following'' is one of the strongest motivations for pursuing a quasi-model-independent search.

\subsection{$\lowercase{e}\mu X$}
\label{section:SignalsThatMightAppearInemuX}

In supersymmetric models (denoting the supersymmetric particles as in Ref.~\cite{SusyReview}), the process $q\bar{q}\rightarrow Z/\gamma^* \rightarrow \gaugino_1^\pm \gaugino_1^\mp \rightarrow e \mu \nu \nu \gaugino_1^0 \gaugino_1^0$ can produce events appearing in the $e\mu\met$ final state.  More generally, any process involving the production of two charginos has the potential for producing a final state containing an electron, a muon, and $\met$.  This final state may also be reached through the leptonic decays of two taus, obtained (for example) from the production of two $\stau$ particles that each decay to $\tau\gaugino_1^0$, or from the production of a heavy $Z$-like object that couples strongly to the third generation.  
An anomalous correction to the standard model $WW\gamma$ vertex or anomalies involving the top quark could also appear in these final states.

\subsection{Final states already considered}
\label{section:SignalsThatMightAppearOtherFinalStates}

A sampling of the types of new physics that might appear in a few of the final states described in Sec.~\ref{section:finalStatesAlreadyConsidered} is provided here.

{$\jj$.}
The dijet final state could contain hints of a massive object (such as an additional neutral gauge boson) produced through $q\bar{q}$ annihilation and decaying back into $q\bar{q}$.  It could also contain indications that quarks are in fact composite objects, interacting through terms in an effective Lagrangian of the form $\frac{c}{\Lambda^2}q\bar{q}q'\bar{q}'$, where $\Lambda \gtapprox 1$~TeV is a compositeness scale and $c$ is a constant of order unity.

{$e\met$.}
Models containing symmetry groups larger than the SU(3)$_C\times$SU(2)$_L\times$U(1)$_Y$ group of the standard model often contain an additional SU(2) group, suggesting the existence of a heavy $W$-like gauge boson ($W'$) that would decay into the $e\met$ final state, with the transverse mass of the electron and neutrino greater than that expected for the standard model $W$.  Production of $\slepton\sneutrino$ decaying to $\ell\gaugino_1^0\nu\gaugino_1^0$ could also produce events in this final state, as could production of $\gaugino_1^\pm \gaugino_2^0$ decaying to $\ell\nu\gaugino_1^0 \nu\nu\gaugino_1^0$.

{$ee$.}
If both quarks and leptons are composite objects, there will be four-fermion contact terms of the form $\frac{c}{\Lambda^2} q\bar{q}\ell^+\ell^-$ in addition to the $\frac{c}{\Lambda^2} q\bar{q}q'\bar{q}'$ terms postulated in the discussion of the $\jj$ final state above.  Such an interaction would produce events with large transverse momentum, opposite-sign leptons, and should appear in the $ee$ and $\mu\mu$ final states.  Some models that employ a strong dynamics to break electroweak symmetry predict the existence of composite ``techni-''particles, such as the $\techniomega$, $\technirho$, and $\technipi$, that are analogous to the composite $\omega$, $\rho$, and $\pi$ mesons that arise from confinement in QCD.  The technirho ($\technirho$) and techniomega ($\techniomega$), if produced, will decay into an $\ell^+\ell^-$ pair if their preferred decay mode to technipions ($\technipi$) is kinematically forbidden.  Such events will appear as a bump in the tail of the $ee$ invariant mass distribution and as an excess in the tail of the electron $p_T$ distribution.  Models containing symmetry groups larger than that of the standard model typically contain a heavy neutral boson (generically called a $Z'$) in addition to the $W'$ boson described above.  If this $Z'$ boson couples to leptons, the process $q\bar{q} \rightarrow Z' \rightarrow \ell\ell$ could produce a signature similar to that expected from the decay of a $\technirho$ or $\techniomega$.

\subsection{$W$+jets-like final states}
\label{section:SignalsThatMightAppearInWjj(nj)}

A variety of new signals have been predicted that would manifest themselves in the $W$+jets-like final states --- those final states containing events with a single lepton, missing transverse energy, and zero or more jets.  A plethora of supersymmetric signatures could appear in these states.  A chargino and neutralino, produced from $q\bar{q}$ through an $s$-channel $W$ boson, can proceed to decay as $\gaugino_1^\pm \rightarrow \ell\nu\gaugino_1^0$ and $\gaugino_2^0 \rightarrow q\bar{q}\gaugino_1^0$, leaving an event that will be partitioned into either the $e\met \jj$ or $W\jj$ final state.  Pair production of top squarks, with $\scalartop\rightarrow b \gaugino_1^\pm$ and subsequent decays of the charginos to $e\nu\gaugino_1^0$ and $qq'\gaugino_1^0$, will produce events likely to fall into the $e\met \jjjj$ or $W\jjjj$ final states.  Depending upon the particular model, even gluino decays can give rise to leptons.  Events with gluinos that are pair-produced and decay, one into $qq' \gaugino_1^\pm$ and the other into $q\bar{q}\gaugino_1^0$, can also find themselves in the $e\met \jjjj$ or $W \jjjj$ final state.  Other possible decays of the supersymmetric spectrum allow many more signals that might populate these final states.  

 The decay of a $\technirho^+$, produced by $q\bar{q}$ annihilation, can produce a $W^+$ boson and a $\technipi^0$, which in turn may decay to $b\bar{b}$ or $gg$.  Such an event should appear in the high tails of the $p_T^W$ and $\sum'{p_T^j}$ distributions in our analysis of the $W\jj$ final state if the technipion is sufficiently massive.  The same final state may also be reached by the process $q\bar{q} \rightarrow \technirho^0 \rightarrow W^- \technipi^+ \rightarrow \ell^- \nu c\bar{b}$.  A neutral color-octet technirho ($\rho_{T8}^0$) produced by $q\bar{q}$ annihilation can decay to two technipions carrying both color and lepton quantum numbers ($\pi_{\rm LQ}$), each of which in turn decays preferentially into a massive quark and a massive lepton.  If the technipion is heavier than the top quark then the decay $\pi_{\rm LQ}\rightarrow t\tau$ or $t\nu_\tau$ is kinematically allowed.  Appropriate decays of the $W$ bosons from the two top quarks leave the event containing one high transverse momentum lepton, substantial $\met$, and several energetic jets. 

The standard model contains three generations of quarks and leptons, but there appears to be no fundamental reason that Nature should choose to stop at three.  A massive charge $-1/3$ fourth-generation quark ($b'$), which could be pair-produced at the Tevatron, would be apt to decay weakly into a $W$ boson and a top quark.  Events in which one of the four $W$ bosons then decays leptonically will result in a final state containing one lepton, substantial missing transverse energy, and many jets.  

Leptoquarks, a consequence of many theories that attempt to explain the peculiar symmetry between quarks and leptons in the standard model, could also be pair-produced at the Tevatron.  If their branching ratio to charged leptons $\beta=0.5$ then the pair will decay to $\ell\nu q\bar{q}$ 50\% of the time, resulting in events that will be classified either as $e\met \jj$ or $W\jj$.  

Models invoking two Higgs doublets predict a charged Higgs that may appear in occasional decays of the top quark.  In such models a top quark pair, produced by $q\bar{q}$ or $gg$ annihilation, can decay into $H^+b W^-\bar{b}$.  Depending upon the mass of the charged Higgs particle, it may decay into $W^+b\bar{b}$, $c\bar{s}$, or $\tau^+\nu$.  Appropriate decay of the $W$ boson(s) in the event will result in the event populating one of the $W\jj(nj)$ final states.  Other predictions abound.

\subsection{$Z$+jets-like final states}
\label{section:SignalsThatMightAppearInZjj(nj)}

Just as in the $W$+jets-like final states, there are a host of theoretical possibilities for new physics in the $Z$+jets-like final states.  Although some of these processes involve the production of two same-flavor, opposite-sign leptons via the production of a standard model $Z$ boson, many others involve particles that decay to leptons of different flavor, or with the same charge.  These different possibilities typically are partitioned into different final states according to our prescription:  events that contain leptons of different flavor (those within $e\mu X$) are considered in Sec.~\ref{section:emuX}; events containing leptons of similar charge (e.g., an $e^+e^+\jj$ event) would in principle be partitioned into different final states than events containing leptons of opposite charge (e.g., an $e^+e^-\jj$ event) if D\O\ distinguished electron charge; and events in which the leptons have an invariant mass consistent with the hypothesis that they are the decay products of a $Z$ boson are partitioned into different final states than those with a dilepton invariant mass outside the $Z$ boson mass window.    

Models containing supersymmetry and imposing conservation of $R$-parity predict signatures containing substantial missing transverse energy.  Such events might therefore populate the $ee\met \jj(nj)$ or $\mu\mu\met \jj(nj)$ channels.  Final state leptons may be obtained in supersymmetric models from the decays of neutralinos (which can produce two same-flavor, oppositely-charged leptons), or charginos or sleptons (which decay into a single charged lepton and missing transverse energy).  The process $q\bar{q}' \rightarrow W^* \rightarrow \gaugino_1^\pm \gaugino_2^0$, with subsequent decay of the chargino to $qq' \gaugino_1^0$ and the neutralino to $\ell^+\ell^- \gaugino_1^0$, results in an event with two same-flavor, opposite-sign leptons, two jets, and missing transverse energy, and would appear in our $ee\met \jj$ or $\mu\mu \jj$ final states.  Events in which gluinos are pair-produced and decay via $\gluino\rightarrow qq'\gaugino_1^\pm$ will appear in the $ee\met \jjjj$ and $\mu\mu \jjjj$ final states when the gaugino decays to $\ell\nu\gaugino_1^0$.  Pair production of scalar top quarks ($q\bar{q}/gg \rightarrow g \rightarrow \scalartop\scalartop^*$) that decay via $\scalartop\rightarrow b \gaugino_1^\pm$ and $\gaugino_1^\pm \rightarrow \ell\nu\gaugino_1^0$ again produce events that populate the $ee\met \jj$ and $\mu\mu \jj$ final states, in addition to the $e\mu\met \jj$ final states already considered.  If $R$-parity is violated, then supersymmetric signals could populate final states without missing transverse energy.  Pair production of gluinos decaying to $\bar{c}\tilde{c}_L$ could produce events that land in the $ee\jjjj$ final state if the $R$-parity-violating decay $\tilde{c}_L\rightarrow e^+d$ is allowed.

Color-octet models predict the existence of a color-octet technirho, which can decay to $\pi_{\rm LQ} \pi_{\rm LQ}$.  These technipions decay preferentially to massive particles, like the color-singlet $\technipi$, but their decay products will carry both color and lepton quantum numbers.  Events in which each $\pi_{\rm LQ}$ decays to a $b$ quark and a $\tau$ lepton will populate $ee\met \jj$ and $\mu\mu \jj$ final states, among others.  Leptoquarks motivated by grand unified theories could be pair-produced at the Tevatron via $q\bar{q} \rightarrow Z/\gamma^* \rightarrow LQ\overline{LQ}$, and might populate the final states $ee\jj$ and $\mu\mu \jj$. Again, other examples abound. 

\subsection{$\ellgamma\ellgamma\ellgamma X$}
\label{section:SignalsThatMightAppearInlllX}

There are few standard model processes that produce events in which the sum of the numbers of electrons, muons, and photons is $\ge3$.  The $\ellgamma\ellgamma\ellgamma X$ final states are therefore quite clean, and the presence of even a few events in any of these states could provide a strong indication of new physics.

Supersymmetric models predict a variety of possible signatures in these states.  Those models in which $R$-parity is conserved produce events with missing transverse energy in addition to three $\ellgamma$ objects.  Models in which the lightest neutralino ($\gaugino_1^0$) is the lightest supersymmetric particle (LSP) usually produce final states without photons.  This case occurs for many models in which the supersymmetry is broken in a hidden sector and communicated to the visible sector through gravitational forces (gravity-mediated supersymmetry breaking).  Models in which the gravitino ($\gravitino$) is the LSP often produce final states with photons from the decay of the next-to-lightest supersymmetric particle via (for example) $\gaugino_1^0 \rightarrow \gamma\gravitino$.  This case, in turn, obtains for many models in which the breaking of the supersymmetry is mediated by gauge fields (gauge-mediated supersymmetry breaking).  For example, the production of a chargino and neutralino through $q\bar{q}$ annihilation into a virtual $W$ boson can produce events in these final states through the decays $\gaugino_1^\pm\rightarrow\ell\nu\gaugino_1^0$ and $\gaugino_2^0\rightarrow\ell\ell\gaugino_1^0$ if the lightest neutralino is the LSP, or through the decays $\gaugino_1^\pm \rightarrow e\nu\gaugino_1^0$, $\gaugino_2^0 \rightarrow q\bar{q}\gaugino_1^0$, and $\gaugino_1^0\rightarrow \gamma\gravitino$ if the gravitino is the LSP.

Charginos can be pair-produced in the reaction $q\bar{q}\rightarrow Z/\gamma^*\rightarrow \gaugino_1^\pm \gaugino_1^\mp$.  If they decay to $e \nu \gaugino_2^0$, and if $\gaugino_2^0$ in turn decays to $\gamma \gaugino_1^0$, these events will populate the final state $ee\gamma\gamma\met$.  
The production of slepton pairs can also result in events falling into the  final state $ee\gamma\gamma\met$, since a typical decay of a selectron in a model with gravity-mediated supersymmetry breaking is $\selectron \rightarrow e\gaugino_2^0$, with $\gaugino_2^0\rightarrow \gamma\gaugino_1^0$.    If a pair of sufficiently massive sleptons are produced, each can decay into the corresponding standard model lepton and the second-lightest neutralino ($\gaugino_2^0$), which in turn could decay into $\ell\ell\gaugino_1^0$.  A similar production of $\slepton\sneutrino$ can easily lead to a final state with one fewer charged lepton, through the decay chain $\sneutrino \rightarrow \ell\gaugino_1^\pm$, and $\gaugino_1^\pm\rightarrow \ell\nu\gaugino_1^0$.  The standard model backgrounds to such events, containing five or more charged leptons and substantial missing transverse energy, are vanishingly small.  Events with four charged leptons and substantial $\met$ could result from the decay of a $\gaugino_2^0 \gaugino_2^0$ pair, in which each $\gaugino_2^0$ decays to $\ell\ell\gaugino_1^0$.  Even pair production of gluinos, each decaying to $q\bar{q}\gaugino_2^0$, with one neutralino decaying to $ee\gaugino_1^0$ and the other to $\gamma\gaugino_1^0$, could produce events in these final states.  With this particular decay, such events would appear in the final state $ee\gamma \jj$.

If leptons exist in excited states several hundred GeV above their ground state, just as hadrons exist in excited states at energy scales a thousand times smaller, they could be produced in the process $q\bar{q}\rightarrow Z/\gamma^* \rightarrow \ell^* \ell^*$ or $q\bar{q}'\rightarrow W^* \rightarrow \ell^*\nu^*$.  The excited leptons can decay by emitting a photon, so that $\ell^* \rightarrow \ell\gamma$ and $\nu^* \rightarrow \nu\gamma$.  Such events would populate the $\ell\ell\gamma\gamma$ and $\ell\met\gamma\gamma$ final states.  If the technirho exists and is sufficiently massive, it can decay to $WZ$.  Roughly one time in fifty both the $W$ and $Z$ bosons will decay to leptons, producing a $\ell^+\ell^-\ell'\met$ event.  More generally, any process producing anomalous triboson couplings will affect the $\ellgamma\ellgamma\ellgamma X$ final states, and (as we show in Sec.~\ref{section:lllSensitivityCheck}) our method is likely to be sensitive to such a signal.  

\section{Comparison of distributions}
\label{section:Distributions}

In this appendix we show kinematic distributions of the data and expected backgrounds for the most heavily populated final states that we have considered.  Figures~\ref{fig:wjj_prd_plot}--\ref{fig:wjjjj_prd_plot} show good agreement between data and the expected background in a number of distributions for the heavily populated $W$+jets-like final states $W\jj$, $W\jjj$, and $W\jjjj$.  Figures~\ref{fig:zjj_prd_plot} and~\ref{fig:zjjj_prd_plot} serve the same function for the final states $Z\jj$ and $Z\jjj$.

{\dofig{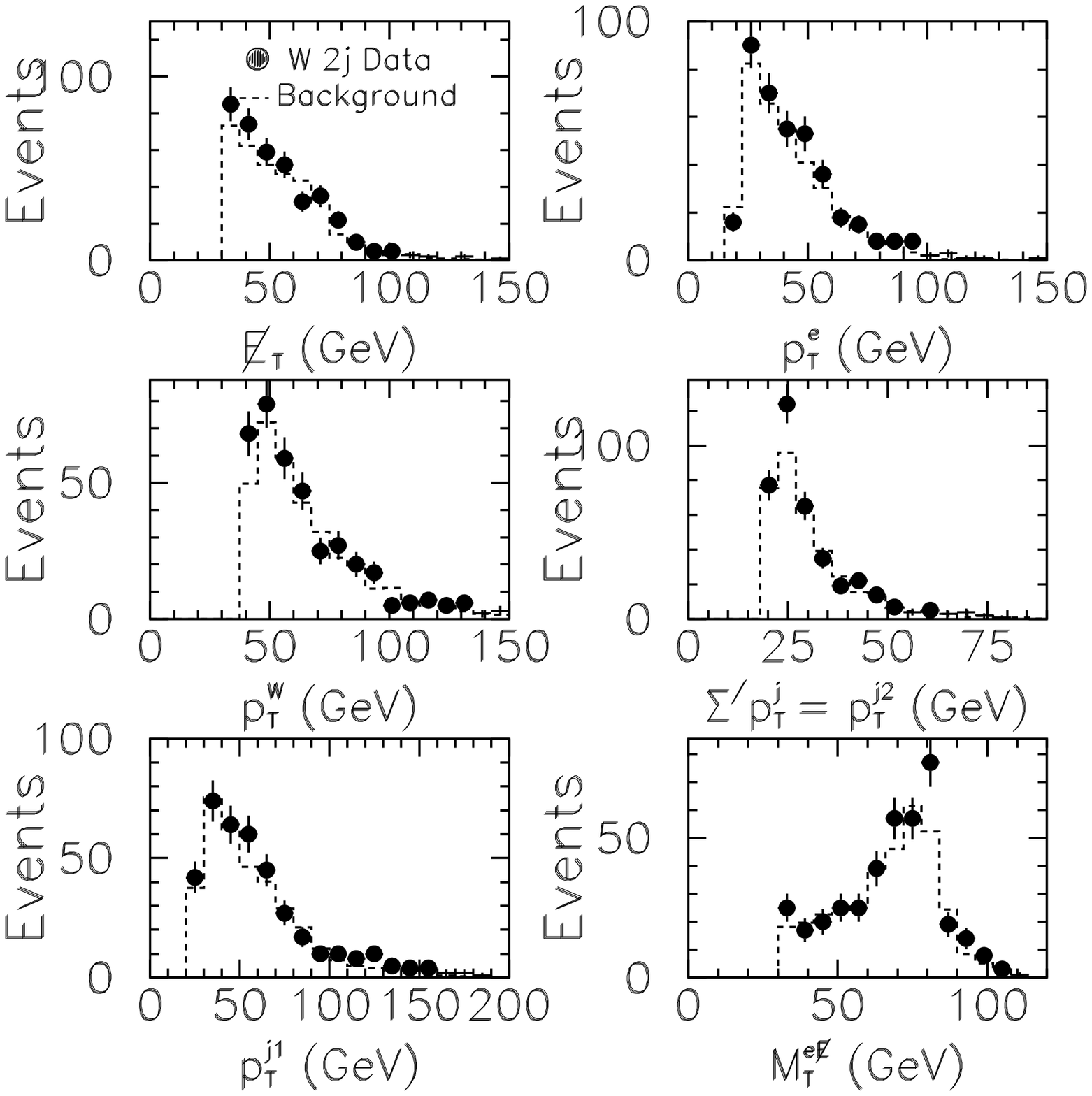} {3.5in} {Comparison of background to data for $W\jj$.} {fig:wjj_prd_plot} }
{\dofig{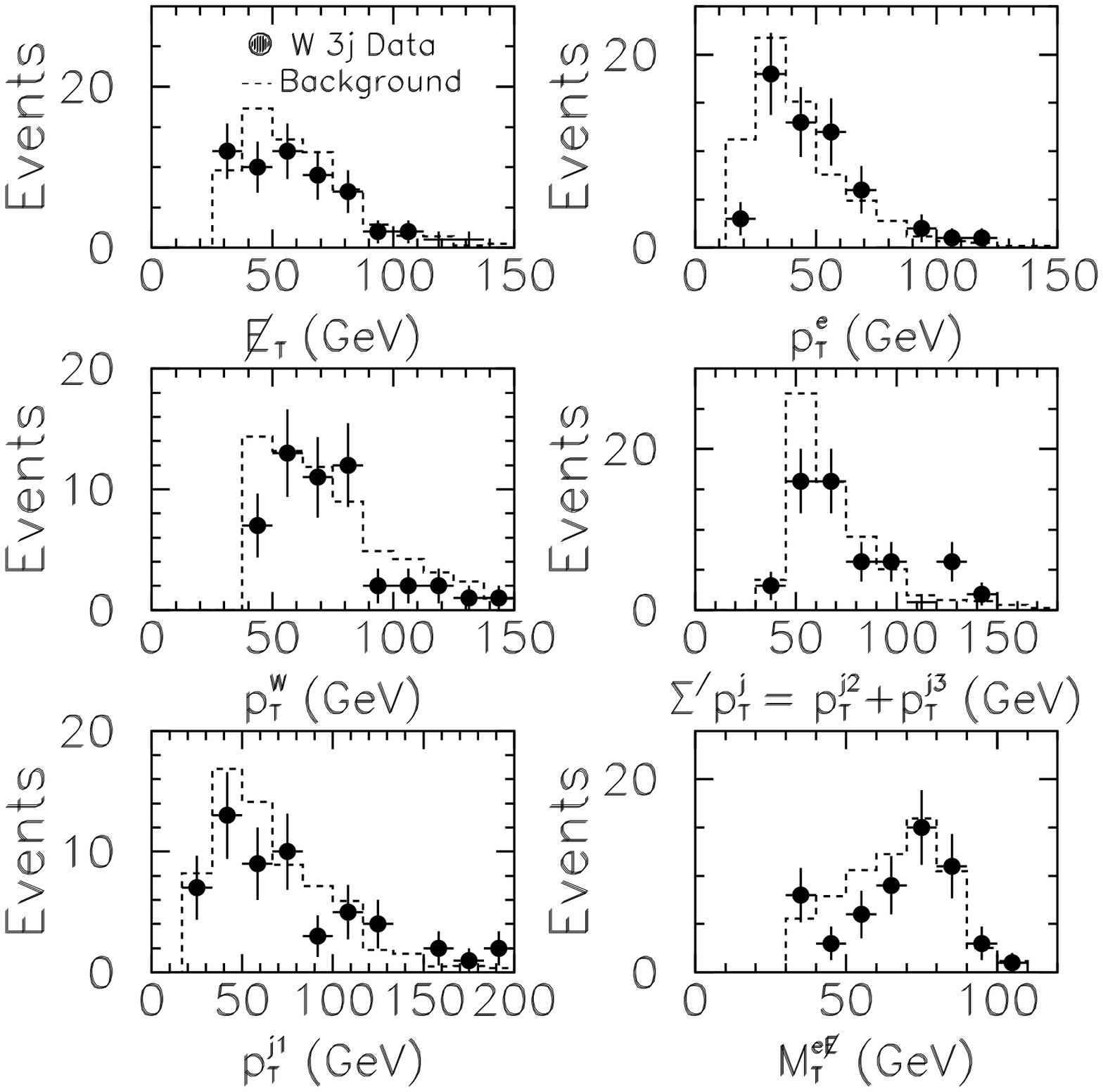} {3.5in} {Comparison of background to data for $W\jjj$.} {fig:wjjj_prd_plot} }
{\dofig{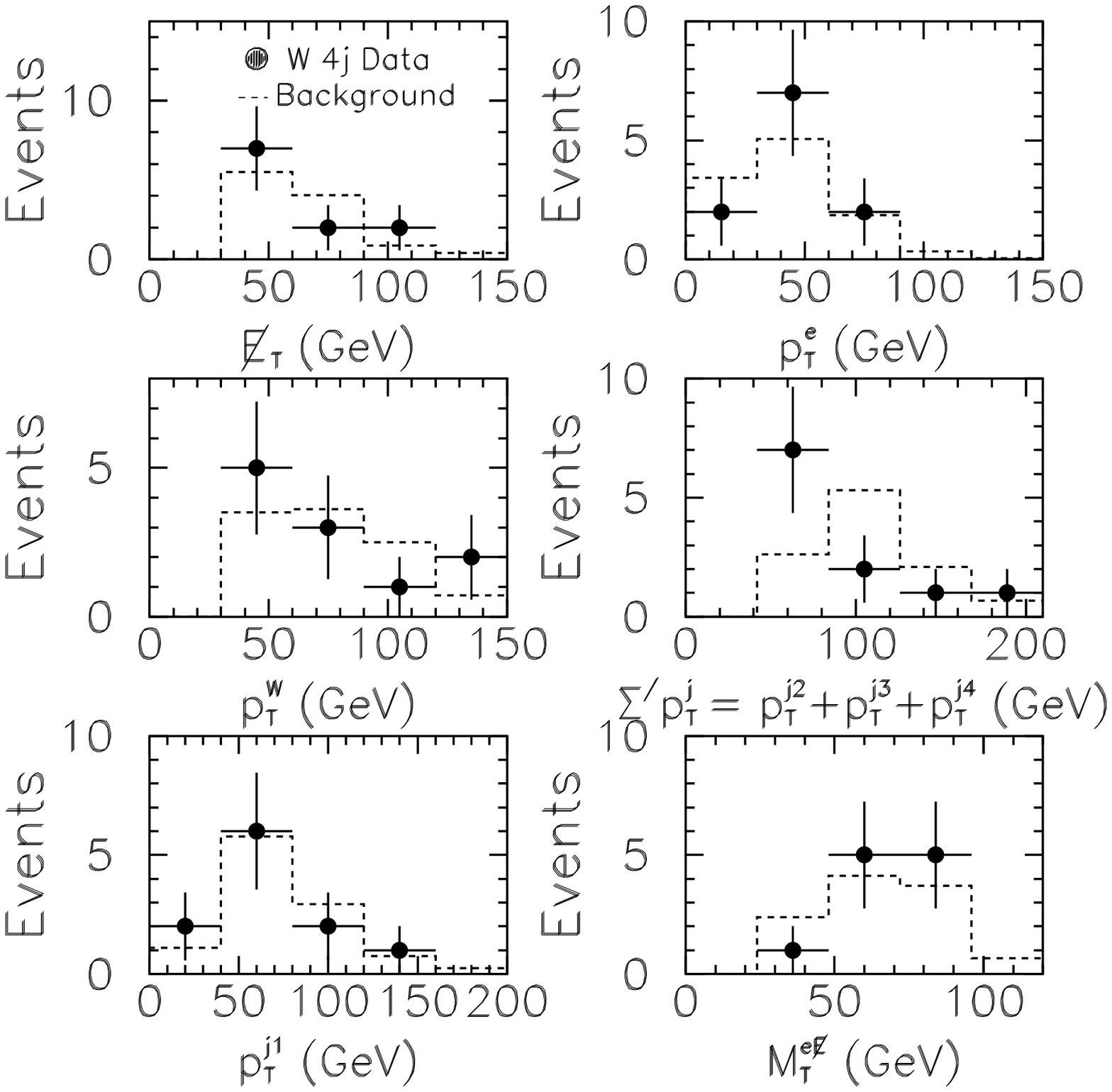} {3.5in} {Comparison of background to data for $W\jjjj$.} {fig:wjjjj_prd_plot} }

{\dofig{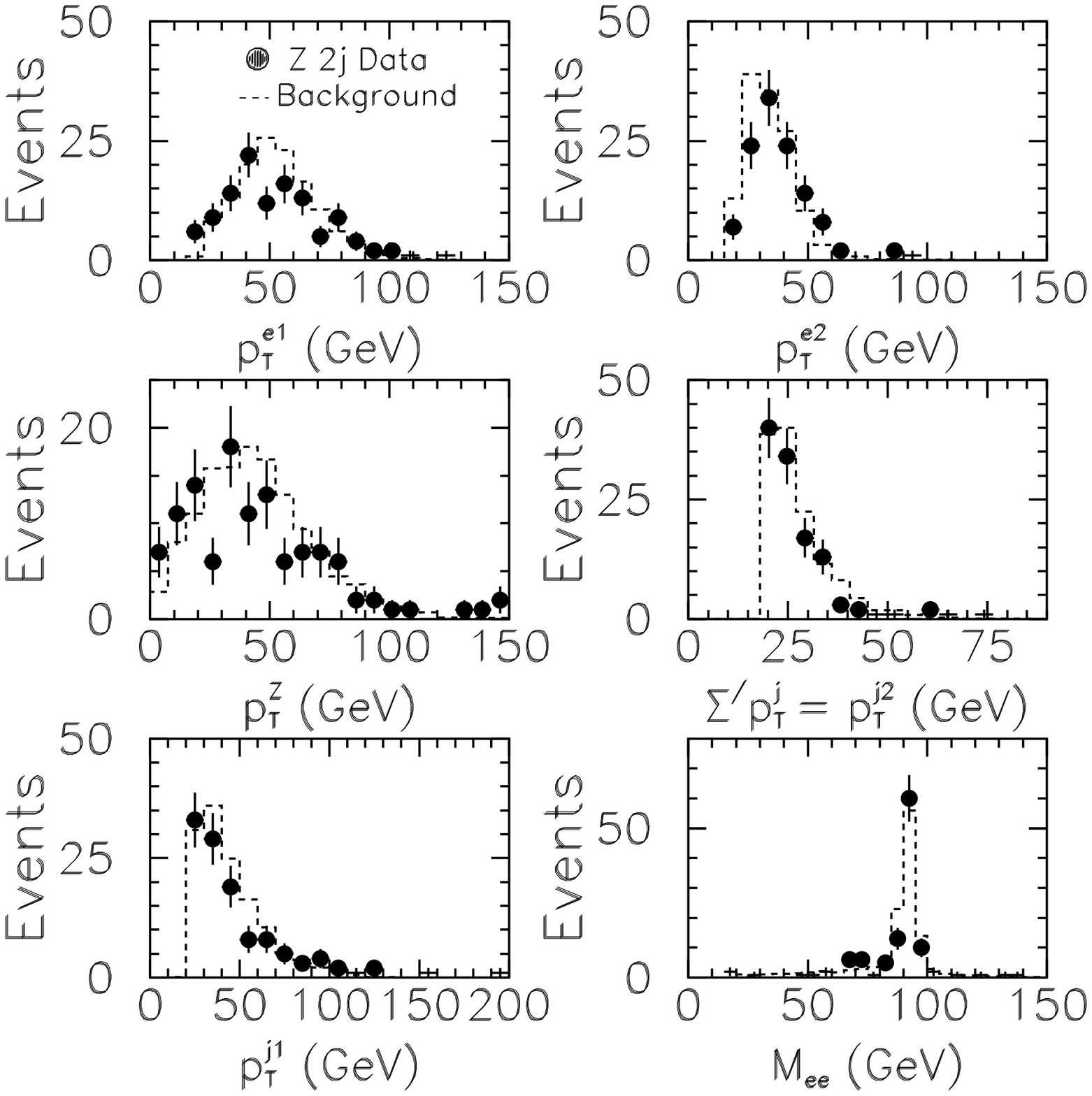} {3.5in} {Comparison of background to data for $Z\jj$.} {fig:zjj_prd_plot} }
{\dofig{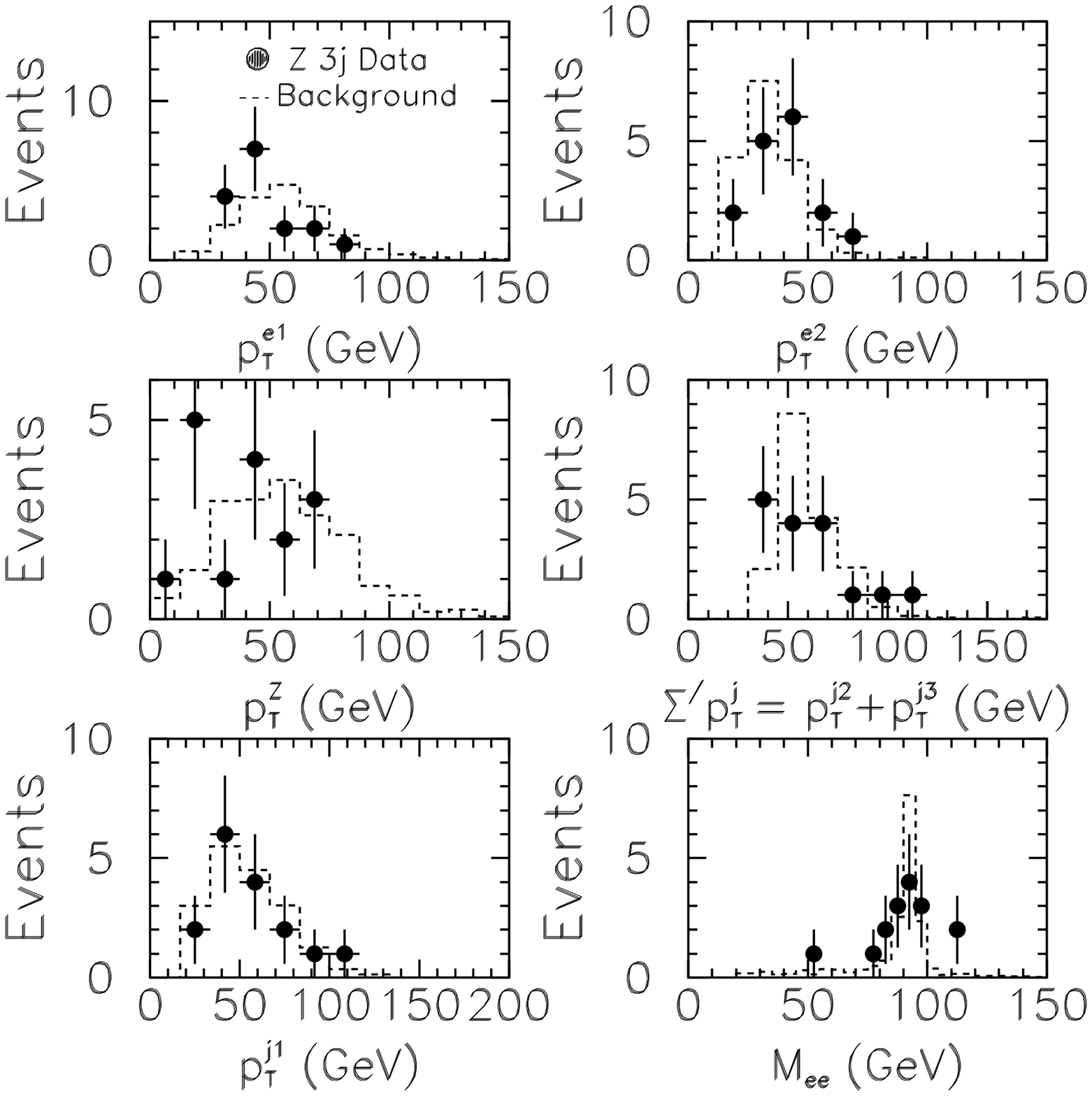} {3.5in} {Comparison of background to data for $Z\jjj$.} {fig:zjjj_prd_plot} }

\section{{\boldmath $\met$} significance}
\label{section:pmet}

We determine the significance of any missing transverse energy in an event in the $Z$+jets-like final states by computing a probability density $p(\met)$.  This is a true probability density in the sense that, for a given event, the probability that the actual missing transverse energy in that event is between $\met$ and $\met \! + \! \delta \met$ is given by $p(\met) \delta \met $.  This density is computed with a Monte Carlo calculation.  For each data event we generate an ensemble of events similar to the original but with the energies of the objects smeared according to their resolutions.  Jets are smeared with a Gaussian of width $\sigma = 80\%\sqrt{E}$, and electrons are smeared with a Gaussian of width $\sigma = 20\%\sqrt{E}$ (a slight inflation of the measured resolution of $15\%\sqrt{E}$), where $E$ is the energy of the object in GeV.  The component of the missing transverse energy $\met_a$ along the direction of the original $\met$ is recalculated for each smeared event, and the values that are obtained are histogrammed.  The histogram is then smoothed, and the likelihood
\begineq
{\cal L}_\met = \frac{p(\met_a)_{\rm max}}{p(\met_a=0)}
\endeq
is calculated.  Studies have shown that a cut of $\log_{10}{{\cal L}_\met}>3$ does an excellent job of retaining events with true $\met$ while rejecting QCD background.

\section{Kinematics of interesting events}
\label{section:InterestingEvents}

Table~\ref{tbl:interestingEvents_kinematics} provides information about the events in the most interesting final states seen in the course of this analysis.  Invariant masses of objects in these events are given in Table~\ref{tbl:interestingEvents_invariantMasses}.

\begin{table}[htb]
\centering
\begin{tabular}{ccccr}
run:event 	& object & $p_T$ (GeV) & $\phi$ & \multicolumn{1}{c}{$\eta$} \\ \hline
~\\
\multicolumn{5}{l}{$ee\jjjj$} \\   
~\\
85918:12437 	& $e$	& 58.0  &   0.74  &  $-0.42$    \\
		& $e$	& 37.9  &   0.30  &  $-1.51$    \\
		& $j$	& 89.0  &   3.94  &  $-0.10$    \\
		& $j$	& 26.0  &   4.20  &  $-0.98$    \\
		& $j$	& 21.3  &   2.55  &  $-1.25$    \\
		& $j$	& 21.2  &   2.07  &  $ 0.77$	\\
90278:31411 	& $e$	& 53.1  &   4.15  &  $ 0.00$    \\
		& $e$	& 33.6  &   0.28  &  $-1.85$    \\
		& $j$	& 80.2  &   0.78  &  $ 1.24$    \\
		& $j$	& 39.9  &   4.46  &  $ 1.81$    \\
		& $j$	& 34.0  &   2.94  &  $-1.55$    \\
		& $j$	& 24.2  &   2.92  &  $ 0.05$    \\
92746:25962 	& $e$	& 64.6  &   1.99  &  $ 0.99$    \\
		& $e$	& 40.6  &   5.72  &  $ 0.55$    \\
		& $j$	& 26.8  &   3.84  &  $-2.13$    \\
		& $j$	& 25.6  &   4.83  &  $ 0.49$    \\
		& $j$	& 20.0  &   5.73  &  $-1.12$    \\
		& $j$	& 21.5  &   1.86  &  $ 2.62$   \\
~\\
\multicolumn{5}{l}{$ee\met \jjjj$} \\   
~\\
89815:17253 	& $e$	& 87.7  &   5.93  &  $ 1.00$    \\
		& $e$	& 22.5  &   4.19  &  $ 1.33$    \\
		& $\met$ & 59.8 &   0.97  &   $-$      \\
		& $j$	& 69.8  &   2.42  &  $-1.33$    \\
		& $j$	& 53.1  &   2.88  &  $ 0.36$    \\
		& $j$	& 52.2  &   4.27  &  $-1.30$    \\
		& $j$	& 25.4  &   5.81  &  $-0.18$    \\
\end{tabular}
\caption{Kinematic properties of the most interesting events seen in this analysis.}
\label{tbl:interestingEvents_kinematics}
\end{table}

\begin{table}[htb]
\centering
\begin{tabular}{ccccc}
run:event & $m_{ee}$ & \multicolumn{2}{c}{$m_T^{e\met}$} & $m_{\jjjj}$ \\ \hline
~\\ $ee\jjjj$ 	\\ ~\\
85918:12437 	& 57.4 	& && 149  \\
90278:31411 	& 119.5 & && 342  \\
92746:25962 	& 100.6 & && 323  \\ 
~\\ $ee\met \jjjj$	\\ ~\\
89815:17253 	& 69.4 	&  89.0	& 73.3 & 239 \\ 
\end{tabular}
\caption{Invariant masses (in units of GeV) of objects in the most interesting events seen in this analysis.}
\label{tbl:interestingEvents_invariantMasses}
\end{table}

\bibliographystyle{unsrt}

\begin{thebibliography}{2}
\bibitem{PseudoRapidity}
{We distinguish between ``physics $\eta$,'' defined as $\eta = -\ln\tan(\theta/2)$ where $\theta$ is the polar angle as measured from the interaction point, and ``detector $\eta$,'' defined as $\eta_{\rm det} = -\ln\tan(\theta_{\rm det}/2)$ where $\theta_{\rm det}$ is the polar angle as measured from the center of the detector.}
\end{thebibliography}

\end{document}